\documentclass[12pt,letterpaper]{article}
\usepackage{jheppub}
\usepackage{epsfig}
\usepackage{amssymb,amsfonts}
\usepackage{color}
\usepackage[usenames,dvipsnames]{xcolor}

\newcommand{\MB}[1]{} 
\newcommand{\RD}[1]{} 
\newcommand{\DV}[1]{} 

\newcommand\comment[1]{}

\newcommand\om{\omega}

\newcommand\ov{\over }

\def\le{\left}
\def\ri{\right}

\def\({\left(}
\def\){\right)}
\def\[{\left[}
\def\]{\right]}
\def\<{\langle}
\def\>{\rangle}

\newcommand{\be}{\begin{equation}}
\newcommand{\ee}{\end{equation}}
\newcommand{\bea}{\begin{eqnarray}}
\newcommand{\eea}{\end{eqnarray}}
\newcommand{\bwt}{\begin{widetext}}
\newcommand{\ewt}{\end{widetext}}

\newcommand{\bi}{\begin{itemize}}
\newcommand{\ei}{\end{itemize}}
\newcommand{\ben}{\begin{enumerate}}
\newcommand{\een}{\end{enumerate}}
\newcommand{\bca}{\begin{cases}}
\newcommand{\eca}{\end{cases}}
\newcommand{\bln}{\begin{align}}
\newcommand{\eln}{\end{align}}
\newcommand{\bst}{\begin{split}}
\newcommand{\est}{\end{split}}

\preprint{MIT-CTP/5116
\begin{flushright}
\vskip -0.3cm
 QMUL-PH-19-26
\end{flushright}
}

\title{\large  Horizon constraints on holographic Green's functions}

\author{Mike Blake,$^{1, 2}$ Richard A.~Davison,$^{3, 4}$ David Vegh$^{5}$}

\affiliation{ $^{1}$ Center for Theoretical Physics, Massachusetts Institute of Technology,
Cambridge, MA 02139, USA}

\affiliation{ $^{2}$ School of Mathematics, University of Bristol, Bristol BS8 1UG, UK}

\affiliation{ $^{3}$ Department of Applied Mathematics and Theoretical Physics, University of Cambridge,
 Cambridge CB3 0WA, UK}
 
 \affiliation{ $^{4}$ Department of Mathematics and Maxwell Institute for Mathematical Sciences, Heriot-Watt University,
Edinburgh EH14 4AS, UK}

\affiliation{ $^{5}$ Centre for Research in String Theory, School of Physics and Astronomy, Queen Mary University of London,
327 Mile End Road, London E1 4NS, UK}

\abstract{We explore a new class of general properties of thermal holographic Green's functions that can be deduced from the near-horizon behaviour of classical perturbations in asymptotically anti-de Sitter spacetimes. We show that at negative imaginary Matsubara frequencies and appropriate complex values of the wavenumber the retarded Green's functions of generic operators are not uniquely defined, due to the lack of a unique ingoing solution for the bulk perturbations. From a boundary perspective these `pole-skipping' points correspond to locations in the complex frequency and momentum planes at which a line of poles of the retarded Green's function intersects with a line of zeroes. As a consequence the dispersion relations of collective modes in the boundary theory at energy scales $\omega\sim T$ are directly constrained by the bulk dynamics near the black-brane horizon. For the case of conserved $U(1)$ current and energy-momentum tensor operators we give examples where the dispersion relations of hydrodynamic modes pass through a succession of pole-skipping points as real wavenumber is increased. We discuss implications of our results for transport, hydrodynamics and quantum chaos in holographic systems.}

\begin{document}
\maketitle


\section{Introduction}

\paragraph{} One of the great advantages of the holographic correspondence is that it provides tools to calculate the properties of quantum field theories (QFTs) in the limit of strong interactions that would otherwise be intractable. This approach has been used extensively to investigate the real-time thermal Green's functions of strongly interacting quantum field theories with classical asymptotically anti-de Sitter (AdS) gravity duals. These Green's functions characterize the near-equilibrium physics of systems, including their transport properties and spectrum of collective excitations.

\paragraph{}The real-time formulation of holography initiated in \cite{Son:2002sd} (see also \cite{Maldacena:1997re, Gubser:1998bc,Witten:1998qj,Horowitz:1999jd,Herzog:2002pc,Skenderis:2008dh,Skenderis:2008dg,Son:2009vu,Glorioso:2018mmw,Liu:2018crr}) relates the Fourier space retarded Green's functions $G^R(\omega,k)$ of boundary operators to the solutions of classical bulk equations for perturbations obeying ingoing boundary conditions at the black hole horizon. While applying the prescription of \cite{Son:2002sd} is in principle straightforward, in practice it can be difficult to evolve the ingoing solution from the horizon to the AdS boundary in order to determine $G^R(\omega,k)$. This evolution requires numerical methods even for simple spacetimes like Schwarzschild-AdS, and also makes it clear that generically the retarded Green's functions depend in a complicated way on the details of the particular QFT state (i.e.~the particular spacetime) under consideration.

\paragraph{} However, there are elements of certain retarded Green's functions for which one can obtain simple and general results for holographic theories. One example of this is the observation that a holographic theory's shear viscosity (set by the $k,\omega\rightarrow0$ limit of the retarded Green's function of the stress tensor) is set by its entropy density \cite{Kovtun:2004de}. This general result arises because the radial evolution of the corresponding bulk perturbation is extremely simple for $k,\omega\rightarrow0$. The corresponding limit of the retarded Green's function can therefore be expressed solely in terms of the near-horizon region of the gravitational solution, and is insensitive to the details of the rest of the spacetime \cite{Iqbal:2008by}. Simplifications of this type occur for the $k,\omega\rightarrow0$ limits of retarded Green's functions of conserved charges in general, and as a consequence the dissipative d.c.~transport properties of holographic systems are sensitive only to the region of the spacetime near the horizon.

\paragraph{} In \cite{Blake:2018leo} it was shown that the near-horizon dynamics of the gravitational description are also directly responsible for certain features of the retarded Green's function of energy density $\varepsilon$ far from the origin of (complex) Fourier space. In other words there are elements of the response of a holographic QFT at $\omega$ of order the temperature $T$ that can be easily determined by examining only the properties of perturbations near the horizon, and are therefore independent of many details of the particular theory. Specifically, it was shown in \cite{Blake:2018leo} that near the points $\omega_*=+i2\pi T$, $k_*=\pm i2\pi T/v_B$ (where $v_B$ is a specific velocity set by the near-horizon metric\footnote{$v_B$ is the `butterfly velocity' \cite{Roberts:2014isa, Roberts:2016wdl,Blake:2016wvh} of the dual quantum field theory.}), the retarded Green's function of energy density takes the special form
\begin{equation}
\label{eq:introequation}
G^R_{\varepsilon\varepsilon}(\omega_*+\delta\omega,k_*+\delta k)=C\frac{\delta \omega - v_{z} \delta k}{\delta \omega - v_{p} \delta k},
\end{equation}
for a large class of holographic systems dual to Einstein gravity with general matter content. The form \eqref{eq:introequation} of the Green's function was called `pole-skipping' in \cite{Grozdanov:2017ajz, Blake:2017ris, Blake:2018leo,Grozdanov:2018kkt}: the retarded Green's function has a pole and a zero intersecting at $(\omega_*,k_*)$ and as a result is not uniquely defined at this location. The fact that the location $(\omega_*,k_*)$ at which there is `pole-skipping' in $G^{R}_{\varepsilon \varepsilon}(\omega,k)$ is universally related to the form of the out-of-time ordered correlator in holographic theories has led to the suggestion that this instance of pole-skipping is a signature of many-body quantum chaos \cite{Grozdanov:2017ajz, Blake:2017ris, Blake:2018leo,Grozdanov:2018kkt}. Pole-skipping in  $G^{R}_{\varepsilon \varepsilon}(\omega,k)$ at $(\omega_*,k_*)$ was first seen numerically in a holographic theory dual to pure Einstein gravity in AdS$_5$ in \cite{Grozdanov:2017ajz}, and also arises as a prediction of the effective theory of chaos proposed in\footnote{Pole-skipping in $G^R_{\varepsilon\varepsilon}$ at $(\omega_*,k_*)$ was also shown to hold in holographic theories dual to higher derivative gravity in \cite{Grozdanov:2018kkt}.} \cite{Blake:2017ris}.

\paragraph{} As we emphasised earlier in the introduction the radial evolution of the perturbations is generically complicated and spacetime-dependent at the scales $(\omega,k)\sim(\omega_*,k_*)$. Nevertheless it was possible in \cite{Blake:2018leo} to show that the energy density Green's function in general takes the `pole-skipping' form \eqref{eq:introequation}  because the boundary conditions of perturbations near the horizon are very special for this choice of $\omega, k$. In particular, for given asymptotic boundary conditions, one finds that there is not a unique solution for the perturbations at $(\omega_*,k_*)$ that is ingoing at the horizon \cite{Blake:2018leo}. Close to this location there is a unique ingoing solution but this solution now depends on the direction $\delta\omega/\delta k$, leading to the form \eqref{eq:introequation} for the Green's function.

\paragraph{} Our intention in this paper is to illustrate that the approach of  \cite{Blake:2018leo} can be generalised to provide constraints of the form \eqref{eq:introequation} on the retarded Green's functions of generic operators in thermal quantum field theories with classical AdS black brane descriptions. Specifically we will show that at the negative imaginary Matsubara frequencies\footnote{The  symbol $\omega_n$ is more conventionally used to refer to the real set of Matsubara frequencies $\omega_E = 2 \pi T n$ at which the Euclidean Green's function 
is defined. Here we are interested in studying the real time correlator $G^{R}(\omega,k)$ and will abuse the more common notation somewhat by using the symbol $\omega_n$ to refer to the special pure imaginary frequencies $\omega = \omega_n = - i 2 \pi T n $. These are referred to as the `negative imaginary Matsubara frequencies' for obvious reasons.} $\omega_n=-i2\pi Tn$ ($n=1,2,3,\ldots$) and appropriate complex values of the wavenumber $k_n$, the retarded Green's functions of scalar, $U(1)$ current, and energy-momentum tensor operators take the pole-skipping form \eqref{eq:introequation}. The locations $k_n^2$ of the pole-skipping points can be easily and systematically determined by an analysis of the near-horizon region of the gravitational solution. They are operator-dependent and the number of them typically grows linearly with $n$. The values of $k_n^2$ become progressively more sensitive to the gravitational solution further from the horizon as $n$ is increased. The relation between the pole-skipping frequencies $\omega_n$ and the Matsubara frequencies is a consequence of the near-horizon Rindler metric seen by the perturbations.

\paragraph{} The existence of pole-skipping points constrains the spectrum of poles and zeroes of $G^R(\omega,k)$, as one of each must pass through each pole-skipping point $(\omega_n,k_n)$. Our analysis therefore provides exact constraints on the dispersion relations $\omega(k)$ of the collective modes (i.e.~the poles of the retarded Green's functions) of holographic systems. These constraints are encoded in a direct way in the properties of perturbations in the near-horizon region of the gravitational solution. We will provide simple examples in which the dispersion relation of a single hydrodynamic collective mode passes through a sequence of pole-skipping points $(\omega_n,k_n)$ as real $k$ is progressively increased to access shorter and shorter distance and time scales. The `UV completion' of these hydrodynamic dispersion relations can therefore be understood in a direct way from a near-horizon analysis of the corresponding perturbation.\footnote{A contemporaneous study of pole-skipping has been performed in \cite{Grozdanov:2019uhi}. Where our results overlap, they agree.}

\paragraph{} There are two important differences between the instances of pole-skipping we describe in this paper, and that identified in the energy density retarded Green's function in \cite{Grozdanov:2017ajz, Blake:2017ris, Blake:2018leo,Grozdanov:2018kkt}. Firstly, the pole-skipping point identified in \cite{Grozdanov:2017ajz, Blake:2017ris, Blake:2018leo,Grozdanov:2018kkt} is the only example for which the frequency is in the upper half of the complex plane i.e.~it is the only example related to a mode that grows exponentially in time. Secondly, the momenta $k_n$ where lower half-plane pole-skipping occurs depend not only on the spacetime metric near the horizon, but also on the action and profiles for the matter fields. Therefore the values of $k_n$ are in general unrelated to the speed $v_B$ that universally controls the upper half-plane pole-skipping wavenumber for $G^R_{\varepsilon\varepsilon}$ described in \cite{Grozdanov:2017ajz, Blake:2017ris, Blake:2018leo,Grozdanov:2018kkt}. Therefore unlike the case described in \cite{Grozdanov:2017ajz, Blake:2017ris, Blake:2018leo,Grozdanov:2018kkt}, we believe it is unlikely that the pole-skipping phenomena that we describe in this paper are related in a straightforward way to the underlying quantum chaotic properties of holographic systems.

\paragraph{} The paper is organised as follows. In Section~\ref{sec:scalarfield} we derive the existence of pole-skipping at $\omega=-i2\pi T$ for the simple case of a minimally coupled scalar field, before systematically generalising this in Section~\ref{sec:higherpoleskipping} to derive the existence of pole-skipping for a scalar field at higher frequencies $\omega=-i2\pi Tn$. In Section~\ref{sec:scalarexamples} we turn to the explicit examples of scalar fields in BTZ and planar AdS-Schwarzschild spacetimes, and confirm that pole-skipping occurs as we predict using exact analytic and numerical results for $G^R(\omega,k)$ these cases. In Section~\ref{sec:hydrocorrelators} we further generalise our pole-skipping analysis beyond scalar operators to the retarded Green's functions of conserved $U(1)$ currents and the energy-momentum tensor, and illustrate (in simple cases) that the real pole-skipping wavenumbers $k_n$ constrain the short distance properties of hydrodynamic excitations. Finally, in Section~\ref{sec:discussion} we close with an extended discussion of the implications of our results for quantum chaos, hydrodynamics and transport, as well as of a number of interesting open questions and future research directions.

\section{Minimally coupled scalar field}
\label{sec:scalarfield}

\paragraph{} We begin by studying the pedagogically simple case of a minimally coupled scalar field $\varphi$ with bulk action
\begin{equation}
\label{action0}
S = \int d^{d+2} x \sqrt{-g}\bigg(R - 2 \Lambda - \frac{1}{2} (g^{\mu \nu} \partial_{\mu} \varphi \partial_{\nu} \varphi + m^2 \varphi^2 ) \bigg) +  S_{\mathrm{matter}},
\end{equation}
where $\Lambda = -d(d+1)/2L^2$. In $S_{\mathrm{matter}}$ we have allowed for extra matter fields in the theory besides $\varphi$. In the standard quantization of the scalar field, $\varphi$ is dual to a scalar boundary operator $\cal O$ of dimension $\Delta$, given by the larger of the two roots to
\begin{equation}
\label{dimension}
\Delta(\Delta - d - 1) = m^2 L^2.
\end{equation}
Our goal in this Section is to derive the phenomenon of pole-skipping by computing the Fourier-transformed retarded Green's function $G^R_{\cal O \cal O}(\omega,k)$ of the scalar boundary operator $\cal O$ in the thermal state of the dual quantum field theory. Henceforth we will set the AdS radius $L = 1$.

\paragraph{}We assume that the action admits a planar black hole solution (with $\varphi=0$) that can be written in the form
\begin{equation}
\label{backgroundmetricrt}
ds^2 = - r^2 f(r) dt^2  + \frac{1}{r^2 f(r)} dr^2 + h(r) d \vec{x}^2,
\end{equation}
with $t,\vec{x}$ giving coordinates on the asymptotically planar AdS boundary as $r \to \infty$. We assume that $f(r)$ and $h(r)$ can be expanded in Taylor series around a horizon located at $r=r_0$ (i.e. $ f(r_0) =0$) with Hawking temperature $4 \pi T = r_0^2 f'(r_0)$. The precise form of $f(r)$ and $h(r)$ will depend on $S_{\mathrm{matter}}$ and we will leave them unspecified in much of what follows. For $S_{\mathrm{matter}} = 0$, the appropriate solution is just the planar AdS$_{d+2}$-Schwarzschild metric
\begin{equation}
\label{eq:schwarzsec2}
f(r) = 1 - \bigg(\frac{r_0}{r}\bigg)^{d+1}, \;\;\;\; h(r) = r^2.
\end{equation}
\paragraph{} To calculate the retarded Green's function for ${\cal O}$, it is convenient to introduce the ingoing Eddington-Finkelstein coordinate $v$
\begin{equation}
v = t + r_*, \;\;\;\;\;\; \frac{d r_*}{d r} = \frac{1}{r^2 f(r)},
\end{equation}
in terms of which the metric is
\begin{equation}
\label{backgroundmetric}
ds^2 = - r^2 f(r) dv^2  + 2 dv dr + h(r) d \vec{x}^2.
\end{equation}

\paragraph{} The retarded Green's function for the boundary operator ${\cal O}$ dual to $\varphi$ can be extracted by finding solutions to the equation of motion
\begin{equation}
\label{scalareom}
\partial_\mu\left(\sqrt{-g}g^{\mu\nu}\partial_\nu\varphi\right)-m^2\sqrt{-g} \varphi=0,
\end{equation}
that obey the ingoing wave boundary condition at the horizon. In practice we implement this by Fourier transforming $\varphi = \phi(r) e^{-i \omega v + i k x}$ and then imposing that $\phi(r)$ has a Taylor series expansion near the horizon. For generic $\omega,k$ this boundary condition is sufficient to yield a unique ingoing solution to \eqref{scalareom}, up to an overall normalisation. Expanding this solution as $\phi = \phi_A(\omega, k) r^{\Delta - d - 1} + \phi_B(\omega, k ) r^{-\Delta} + \dots$ near the AdS boundary, the boundary retarded Green's function is then specified uniquely by
\begin{equation}
\label{greensfunction}
G^R_{\cal O \cal O}(\omega,k) =  (2 \Delta - d - 1) \frac{\phi_B(\omega,k)}{\phi_A(\omega,k)},
\end{equation}
up to the possible existence of contact terms.

\paragraph{} The purpose of this paper is to emphasise a simple but general new aspect of holographic Green's functions such as \eqref{greensfunction}. Specifically, at frequencies $\omega_n = -i 2 \pi T n$ and certain complex values of momentum $ k_n $, the imposition of the ingoing boundary condition at the horizon is not sufficient to uniquely specify $\varphi$ (up to an overall normalisation constant). In fact, at these special points in complex Fourier space any solution to \eqref{scalareom} is regular at the horizon in ingoing coordinates. The locations of these special points can easily and systematically be determined by expanding \eqref{scalareom} near the horizon of the black hole, and can be used to obtain highly non-trivial information about the boundary Green's function $G^R_{\cal O \cal O}(\omega,k)$. Generically we will find that $G^R_{\cal O \cal O}(\omega,k)$ is not uniquely defined at $(\omega_n, k_n)$ but rather depends on the slope $\delta \omega/ \delta k$ at which one approaches these special points. Further we will show that the $G^R_{\cal O \cal O}(\omega,k)$ must have both a line of poles and a line of zeroes passing through such points, which have thus recently been christened `pole-skipping' points \cite{Grozdanov:2017ajz, Blake:2017ris, Blake:2018leo,Grozdanov:2018kkt}. As such we will show how to directly obtain non-trivial information about the dispersion relations of poles and zeroes of $G^R_{\cal O \cal O}(\omega,k)$ from a simple analysis of perturbations near the black hole horizon, and will demonstrate this explicitly in several examples.

\subsection{Existence of multiple ingoing solutions}
\label{sec:scalarmultisols}

\paragraph{} We first explain why there are certain special values of $\omega, k$ at which imposing ingoing boundary conditions is not sufficient to uniquely specify a solution to \eqref{scalareom} (up to an overall normalisation). Following the Fourier transform, the equation \eqref{scalareom} in the coordinate system \eqref{backgroundmetric} is
\begin{equation}
\label{scalareqEF}
\frac{d}{dr}\left[h^{d/2}\left(r^2f\partial_r\phi-i\omega\phi\right)\right]-i\omega h^{d/2}\partial_r\phi-h^{d/2-1}\left(k^2+m^2h\right)\phi=0,
\end{equation}
and we are interested in solutions that are regular around the horizon i.e.~those with a Taylor series expansion 
\begin{equation}
\label{regularexp}
\phi(r) = \sum_{p = 0}^{\infty} \phi_p (r - r_0)^p =  \phi_0 + \phi_1 (r - r_0) + \dots.
\end{equation}
 \paragraph{} For generic $\omega$, the two independent power law solutions $\phi=(r-r_0)^\alpha$ to \eqref{scalareqEF} near the horizon are\footnote{Because we are working in Eddington-Finkelstein coordinates $(r,v)$ these are shifted from the usual power laws $\pm i \omega/4 \pi T$ one finds in $(r,t)$ coordinates}
\begin{equation}
\label{powerlaws}
\alpha_1 = 0,  \;\;\;\;\;\;\;\;  \alpha_2 = \frac{i \omega}{2 \pi T}.
\end{equation}
These are independent of $k,m$ and are only sensitive to the metric through the value of $T$ because they are set by the perturbation equation in the near-horizon region, where the metric looks like that of Rindler space. The solution with exponent $\alpha_1$ is the `ingoing' solution as it is of the form \eqref{regularexp}, while the `outgoing' solution with exponent $\alpha_2$ generically is not. The choice of ingoing boundary conditions therefore generically picks out a solution of the form $\phi = \phi_0  + \dots $ near the horizon, which is unique up to an overall normalisation constant.

\paragraph{} However at the special frequencies $\omega_n = - i 2 \pi T n $ $(n=1,2,3,\ldots)$ both power laws $\alpha_1 = 0$ and $\alpha_2 = n$ naively appear to correspond to regular ingoing solutions. In fact, a more careful analysis shows that logarithmic corrections to the leading power law solutions generically destroy the regularity of one solution (see Appendix~\ref{app:logarithms}) such that there is still a unique ingoing solution. But we will focus on the non-generic case and show that at certain complex values of the wavevector $k_n$, logarithmic corrections are absent and therefore there are two independent ingoing solutions, which take the form
\begin{equation}
\begin{aligned}
\label{solutionslogs21}
\phi &=\phi_0\left[1+c_1(r-r_0)+\ldots \right]+\phi_n(r-r_0)^n\left[1+d_1(r-r_0)+\ldots\right],
\end{aligned}
\end{equation}
where $\phi_0, \phi_n$ are independent parameters in the expansion \eqref{regularexp} and $c_1,d_1$ etc are constants fixed by the background spacetime and by the mass $m$. As such we find that at these locations $(\omega_n,k_n)$ there is not a unique ingoing solution to \eqref{scalareqEF} and hence there is an ambiguity in defining the Green's function $G^R_{\cal O \cal O}(\omega_n,k_n)$.

\paragraph{}To demonstrate the existence of ingoing solutions of the form \eqref{solutionslogs21}, we will explicitly construct them order-by-order in the near-horizon expansion \eqref{regularexp}. This can be achieved by inserting \eqref{regularexp} into \eqref{scalareqEF} and then expanding the scalar equation of motion in powers of $(r - r_0)$. Denoting the scalar equation \eqref{scalareqEF} as ${\cal S} =0$ with
\begin{equation}
\label{scalarexp}
{\cal S} = \sum_{p = 0}^{\infty} {\cal S}_p (r - r_0)^p = {\cal S}_0 + {\cal S}_1 (r - r_0) + \dots,
\end{equation}
we then obtain a series of equations ${\cal S}_p = 0$ that are recursion relations for the parameters $\phi_p$ in the expansion \eqref{regularexp}.

\paragraph{} For now we will focus on the simplest example of pole-skipping, which occurs at $\omega_1  = -i 2 \pi T$. For this case it will be sufficient to focus just on the equation ${\cal S}_0=0$, which is equivalent to evaluating the scalar equation of motion \eqref{scalareqEF} on the horizon. This equation is
\begin{equation}
\label{horizonequation}
- \bigg(k^2 + m^2 h(r_0) + \frac{i \omega d h'(r_0)}{2} \bigg) \phi_0 + (4 \pi T - 2 i \omega) h(r_0) \phi_1 = 0.
\end{equation}
For a generic $\omega, k$ it is clear that \eqref{horizonequation} fixes $\phi_1$ in terms of the initial value $\phi_0$ on the horizon. After solving \eqref{horizonequation} for $\phi_1$ it is then possible at generic $\omega,k$ to iterate this process using the equations of motion ${\cal S}_p = 0$ to solve for the higher order coefficients $\phi_p$ uniquely in terms of $\phi_0$ and thus construct a regular solution to \eqref{scalareqEF} that is unique up to the overall normalisation $\phi_0$.

\paragraph{} At $ \omega_1 = - i 2 \pi T$ we are unable to construct the solution in this manner. Precisely at $\omega = \omega_1$ the coefficient of the $\phi_1$ term in \eqref{horizonequation} vanishes, and hence $\phi_1/\phi_0$ is no longer fixed by this equation. Instead, at $\omega = \omega_1$ \eqref{horizonequation} reduces to
\begin{equation}
\label{horizonequation2}
 \bigg( k^2 + m^2 h(r_0) +  d \pi T h'(r_0)  \bigg) \phi_0 = 0.
\end{equation}
For a generic value of $k^2$, \eqref{horizonequation2} therefore sets $\phi_0 =0$ in the near-horizon solution \eqref{regularexp} and $\phi_1$ then becomes the free parameter. The remaining equations $S_{p} = 0$ can then be solved iteratively to determine the higher order coefficients $\phi_p$ in terms of $\phi_1$ and produce an ingoing solution that is unique up to the normalisation $\phi_1$.

\paragraph{} However it is now possible to see that there is a very special location in complex Fourier space given by
\begin{equation}
\label{kscalar}
\omega =\omega_1 = -2\pi Ti,\quad\quad\quad k = k_1, \;\;\;\;\;\;\;\;\;\; k_1^2 =-m^2h(r_0)- d \pi T h'(r_0).
\end{equation}
At this location, \eqref{horizonequation} is trivially satisfied by any value of $\phi_0$ and $\phi_1$ and thus both coefficients are free parameters in the general series solution \eqref{regularexp}. One can then iteratively solve the remaining equations ${\cal S}_p = 0$ to yield a family of regular ingoing solutions to \eqref{scalareqEF} in terms of the two independent parameters $\phi_0$ and $\phi_1$. As such we conclude that both independent solutions to the differential equation \eqref{scalareqEF} are consistent with ingoing boundary conditions at \eqref{kscalar}, and can be expanded near the horizon in a Taylor series expansion of the form \eqref{regularexp}.\footnote{Note that by constructing a two parameter family of solutions of the form \eqref{regularexp} we have demonstrated that the regularity of one of the solutions in \eqref{powerlaws} is not destroyed by subleading logarithmic corrections. Indeed in Appendix~\ref{app:logarithms} we provide another perspective on the location in \eqref{kscalar} by showing that the wavenumber $k_1$ is precisely the value at which the logarithmic corrections to \eqref{powerlaws} vanish.}

Note that the special value of the wavenumber $k_1^2$ is sensitive only to the near-horizon region of the black hole. In general $k_1^2$ does not have to be positive and thus the special locations can be at complex values of $k_1$. When we examine specific cases in Sections~\ref{sec:scalarexamples} and \ref{sec:hydrocorrelators} we will find examples with both real and complex values of $k_n$.

\subsection{Green's functions near special location}
\label{sec:matching}

\paragraph{} We have just demonstrated that at the special location in \eqref{kscalar} there are two independent ingoing solutions to \eqref{scalareqEF}, rather than the one found at generic points $(\omega,k)$. The existence of an extra ingoing solution for metric perturbations was recently observed in \cite{Blake:2018leo}, where it was argued to have dramatic consequences for the boundary retarded Green's function (of energy density). Here we will demonstrate that a similar analysis applies to the scalar Green's function near \eqref{kscalar}. In particular we will argue that generically there must be both a line of poles and a line of zeroes in $G^R_{\cal O \cal O}(\omega,k)$ that pass through the locations \eqref{kscalar}, a phenomenon known as `pole-skipping'.

\paragraph{} In particular as there are two independent ingoing solutions at \eqref{kscalar}, it is clear that $G^R_{\cal O \cal O}(\omega,k)$ cannot be uniquely defined by working at this location. In order to define $G^R_{\cal O \cal O}(\omega,k)$ it is necessary as in \cite{Blake:2018leo}  to move infinitesimally away from \eqref{kscalar} to $\omega = - i 2 \pi T + \epsilon \delta \omega$, $k = k_1 + \epsilon \delta k$. After doing so, the horizon equation \eqref{horizonequation} becomes non-trivial in the limit $\epsilon \to 0$ and is given by
\begin{equation}
\label{nearlocation}
- \bigg(\frac{i \delta \omega d h'(r_0)}{2} + 2 k_1 \delta k\bigg) \phi_0 - 2 i \delta \omega h(r_0)   \phi_1= 0.
\end{equation}
The horizon equation \eqref{nearlocation} is now well-defined and fixes $\phi_1$ in terms of $\phi_0$. One can then construct a solution of the form \eqref{regularexp} that depends only on the overall normalisation $\phi_0$. However the ingoing solution $\phi(r)$ obtained by solving \eqref{nearlocation} for $\phi_1$ will clearly depend on the slope $\delta \omega/\delta k$ with which we move away from the special location \eqref{kscalar}. The retarded Green's function one extracts using \eqref{greensfunction} therefore also depends on the slope $\delta \omega/\delta k$. $G^R_{\cal O \cal O}(\omega_n,k_n)$ is therefore not uniquely defined but is infinitely multivalued, depending on how the point \eqref{kscalar} is approached.

\paragraph{} Furthermore, the slope $\delta \omega/\delta k$ now plays the role of the aforementioned extra free parameter in the ingoing solution, and so an arbitrary solution to \eqref{scalareqEF} obeys ingoing boundary conditions for an appropriate choice of slope. In particular  we can always pick a slope $(\delta \omega/\delta k)_{p}$ so that the ingoing solution is normalisable in the UV -- i.e.~is a solution $\phi^{(n)}$ to \eqref{scalareqEF} for which $\phi_A = 0$ as $r \to \infty$. Near the horizon the normalisable solution to \eqref{scalareqEF} at \eqref{kscalar} can formally be expanded as
\begin{equation}
\phi^{(n)} = \phi_0^{(n)} + \phi_1^{(n)} (r - r_0) + \dots,
\end{equation}
for some fixed coefficients $\phi_0^{(n)}, \phi_1^{(n)}$ determined by solving \eqref{scalareqEF} subject to the normalisable boundary condition in the UV. We can therefore ensure that the ingoing solution is normalisable simply by moving away from \eqref{kscalar} infinitesimally along the direction
\begin{equation}
\label{slope}
\bigg(\frac{\delta \omega}{\delta k}\bigg)_p =  \frac{ 4 i k_1 \phi_0^{(n)}}{4  h(r_0) \phi_1^{(n)} +  d h'(r_0)  \phi_0^{(n)}}.
\end{equation}
Since the normalisable solution corresponds to a pole in the Green's function we therefore conclude that $G^R_{\cal O \cal O}(\omega,k)$ must contain a line of poles passing through \eqref{kscalar} with a slope $(\delta \omega/\delta k)_p$ given by \eqref{slope}.

\paragraph{} Alternatively we could instead move away from \eqref{kscalar} along a different slope such that the ingoing solution instead matches on to the solution $\phi^{(nn)}$ with no normalisable component in the UV (i.e.~the ingoing solution has $\phi_B = 0$ as $r \to \infty$). This implies there must also be a line of zeroes in $G^R_{\cal O \cal O}(\omega,k)$ passing through \eqref{kscalar} with a slope $(\delta \omega/\delta k)_{z}$ that will just be given as in \eqref{slope} but where $\phi^{(n)}_0, \phi^{(n)}_1$ are replaced by the corresponding coefficients for the near-horizon expansion of $\phi^{(nn)}$.

\paragraph{} For a general choice of $\delta \omega/\delta k$ the ingoing solution is a linear combination of $\phi^{(n)}$ and $\phi^{(nn)}$ that depends on the slope (see Appendix~\ref{app:greensfunction}). The retarded Green's function extracted from such a solution takes the form
\begin{equation}
\label{greensfunctionnear}
G^R_{\cal O \cal O}(\omega_1 + \epsilon \delta \omega, k_1 + \epsilon \delta k ) \propto \frac{\delta \omega - (\delta \omega/\delta k)_{z} \delta k}{\delta \omega - (\delta \omega/\delta k)_{p} \delta k},
\end{equation}
which manifestly displays both a line of poles and a line of zeroes passing through \eqref{kscalar}. This is the same as the `pole-skipping' form described in \cite{Blake:2018leo}. The values of $(\delta \omega/\delta k)_{p}$ and $(\delta \omega/\delta k)_{z}$ cannot be deduced from our near-horizon analysis alone: they depend on the radial evolution of the normalisable and non-normalisable solutions from the boundary to the horizon.

\paragraph{} Whilst the phenomenon of `pole-skipping' in $G^R_{\cal O \cal O}(\omega,k)$ was easy to deduce from analysing the properties of perturbations near the horizon, it has provided us with highly non-trivial information about properties of the retarded Green's function. In particular, as a consequence of the additional ingoing solution we have deduced that there must be a line of poles (and zeroes) with dispersion relation $\omega(k)$ that pass through the point \eqref{kscalar}. Moreover, we will shortly see that the existence of an extra ingoing solution also occurs at higher frequencies $\omega_n = - i 2 \pi T n$ and appropriate wavevectors $k_n^2$ that can be similarly determined. From the locations $(\omega_n,k_n)$ of these higher `pole-skippings'  we are therefore able to obtain a whole tower of constraints on the dispersion relations of poles in $G^R_{\cal O \cal O}(\omega,k)$. Note that the locations of pole-skipping points, and the slope $(\delta \omega/\delta k)_{p}$ of the line of poles passing through \eqref{kscalar}, are generically independent of contact terms. In contrast, the slope $(\delta \omega/\delta k)_{z}$ of the line of zeroes passing through \eqref{kscalar} is sensitive to any contact terms added to \eqref{greensfunction}.

\paragraph{} Finally we note that although the above discussion generically applies to the retarded Green's function near \eqref{kscalar}, our analysis breaks down if the location at which multiple ingoing solutions exists is $k_1 = 0$. In this case we see from \eqref{nearlocation} that we can no longer generate an arbitrary solution by varying the slope $\delta \omega/\delta k$ in \eqref{nearlocation}. As such the Green's function near \eqref{kscalar} will not have the pole-skipping form \eqref{greensfunctionnear} if $k_1 = 0$, even though there are multiple ingoing solutions.  In this paper we will refer to such locations at which multiple ingoing solutions exists but for which the Green's function does not take the form \eqref{greensfunctionnear} as `anomalous points', and will shortly see that they can also arise at higher $\omega_n = - i 2 \pi T n$. Whilst such `anomalous points' are not generic we will discuss several explicit examples of them in Sections~\ref{sec:btz},  Appendix~\ref{app:higherorderdetails} and Appendix~\ref{app:integerdeltaBTZ}. Interestingly we will find that in all these explicit examples there are still poles whose dispersion relations pass through the anomalous points, even though the form of the Green's function near these locations is not that of \eqref{greensfunctionnear}.

\section{Pole-skipping at higher Matsubara frequencies}
\label{sec:higherpoleskipping}
\paragraph{}In the last section we demonstrated that for a minimally coupled scalar field there can be pole-skipping in the boundary retarded Green's function $G^R_{\cal O \cal O}(\omega,k)$ at a frequency $\omega_1 = - i 2 \pi T$ and appropriate wavenumber $k_1$. Here we extend our analysis and show that the same phenomenon can also occur at higher Matsubara frequencies $\omega_n = - i 2 \pi T n$. In particular at $\omega = \omega_n$ we find that there are generically $n$ wavenumbers $k_n^2$ at which there will be pole-skipping in $G^R_{\cal O \cal O}(\omega,k)$. The locations $k_n^2$ at which pole-skipping occurs follow from the determinant of an $n$ by $n$ matrix ${\cal M}^{(n)}(\omega,k)$, whose coefficients are determined by the near-horizon expansion \eqref{scalarexp} of the scalar equation of motion. This prescription therefore allows us to systematically identify a whole tower of pole-skipping points $(\omega_n, k_n)$ that constrain the dispersion relations of poles at frequencies $\omega_n = - i 2 \pi T n$.
\subsection{Multiple ingoing solutions at $\omega_n = - i 2 \pi T n$}
\label{sec:multiple1}

\paragraph{} We first demonstrate that at Matsubara frequencies $\omega_n = - i 2 \pi T n$ there are certain choices of (complex) wavenumber $k_n$ for which the general ingoing solution to the equation of motion \eqref{scalareqEF} is not uniquely specified by the overall normalisation. That is at the locations $(\omega_n, k_n)$ we show that there is a two-parameter family of regular ingoing solutions of the form \eqref{regularexp}, labelled by independent parameters $\phi_0$ and $\phi_n$.

\paragraph{} In Section \ref{sec:scalarfield} we were able to see the existence of multiple ingoing solutions at $(\omega_1, k_1)$ solely from the horizon equation of motion ${\cal S}_{0} = 0$. At higher $n$ it is also necessary to look at the equations ${\cal S}_{p} = 0$ that arise from our expansion of \eqref{scalareqEF} around the horizon. We will show that the locations $(\omega_n, k_n)$ at which pole-skipping occurs can be easily extracted from a matrix ${\cal M}^{(n)}(\omega,k)$ defined using the first $n$ equations arising from this expansion around the horizon.

\paragraph{} In order to illustrate how to find these locations it is useful for us to write out the first few equations ${\cal S}_p = 0$ somewhat explicitly. In particular, the first three equations in the expansion of \eqref{scalareqEF} are
\begin{equation}
\begin{aligned}
&0 =  M_{11}(\omega,k^2) \phi_0 + (2 \pi T -  i \omega)  \phi_1, \\
&0 = M_{21}(\omega,k^2)  \phi_0+M_{22}(\omega,k^2) \phi_1+\left(4 \pi T -i \omega\right)\phi_2,  \\
&0 = M_{31}(\omega, k^2) \phi_0+M_{32}(\omega,k^2) \phi_1+M_{33}(\omega,k^2)  \phi_2+\left(6 \pi T -i \omega\right) \phi_3,  \\
\end{aligned}
\label{firstfeweqns}
\end{equation}
where the coefficients $M_{ij}(\omega,k^2)$ take the form
\begin{equation}
\label{elements}
M_{i j}(\omega,k^2) =  i \omega a_{ij}  + k^2 b_{ij} + c_{ij},
\end{equation}
with $a_{ij},b_{ij},c_{ij}$ determined by the background spacetime metric \eqref{backgroundmetric}, its derivatives at the horizon, and $m$. The explicit forms of the coefficients $a_{ij}, b_{ij},c_{ij}$ are rather complicated, and will not be needed for our general discussion in this section. Nevertheless they can be easily computed by the expansion of \eqref{scalareqEF}, and we include explicit expressions for the matrix elements in \eqref{firstfeweqns} in Appendix~\ref{sec:scalarappendix}. Generally $M_{ij}$ is sensitive to the $i^\text{th}$ derivative of the spacetime metric functions $f(r)$ and $h(r)$ at the horizon. In this sense, higher coefficients in the equation's near-horizon expansion are progressively more sensitive to the spacetime metric away from the horizon.

\paragraph{} Although we have only written out the first few equations explicitly, the general structure of the equations \eqref{firstfeweqns} continues at higher order. Constructing an ingoing solution is then equivalent to finding a solution to a set of linear equations of the form
\begin{equation}
\label{eq:matrixequation}
M(\omega,k^2)\cdot\phi\equiv
\begin{pmatrix}
M_{11} & \left(2\pi T-i\omega\right) & 0 & 0 & \ldots \\
M_{21} & M_{22} & \left(4\pi T-i\omega\right) & 0 & \ldots \\
M_{31} & M_{32} & M_{33} & \left(6\pi T-i\omega\right) & \ldots \\
\ldots & \ldots & \ldots & \ldots & \ldots
\end{pmatrix}
\begin{pmatrix}
\phi_0 \\ \phi_1 \\ \phi_2 \\ \ldots
\end{pmatrix}
=0.
\end{equation}
In what follows a key role will be played by the $n$ by $n$ matrix ${\cal M}^{(n)}(\omega,k^2)$ that corresponds to keeping the first $n$ rows and $n$ columns of $M(\omega,k^2)$. Note that this matrix ${\cal M}^{(n)}(\omega,k^2)$ is nothing more than the coefficients of the $\phi_0,...,\phi_{n-1}$ terms in the first $n$ equations \eqref{firstfeweqns} in our expansion of \eqref{scalareqEF} around the horizon.

\paragraph{} In order to explain why ${\cal M}^{(n)}(\omega,k^2)$ is important in characterising the pole-skipping locations, let us first note that at a generic frequency $\omega \ne-i2\pi Tn $ it is straightforward to solve the equations \eqref{eq:matrixequation} iteratively to determine a unique (up to normalisation) ingoing solution  in the manner we outlined in Section~\ref{sec:scalarfield}. One simply starts by solving the first equation in \eqref{firstfeweqns} to determine $\phi_1$ in terms of $\phi_0$.  After inserting this solution into the second equation in \eqref{firstfeweqns} one can then determine $\phi_2$ in terms of $\phi_0$. By repeating this iterative process one can solve for all the coefficients $\phi_p$ in terms of a single $\phi_0$.

\paragraph{} However at frequencies $\omega = \omega_n$ we can see from the structure of \eqref{eq:matrixequation} that it is not possible to construct the solution iteratively in terms of $\phi_0$ in this manner. This is because the coefficient of the parameter $\phi_n$ vanishes in the $n^{th}$ row of \eqref{eq:matrixequation}. This has two important consequences. Firstly it implies that $\phi_n$ can no longer be fixed in terms of the lower coefficients $\tilde{\phi} = (\phi_0, ..., \phi_{n-1})$ by iteratively solving \eqref{eq:matrixequation}, and hence $\phi_n$ becomes a free parameter in the general near-horizon solution. Secondly it implies that the first $n$ equations in the expansion around the horizon \eqref{eq:matrixequation} decouple to form a closed system of equations for the coefficients $\tilde{\phi} = (\phi_0,...\phi_{n-1})$. This equation takes the form
\begin{equation}
\label{matrixreduced}
{\cal M}^{(n)}(\omega_n, k^2)\cdot\tilde{\phi} = 0.
\end{equation}
For a generic choice of $k^2$ the matrix ${\cal M}^{(n)}(\omega_n,k^2)$ will be invertible, and hence \eqref{matrixreduced} has the solution $\tilde{\phi}=0$. In these cases there will be a unique ingoing solution of the form $\phi = \phi_n (r - r_0)^n + \dots$, characterised by the free parameter $\phi_n$.

\paragraph{} However it is immediately clear from above discussion that there will be an extra ingoing solution for certain complex wavevectors $k^2$ for which the matrix ${\cal M}^{(n)}(\omega_n,k^2)$ is not invertible. At such values of $k^2$ there will now be a non-trivial solution $\tilde{\phi} = \tilde{\phi}_a$ to \eqref{matrixreduced}.\footnote{As \eqref{scalareqEF} can have at most two independent solutions, \eqref{matrixreduced} can have only one non-trivial solution.} This extra non-trivial solution will then result in an extra free parameter in our expansion that we can take to be the value of $\phi_0$ in $\tilde{\phi}_a$ We therefore conclude that at the locations
\begin{equation}
\label{multiplelocations}
\omega_n = - i 2 \pi T n, \;\;\;\;\;\;\;\;\; k^2  = k_n^2,  \;\;\;\;\;\;\;\;\; \det\mathcal{M}^{(n)}(\omega_n, k_n^2) =0,
\end{equation}
the regular solutions to \eqref{scalareqEF} are labelled by two independent parameters $\phi_0, \phi_n$ in our expansion \eqref{regularexp}. Note that since the elements of $M_{ij}$ are of the form \eqref{elements} then the equation $\det\mathcal{M}^{(n)}(\omega_n, k^2) =0 $ is a polynomial in $k^2$ of degree $n$. As such there will generically be $n$ distinct complex roots $k_n^2$ to this equation and hence the number of locations in \eqref{multiplelocations} grows with $n$. Furthermore, due to properties of the elements $M_{ij}$ mentioned above, $k_n$ is typically sensitive to the $n^\text{th}$ derivative of the spacetime metric functions $f(r)$ and $h(r)$ on the horizon. In this sense, the locations \eqref{multiplelocations} are progressively more sensitive to the spacetime away from the horizon as $n$ is increased.

\paragraph{}Mathematically, the existence of multiple ingoing solutions is tied to the nature of the differential equation \eqref{scalareqEF} at the horizon $r=r_0$. For a generic Fourier mode $(\omega,k)$ the horizon is a regular singular point of the equation with indicial exponents $0$ and $i\omega/2\pi T$, and thus there is only one analytic solution near $r=r_0$. For the mode $(\omega_1,k_1)$, the regular singular point reduces to simply a regular point of the differential equation and therefore both solutions are analytic. For the higher-order modes $(\omega_{n>1},k_{n>1})$, while the horizon is a regular singular point, it is an \textit{apparent} singularity (as opposed to a real singularity) as both solutions are analytic in the vicinity of this singular point. Sufficient conditions for a singularity to be apparent are that the indicial exponents are non-negative integers and that there are no logarithmic terms in the solution near the singular point \cite{InceBook}. In the above, we have described a procedure by which one can systematically identify values of $(\omega,k)$ at which the singularity at the horizon is only apparent and therefore the ingoing solution is non-unique.

\paragraph{} Whilst the above discussion has been somewhat abstract we wish to emphasise that equation \eqref{multiplelocations} provides a systematic way of identifying the pole-skipping locations $(\omega_n, k_n)$ for any given $n$. In particular the matrix ${\cal M}^{(n)}(\omega,k^2)$ that characterises these locations simply corresponds to reading off the coefficients of $\phi_0,...,\phi_{n-1}$ that appear in the first $n$ equations in the near-horizon expansion \eqref{firstfeweqns}. As such for small $n$ it is straightforward to explicitly compute $\det\mathcal{M}^{(n)}(\omega_n, k^2)$ for a given theory and hence identify these locations. We will shortly discuss several explicit examples of this in detail in Section~\ref{sec:scalarexamples}. However we first examine the form of the Green's functions near \eqref{multiplelocations} and hence argue that generically we should expect pole-skipping in $G^R_{\cal O \cal O}(\omega,k)$ at the locations $(\omega_n, k_n)$.

\subsection{Green's function near special locations}
\label{sec:anomaloussection2}

\paragraph{} We have just argued that there is a two parameter family of ingoing solutions to \eqref{scalareqEF} at the locations $(\omega_n, k_n)$ in \eqref{multiplelocations}. In other words both independent solutions to \eqref{scalareqEF} are consistent with ingoing boundary conditions, and thus it is unclear how to uniquely define $G^R_{\cal O \cal O}(\omega_n,k_n)$. We will now show that near \eqref{multiplelocations} $G^R_{\cal O \cal O}(\omega,k)$ generically takes the pole-skipping form \eqref{greensfunctionnear}. In order to do this it's helpful to first give a slightly different perspective on the origin of the extra ingoing mode at \eqref{multiplelocations}. This will allow us to straightforwardly generalise the matching argument of Section~\ref{sec:matching} to these higher instances of pole-skipping.

\paragraph{} There is a more explicit way to reach the conclusion that there is an extra ingoing solution at the  locations \eqref{multiplelocations}. For a generic $\omega \neq - i 2 \pi T n $ we have commented that a unique ingoing solution (up to overall normalisation) can be constructed by solving \eqref{eq:matrixequation} iteratively. Whilst this iterative process breaks down exactly at the special frequencies $\omega = \omega_n$, near $\omega = \omega_n$ we can always use it to uniquely solve for the solution up to $\phi_{n-1}$ in terms of $\phi_0$. After determining the coefficients $\tilde{\phi} = (\phi_0,...,\phi_{n-1})$ in terms of $\phi_0$ in this manner we can then insert these expressions into the $n^{th}$ line of \eqref{eq:matrixequation} to obtain an equation relating $\phi_n$ to $\phi_0$. The resulting equation can be written as
\begin{equation}
\label{horizonmultiple}
\frac{1}{N^{(n)}(\omega)} \det\mathcal{{M}}^{(n)}(\omega, k^2) \phi_0 + (n 2 \pi T - i \omega) \phi_n = 0,
\end{equation}
where ${\cal M}^{(n)}(\omega,k)$ is the matrix we introduced previously and we have defined\footnote{Note that the factor of $1/N^{(n)}(\omega)$ in \eqref{horizonmultiple} diverges at lower Matsubara frequencies  $ \omega_m = - i 2 \pi T m$ with $m < n$. In writing down \eqref{horizonmultiple} we have assumed we are not at such a frequency. We are predominantly interested in studying \eqref{horizonmultiple} near $\omega = \omega_n$, where it is always well-defined.}
\begin{equation}
N^{(n)}(\omega) = (i \omega - 2 \pi T)(i \omega - 4 \pi T) \dots (i \omega - (n-1) 2 \pi T).
\end{equation}

\paragraph{} The equation \eqref{horizonmultiple} is a direct analogue of the horizon equation \eqref{horizonequation} that we used to demonstrate pole-skipping at $(\omega_1,k_1)$. In particular we see that at generic $\omega,k$ \eqref{horizonmultiple} provides a constraint relating $\phi_n$ to $\phi_0$ that can be used to construct an ingoing solution with a single parameter $\phi_0$. However, precisely at the locations identified in \eqref{multiplelocations} we see that \eqref{horizonmultiple} becomes trivial and is satisfied by any $\phi_0, \phi_n$. As such we again see that at the location in \eqref{multiplelocations} there is a two-parameter family of ingoing solutions.

\paragraph{}  Furthermore it is now straightforward to expand \eqref{horizonmultiple} near the location \eqref{multiplelocations} as in our matching analysis in Section~\ref{sec:matching}. In particular, if we move away from the location \eqref{horizonmultiple} to $\omega = \omega_n + \epsilon \delta \omega$, $k = k_n + \epsilon \delta k$ we find an equation relating $\phi_n$ to $\phi_0$
\begin{equation}
\label{horizonmultipleexp}
\frac{1}{N(\omega_n)} \bigg( \partial_{k} \det\mathcal{{M}}^{(n)}(\omega_n,k_n^2) \delta k + \partial_{\omega} \det\mathcal{{M}}^{(n)}(\omega_n,k_n^2) \delta \omega \bigg) \phi_0  - i \delta \omega \phi_n = 0,
\end{equation}
with $N(\omega_n) = (n-1)!(2 \pi T)^{n-1}$.

\paragraph{} As in Section~\ref{sec:matching} the equation \eqref{horizonmultipleexp} can now be solved to determine $\phi_n$ in terms of $\phi_0$ and continue the iterative construction of the general ingoing solution dependent on a single parameter $\phi_0$. However we see that this solution, and in particular the ratio $\phi_n/\phi_0$, will now generically depend on the slope $\delta \omega/\delta k$ with which we move away from \eqref{multiplelocations}. As such by varying the slope we can ensure that an arbitrary solution to \eqref{scalareqEF} is ingoing. Following a similar logic to in Section~\ref{sec:matching} we then conclude that there will be both a line of poles and and a line of zeroes passing through the locations in \eqref{multiplelocations}, and the Green's function $G^R_{\cal O \cal O}(\omega,k)$ will generically have the pole-skipping form \eqref{greensfunctionnear} near $(\omega_n,k_n)$.

\paragraph{} Whilst generically we expect pole-skipping at the locations in \eqref{multiplelocations} it is worth noting that there can be anomalous cases if we have a location $k_n^2$ which satisfies both $\det\mathcal{{M}}^{(n)}(\omega_n, k_n^2) = 0$ and also the condition\footnote{Note that since the equation $\det\mathcal{{M}}^{(n)}( \omega_n, k^2 )= 0$ is just a polynomial of degree $n$ in $k^2$ then points satisfying \eqref{matchingcondition} correspond to special cases where either we have a solution with $k^2_n = 0$ or for which there is a solution with $k_n^2 \neq 0$ that corresponds to a repeated root of $\det\mathcal{{M}}^{(n)}(\omega_n, k_n^2) = 0$.}
\begin{equation}
\label{matchingcondition}
 \partial_{k} \det\mathcal{{M}}^{(n)}(\omega_n, k_n^2) = 0.
\end{equation}
At such locations there are two independent ingoing solutions to \eqref{scalareqEF}, but from \eqref{horizonmultipleexp} we see that it is no longer possible to match to an arbitrary linear combination of these by moving away from \eqref{multiplelocations} along an appropriate slope $\delta \omega/\delta k$. These cases are further examples of the anomalous points we mentioned at the end of Section~\ref{sec:matching}, and the Green's function $G^R_{\cal O \cal O}(\omega,k)$ will not take the pole-skipping form \eqref{greensfunctionnear} near these points. We will show in Section~\ref{sec:btz} that examples of anomalous points at $n > 1$ arise for the retarded Green's function $G^R_{\cal O \cal O}(\omega,k)$ of a scalar field with integer $\Delta$ in the BTZ spacetime.

\section{Scalar field examples}
\label{sec:scalarexamples}

\paragraph{} Until now we have rather abstractly discussed the phenomenon of pole-skipping for a minimally coupled scalar. In particular we argued that at frequencies $\omega_n = -i 2 \pi T n$ there are special wavenumbers $k_n$ given by \eqref{multiplelocations} at which there are multiple ingoing solutions to the bulk equation \eqref{scalareqEF}. As a result, the retarded Green's function near such locations generically takes the form \eqref{greensfunctionnear}, and in particular there will be both a line of poles and a line of zeroes in $G^R_{\cal O \cal O}(\omega,k)$ passing through these locations. We now wish to illustrate these statements by considering several explicit examples.

\subsection{BTZ black hole}
\label{sec:btz}

\paragraph{} We begin with the simplest example: a minimally coupled scalar field in the BTZ background
\begin{equation}
\label{backgroundmetricbtz}
ds^2 = - (r^2 -r_0^2) dt^2  + \frac{1}{(r^2 - r_0^2)} dr^2 + r^2 dx^2,
\end{equation}
which is dual to a (1+1)-dimensional conformal field theory with temperature $2 \pi T = r_0$. A minimally coupled scalar field of mass $m$ propagating in this spacetime is dual to an operator of conformal dimension $\Delta$ via \eqref{dimension}. For standard quantisation $\Delta$ is the largest root to the equation $\Delta(\Delta -2) = m^2$, whilst for alternative quantisation $\Delta$ is the smaller root of the same equation. %

\subsection*{Predictions from near-horizon analysis}

\paragraph{} We first consider the pole-skipping at $\omega_1 = - i 2 \pi T$ discussed in Section~\ref{sec:scalarfield}. From \eqref{kscalar} we see that for a background of the form \eqref{backgroundmetricbtz} there should be pole-skipping at
\begin{equation}
\label{locationbtz}
\omega_1=- i 2 \pi T ,\quad\quad\quad k_1^2= - r_0^2(\Delta - 1)^2, \;\;\;\;\;\;\;\;\;  r_0 = 2 \pi T.
\end{equation}
To look for instances of pole skipping at higher frequencies $\omega_n = - i 2 \pi T n$ we expand the equation of motion as described in Section~\ref{sec:higherpoleskipping} and compute the determinant of the matrix ${\cal M}^{(n)}$. This computation is straightforward and for the first few values of $n$ yields (up to overall normalisation factors)
\begin{equation}
\label{matrices}
\begin{aligned}
\det\mathcal{M}^{(1)}&=\left[k^2+(\Delta-1)^2r_0^2\right],\\
\det\mathcal{M}^{(2)}&=\left[k^2+\Delta^2r_0^2\right]\left[k^2+\left(\Delta-2\right)^2r_0^2\right],\\
\det\mathcal{M}^{(3)}&=\left[k^2+(\Delta+1)^2r_0^2\right]\left[k^2+\left(\Delta-1\right)^2r_0^2\right]\left[k^2+\left(\Delta-3\right)^2r_0^2\right],\\
\end{aligned}
\end{equation}
 from which we read off the first few pole-skipping locations to be
\begin{equation}
\begin{aligned}
\omega_1 &= -i 2 \pi T, \;\;\;\;\;\;\;\; k_1^2 = - r_0^2 (\Delta - 1)^2, \\
\omega_2 &= -i 4 \pi T, \;\;\;\;\;\;\;\; k_2^2 = - r_0^2 (\Delta - 2)^2,  - r_0^2 \Delta^2,\\
\omega_3 &= -i 6 \pi T, \;\;\;\;\;\;\;\; k_3^2 = - r_0^2(\Delta - 3)^2,  - r_0^2 (\Delta-1)^2, -r_0^2(\Delta + 1)^2.
\end{aligned}
\end{equation}
The same pattern continues at higher $n$ such that $\det\mathcal{M}^{(n)}(\omega_n,k^2)$ takes the form (up to overall normalisation)
\begin{equation}\label{detbtz}
\det\mathcal{M}^{(n)}(\omega_n,k^2) = \prod_{q=1}^{n} (k^2 - k_{n,q}^2), \;\;\;\;\;\;\;\;\;\;  k_{n,q}^2  = -r_0^2 (n-2q+\Delta)^2,
\end{equation}
for any $n \in \{1,2,\ldots\}$ and where $q \in \{ 1, \ldots, n \}$. For the purposes of our discussion in the main text we will assume that $\Delta$ is generic (i.e.~non-integer), for which there are $n$ distinct to solutions to $ \det\mathcal{M}^{(n)}(\omega_n,k^2) = 0$ corresponding to the values $k^2 = k_{n,q}^2 $ in \eqref{detbtz}.\footnote{The case of integer $\Delta$ is described in Appendix~\ref{app:integerdeltaBTZ}.} In turn this then yields $ 2 n$ imaginary wavenumbers $k_n$ at which we expect pole-skipping
\be
\label{locations}
\om_n = -i 2\pi  T n, \qquad k_{n,q} = \pm 2\pi i T(n-2q+\Delta),
\ee
where again $q \in \{ 1, \ldots, n \}$. From our discussion in Sections \ref{sec:scalarfield} and \ref{sec:higherpoleskipping} we then expect that the retarded Green's function near the locations \eqref{locations} should have the pole-skipping form \eqref{greensfunctionnear}. In particular there should be both a line of poles and a line of zeroes passing through each of the locations in \eqref{locations}.

\subsection*{Comparison to exact Green's function}

\begin{figure}
 \begin{tabular}{cc}
\includegraphics[width=.45\textwidth]{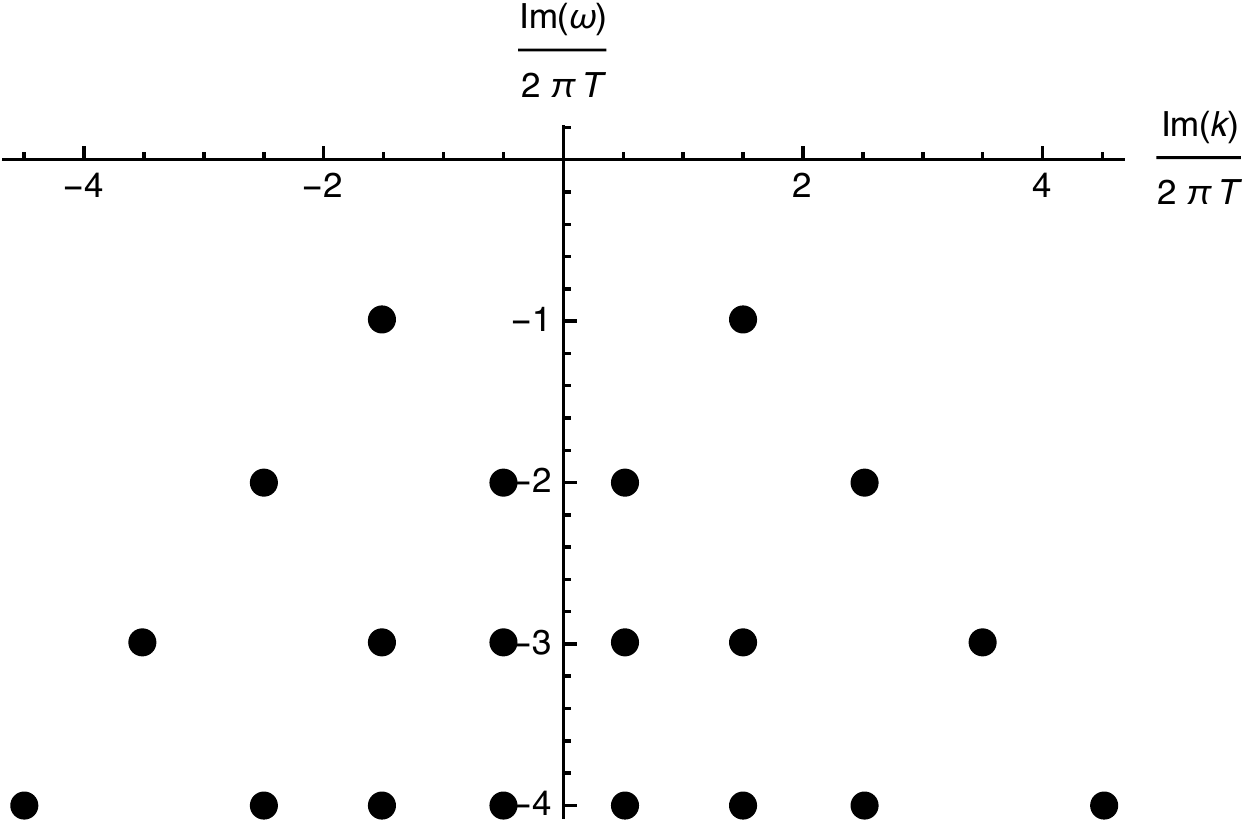}&
\includegraphics[width=.45\textwidth]{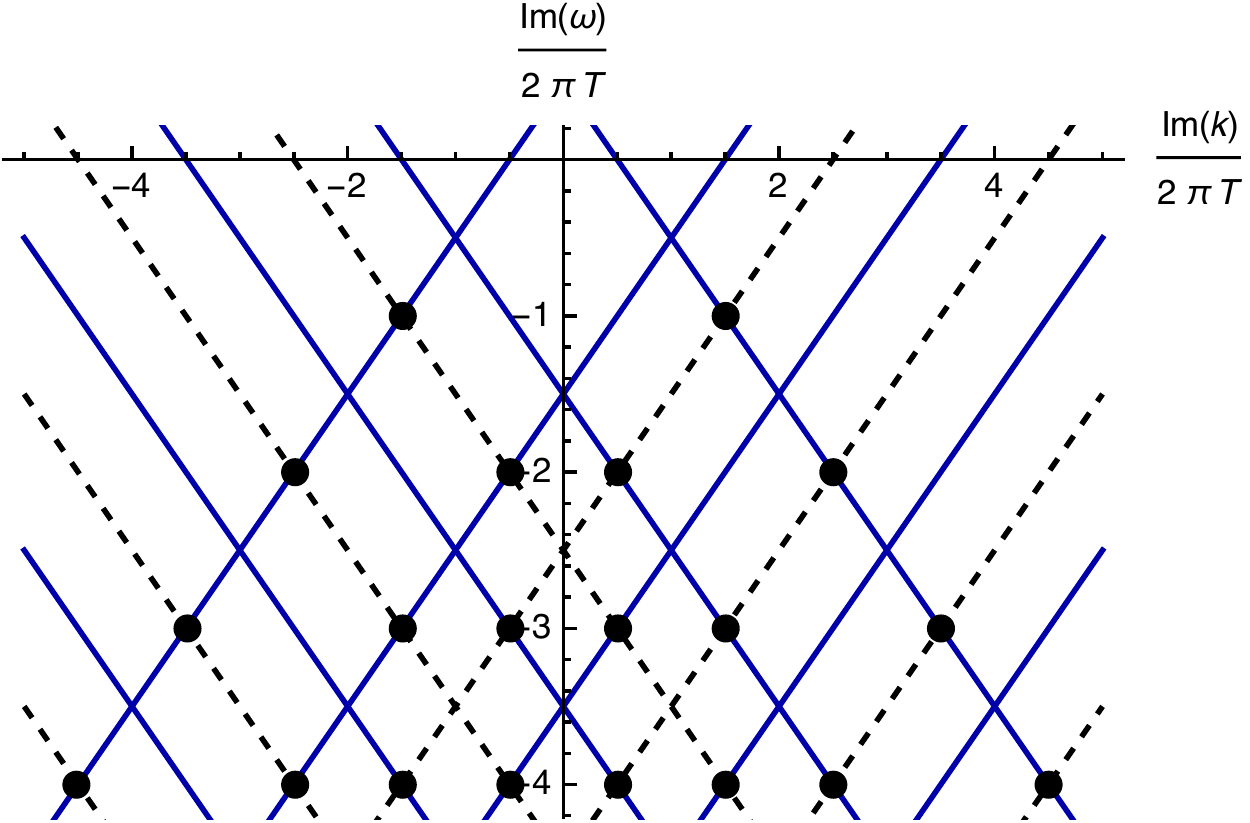}
\end{tabular}
\caption{ The left hand plot shows the locations \eqref{locations} where our study of near horizon perturbations predicts pole-skipping for a field in the BTZ background with $\Delta = 2.5$ and for $n=1,2,3,4$. The right hand plots shows the dispersion relations \eqref{disppoles} and \eqref{dispzeroes} of the lines of poles (dashed) and zeroes (solid) in the $\Delta = 2.5$ Green's function \eqref{btzgreensfunction}. These lines can be seen to intersect precisely at the pole-skipping locations \eqref{locations}, as expected from our analysis in Section~\ref{sec:higherpoleskipping}.}
\label{fig:intersections}
\end{figure}

\paragraph{}For the BTZ metric \eqref{backgroundmetricbtz} exact analytic expressions are available for the entire retarded Green's function $G^R_{\cal O \cal O}(\omega,k)$ for an operator of any dimension $\Delta$, and hence we can easily verify the predictions of our near-horizon analysis. As we discuss in Appendix~\ref{app:btz}, the $\omega$ and $k$ dependence of the retarded Green's function for non-integer $\Delta$ is given by a ratio of Gamma functions
\be
\label{btzgreensfunction}
 G^R_{\cal O \cal O}(\omega,k)\propto {
 \Gamma\le({\Delta \ov 2}+{i(k-\om)\ov 4\pi T}\ri)
 \Gamma\le({\Delta \ov 2}-{i(k+\om)\ov 4\pi T}\ri) \ov
 \Gamma\le(1 - { \Delta \ov 2}+{i(k-\om)\ov 4\pi T}\ri)
 \Gamma\le(1 - { \Delta  \ov 2}-{i(k+\om)\ov 4\pi T}\ri) }.
\ee
The Gamma function never vanishes, and has simple poles at non-positive integer values of its argument. Thus there are poles of $G^R_{\cal O \cal O}(\omega,k)$ at the frequencies
\be
\label{disppoles}
  \om^p_{L, m}(k) = k-i2 \pi T (\Delta + 2 m), \qquad  \om^p_{R, m}(k)  = -k-i2 \pi T (\Delta + 2 m),
\ee
and zeroes at the frequencies
\be
\label{dispzeroes}
  \om^z_{L, m}(k) = k-i2\pi T (2 - \Delta  + 2 m), \qquad  \om^z_{R, m}(k) = -k- i2\pi T (2 - \Delta  + 2 m),
\ee
for any $m \in \{0,1,2,\ldots\}$.
\paragraph{}The first ($n=1$) examples of pole-skipping involve the poles and zeroes closest to the origin ($m=0$). It is simple to see that the dispersion relations of the left (right) moving pole and the right (left) moving zero intersect at the first pole-skipping frequency: $\omega_{L,0}^p(k_1)=\omega_{R,0}^z(k_1)=-i2\pi T$ and $\omega_{R,0}^p(-k_1)=\omega_{L,0}^z(-k_1)=-i2\pi T$ where $k_1=i2\pi T(\Delta-1)$. The retarded Green's function \eqref{btzgreensfunction} therefore has exactly the pole-skipping property predicted by our near-horizon analysis of Section~\ref{sec:scalarfield}.
\paragraph{} Indeed the expression \eqref{btzgreensfunction} exhibits pole-skipping at the entire tower of frequencies $\omega_n = - i 2 \pi T n $. To see this note that we should get examples of pole-skipping whenever one of the lines of poles in \eqref{disppoles} intersects with one of lines of zeroes in \eqref{dispzeroes}. This happens at the locations
\be
  \om_n = -i 2\pi T n, \qquad k_{n,q} = \pm 2\pi i T(n-2q+\Delta),
\ee
for any $n \in \{1,2,\ldots\}$ and $q \in \{ 1, \ldots, n \}$, and hence precisely matches the locations \eqref{locations} indicated by our near-horizon analysis. The intersections between the lines of poles and zeroes in \eqref{btzgreensfunction} are illustrated in Figure \ref{fig:intersections}, where it is easy to see the existence of the whole tower of pole-skipping points.

\paragraph{}A slightly more sophisticated analysis is required for integer $\Delta$, as in this case some of the apparent pole-skipping points are in fact anomalous (in the sense described in Section~\ref{sec:anomaloussection2}). We discuss this case in detail in Appendix~\ref{app:integerdeltaBTZ} and again find that the locations of the pole-skipping predicted from our near horizon analysis agree perfectly with the exact analytic expression for the BTZ Green's function. It is interesting to note that in this example the anomalous points coincide with locations at which two lines of poles intersect. Therefore poles do still pass through the anomalous points in this example, even though the Green's function does not have the pole-skipping form \eqref{greensfunctionnear} there.

\paragraph{} Our pole-skipping analysis is in a sense redundant for the BTZ example, as we already know the exact Green's functions. We present it to demonstrate that there are non-trivial features of these Green's functions that can be exactly determined by a simple analysis of the properties of perturbations near the horizon. In the following sections we will generalise to cases where expressions for the Green's functions are not known.

\subsection{Higher dimensional AdS-Schwarzschild}

\paragraph{} In higher dimensions, or in the presence of matter fields,  it is usually impossible to obtain analytic expressions for the dispersion relations of poles and zeroes of $G^R_{\cal O \cal O}(\omega,k)$ in the theories \eqref{action0}. However, for a given theory these dispersion relations can be computed exactly by numerical evaluation of \eqref{greensfunction}. We will now study the case of a (massless) minimally coupled scalar field in the $AdS_{d+2}$-Schwarzschild spacetime \eqref{eq:schwarzsec2} and verify that the exact Green's functions do have poles passing through the locations we derived from a near-horizon analysis.

\paragraph{} Following the analysis of Section~\ref{sec:scalarfield} for this particular spacetime we find that the first ($n=1$) instance of pole skipping in $G^R_{\cal O \cal O}(\omega,k)$ occurs at the wavenumber
\begin{equation}
\label{eq:specialkhigherd}
k_1^2+r_0^2\left(\Delta\left(\Delta-d-1\right)+\frac{d\left(d+1\right)}{2}\right)=0, \;\;\;\;\;\;\;\;\; (d+1)r_0 = 4 \pi T.
\end{equation}
Similarly by constructing ${\cal M}^{(2)}(\omega_2,k^2)$ as described in Section~\ref{sec:higherpoleskipping} we conclude the location of the $n=2$ pole-skipping wavenumber $k_2$ obeys
\begin{equation}
\label{higherdmultiple}
\begin{aligned}
&k_2^4+2k_2^2r_0^2\left[\Delta\left(\Delta-d-1\right)+d\left(d+1\right)\right]\\
&+r_0^4\bigl\{\left[\Delta\left(\Delta-d-1\right)+d\left(d+1\right)\right]^2-\left(d+1\right)\left[2\Delta\left(\Delta-d-1\right)+d\left(d+1\right)\right]\bigr\}=0.
\end{aligned}
\end{equation}
It is straightforward to determine the polynomial equations governing $k_3$ and higher, but the expressions expressions quickly become rather lengthy and so for conciseness we will not present them here.

\begin{figure}
\begin{tabular}{cc}
\includegraphics[width=0.45\textwidth]{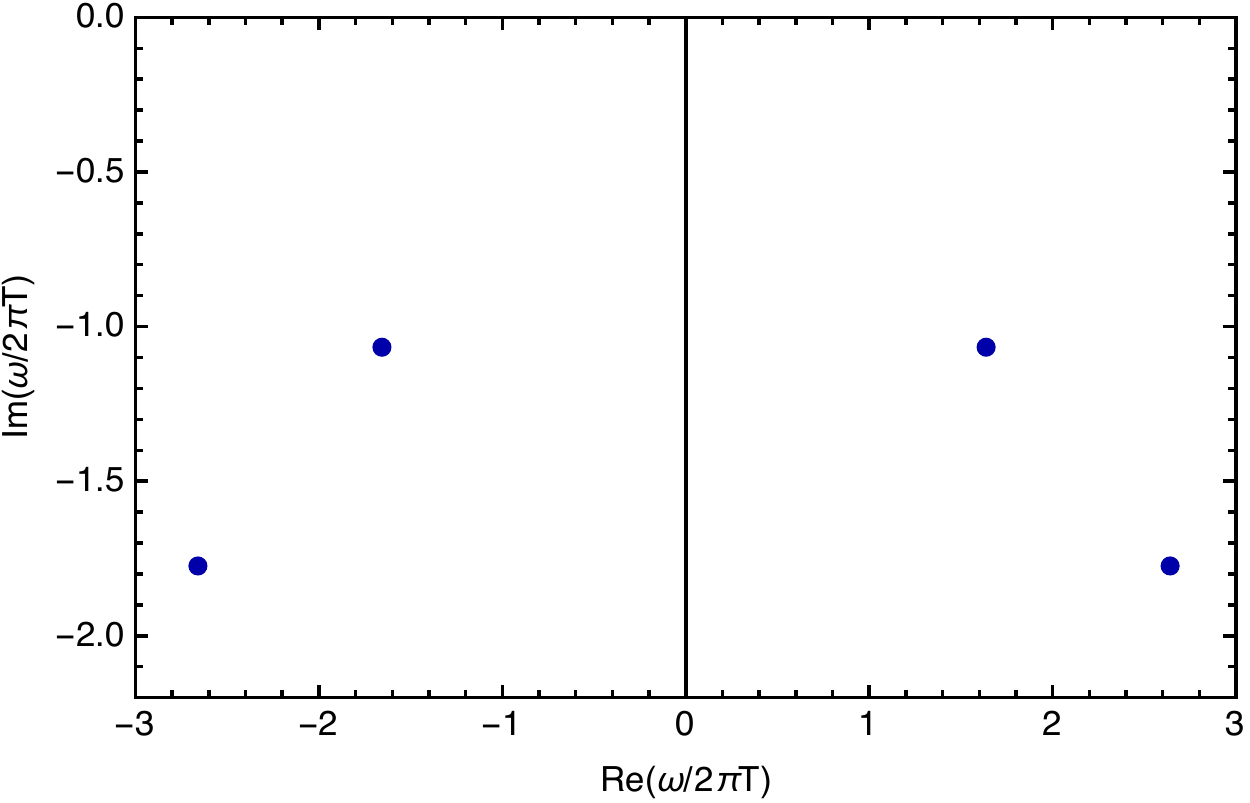}&
\includegraphics[width=0.45\textwidth]{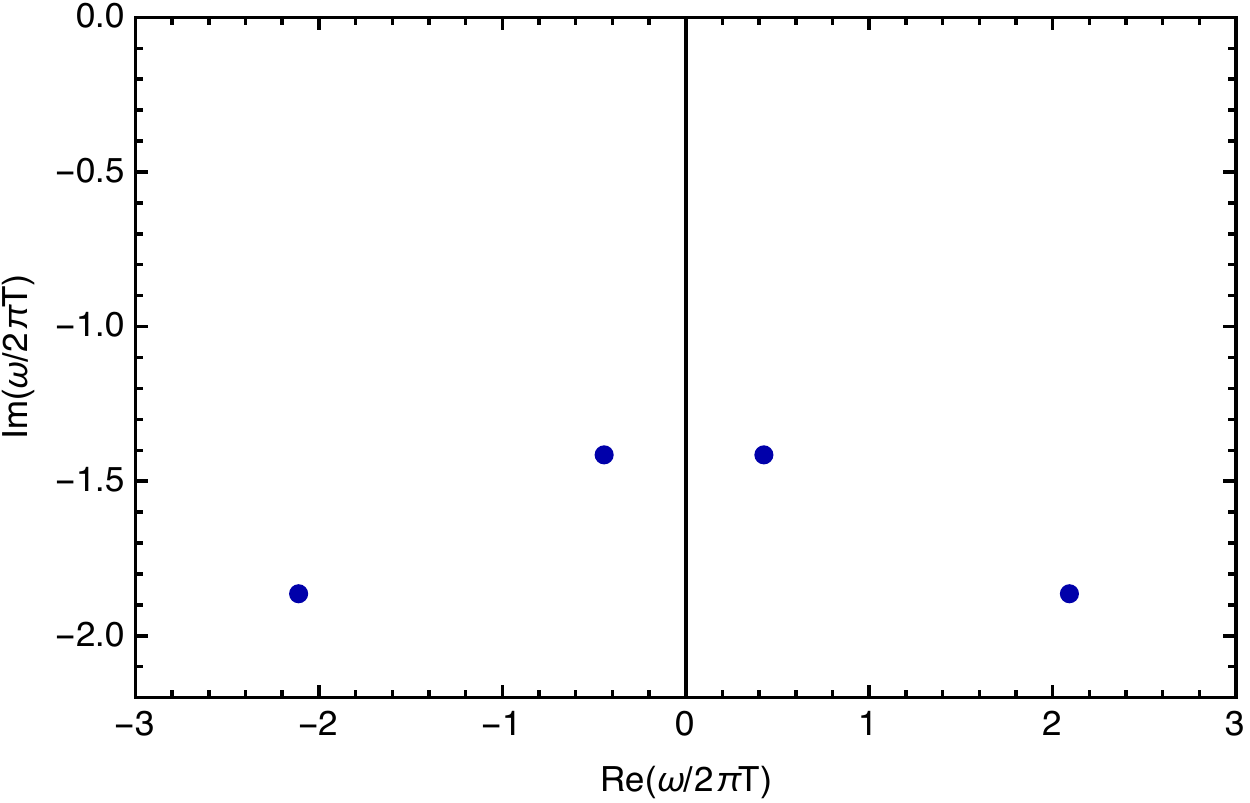}\\
\includegraphics[width=0.45\textwidth]{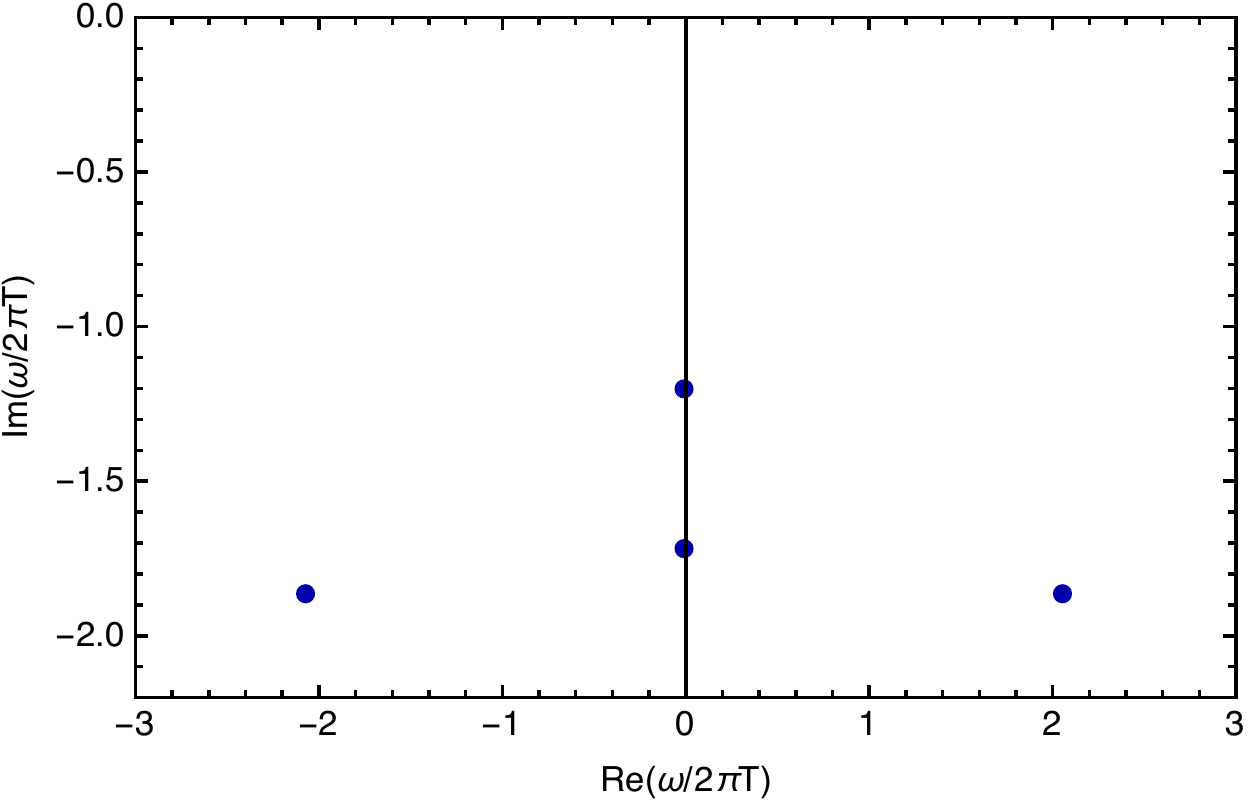}&
\includegraphics[width=0.45\textwidth]{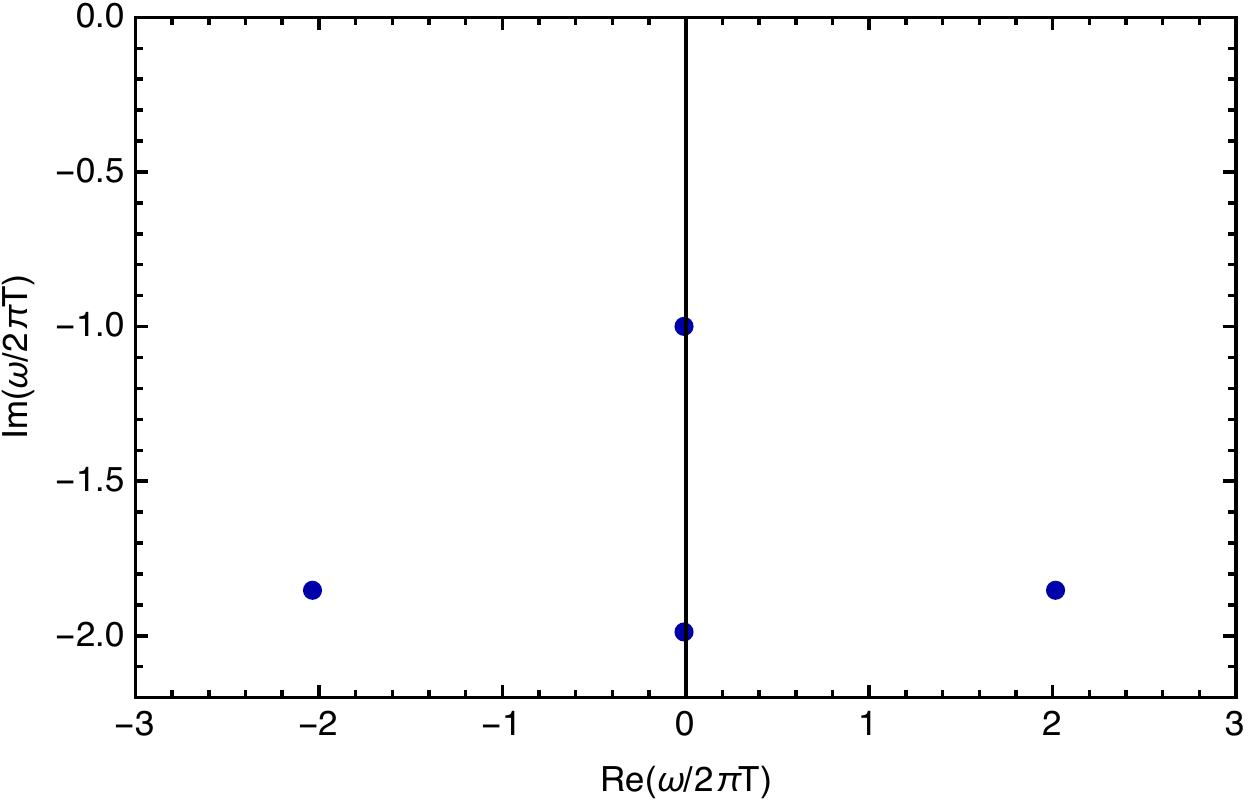}
\end{tabular}
\caption{Blue dots denoting the location of poles of the boundary retarded Green's function for a massless scalar field in AdS$_6$-Schwarzschild. The locations were determined numerically (using the procedure described in Section 4.2 of \cite{Denef:2009yy}) for four values of $k/r_0$: $0$ (top left), $3.0i$ (top right), $3.1i$ (bottom left) and $3.16i$ (bottom right). The pole locations are consistent with our near-horizon analysis, which indicates that $k_1/r_0=k_2/r_0=\sqrt{10}i\sim 3.16i$ is a pole-skipping wavenumber for the first two pole-skipping frequencies $\omega_1$ and $\omega_2$.}
\label{fig:masslessscalar}
\end{figure}

\paragraph{} For the special case of a massless scalar field $\Delta=d+1$, the expressions for the pole-skipping wavenumbers simplify to
\begin{equation}
\begin{aligned}
\label{simplelocation}
k_1^2=-r_0^2\frac{d\left(d+1\right)}{2},\quad\quad\quad k_2^2=-r_0^2d\left(d+1\right)\left(1\pm d^{-1/2}\right).
\end{aligned}
\end{equation}
As in the BTZ example, this corresponds to imaginary values of $k_1$ and $k_2$. At $k=0$, the locations of the poles of the $\Delta=d+1$ retarded Green's functions have been calculated numerically. They form a `Christmas tree'-like pattern in the complex $\omega$ plane (see e.g.~\cite{Morgan:2009pn} or the top left panel of Figure~\ref{fig:masslessscalar}) and as a consequence, must move significantly as imaginary $k$ is increased in order that they pass through the pole-skipping locations we have predicted. This is in fact what happens.

\paragraph{}In Figure~\ref{fig:masslessscalar} we present the result of a numerical calculation showing how the poles of a massless scalar in AdS$_6$-Schwarzschild move in the complex $\omega$ plane as imaginary $k$ is increased from $0$ to $k_1$. Two poles approach the imaginary axis and collide, one of which moves up the imaginary $\omega$ axis and passes through $\omega=\omega_1$ exactly at the wavenumber $k=k_1$ predicted by our near-horizon analysis. A similar phenomenon occurs for AdS$_{4,5}$-Schwarzschild. For the particular case of AdS$_6$-Schwarzschild, \eqref{simplelocation} implies that poles should pass through both $\omega_1$ and $\omega_2$ when $k=k_1$, and this is also confirmed by our numerical results in Figure~\ref{fig:masslessscalar}.

\paragraph{}We have thus confirmed that our simple near-horizon analysis of bulk perturbations does precisely capture non-trivial features of the exact Green's functions $G^R_{\cal O \cal O}(\omega,k)$.

\section{Current and energy-momentum tensor Green's functions}
\label{sec:hydrocorrelators}

\paragraph{} So far in this paper we have focused on discussing the phenomenon of pole-skipping in the retarded Green's functions of scalar operators dual to minimally coupled bulk scalar fields. This focus was for pedagogical reasons. We believe that this phenomenon is in fact a generic feature of Green's functions in holographic theories, and in particular that it also occurs in the retarded Green's functions of conserved $U(1)$ current and energy-momentum tensor operators. In this section, we explore these latter examples.

\paragraph{}The Green's functions of these operators characterise energy and charge dynamics and thus the pole-skipping locations provide us with non-trivial information about the collective modes responsible for charge and energy transport in holographic systems. For example, by examining simple cases we will show that a pole-skipping anlaysis provides exact information on how the dispersion relations of long wavelength hydrodynamic modes evolve to shorter distances and timescales.

\paragraph{}In the interests of brevity, in the main text we will focus on the results of the analysis. The mathematical details (see Appendix~\ref{app:higherorderdetails}) are conceptually very similar to the case of a scalar field described in Sections~\ref{sec:scalarfield} and Section~\ref{sec:higherpoleskipping}. We restrict to the cases $d\geq2$ where conventional hydrodynamics (see e.g.~\cite{Kovtun:2012rj}) is valid.

\subsection{$U(1)$ current Green's functions}
\label{sec:gaugefieldpoleskip}

\paragraph{} We first study the Green's function of a conserved $U(1)$ charge current operator $J^\mu$ in a state with $\langle J^\mu\rangle=0$. This is dual to a bulk $U(1)$ gauge field $A_\mu$ in a black hole spacetime \eqref{backgroundmetric}. We assume the following general action for the field strength
\begin{equation}
S_{\mathrm{Maxwell}}  =  \int d^{d+2} x \sqrt{-g} \bigg( -\frac{1}{4} Z(\Phi) F^{\mu \nu} F_{\mu \nu} \bigg),
\end{equation}
where $\Phi(r)$ is a scalar field and where we assume that the black hole solution has a vanishing gauge field.
\paragraph{} The boundary retarded Green's functions $G^R_{J^{\mu} J^{\nu}}(\omega,k)$ can be extracted by solving the following equations of motion for small perturbations of the gauge field
\begin{equation}\label{maxwelleqn}
\partial_{\mu} (\sqrt{-g} Z(\phi) F^{\mu \nu} ) = 0,
\end{equation}
in an analogous manner to our discussion for the scalar field. There are two independent components of these Green's functions, corresponding to whether the current is parallel or perpendicular to the direction of the wavenumber $k$ of the perturbation (which we call the $x$ direction).

\paragraph{} In the perpendicular case, the relevant bulk equation of motion is very similar to that of the minimally coupled scalar field. It is therefore straightforward to apply the analysis of Sections \ref{sec:scalarfield} and \ref{sec:higherpoleskipping} and verify that there will be pole skipping at frequencies $\omega_n=-i2\pi Tn$ and appropriate wavenumbers $k_n$ (which are different from those of the scalar field). Due to its similarity to the scalar field case, we will not discuss this case further.

\paragraph{} We will focus on the more interesting case of the retarded Green's function of the current parallel to the wavenumber $G^R_{J^xJ^x}(\omega,k)$. This is related by a simple Ward identity to the charge density correlator $G^R_{J^tJ^t}(\omega,k)$ and supports a gapless hydrodynamic charge diffusion mode with the small-$k$ dispersion relation
\begin{equation}
\label{chargehydro}
\omega_h(k)=-iD_ck^2+\ldots.
\end{equation}

\paragraph{} The relevant bulk perturbations are $\delta A_x$, $\delta A_v$, $\delta A_r$, which are coupled. After Fourier transforming it is convenient to algebraically solve one equation of motion for $\delta A_r$, leaving the single equation of motion
\begin{equation}
\label{eq:GIlongeq}
\frac{d}{dr}\left[\frac{h^{d/2}Z}{\omega^2h-k^2r^2f}\left(r^2f\psi_1'-i\omega\psi_1\right)\right]+\frac{h^{d/2-1}Z}{\omega^2h-k^2r^2f}\left(-i\omega h\psi_1'-k^2\psi_1\right)=0,
\end{equation}
for the variable $\psi_1\equiv\omega \delta A_x+k\delta A_v$. While \eqref{eq:GIlongeq} is more complicated than the scalar equation \eqref{scalareqEF}, its near-horizon limit is very similar. We can therefore perform a similar analysis to that in Sections \ref{sec:scalarfield} and \ref{sec:higherpoleskipping} (see Appendix \ref{sec:gaugeappendix}) and conclude that there is pole skipping in $G^R_{J^xJ^x}(\omega,k)$ at $\omega_n=-i2\pi T n$ and appropriate values of $k=k_n$. The first instance of pole skipping occurs when
\begin{equation}
\label{eq:longitudinalskipping}
k_1^2=\left(d-2\right)\pi Th'(r_0)+2\pi Th(r_0)\frac{Z'(r_0)}{Z(r_0)},
\end{equation}
where the prime denotes a derivative with respect to $r$. Note that $k_1$ is sensitive not just to the metric near the horizon, but also to the effective Maxwell coupling $Z$.

\paragraph{} While hydrodynamic arguments impose the constraint that there must be a pole of $G^R_{J^xJ^x}(\omega,k)$ passing through $\omega=0$ and $k=0$ (with the dispersion relation \eqref{chargehydro}), our pole skipping analysis is complementary to this and constrains the pole structure at higher frequencies and wavenumbers. To illustrate this, we now look at the particular example of the AdS$_{d+2}$-Schwarzschild black brane metric \eqref{eq:schwarzsec2} with $Z(\Phi)=1$, for which the charge diffusion constant is $4\pi T D_c=\left(d+1\right)/\left(d-1\right)$ (see e.g.~\cite{Iqbal:2008by}). In Appendix \ref{sec:gaugeappendix}, we show that for each pole skipping frequency $\omega_n$, one of the $n$ pole skipping wavenumbers $k_n^2$ is positive. For example, for $d=3$ the first few pole skipping points with positive values of $k_n^2$ are
\begin{equation}
\label{eq:ads5gaugepoints}
k_1^2=2r_0^2,\quad\quad\quad k_2^2=4\left(\sqrt{3}-1\right)r_0^2,\quad\quad\quad k_3^2=\left(4\sqrt{6}-6\right)r_0^2.
\end{equation}
For these cases, the pole skipping analysis therefore produces constraints on dispersion relations $\omega(k)$ of modes at \textit{real} values of $k$, which are those most commonly studied.

\begin{figure}
\begin{tabular}{cc}
\includegraphics[width=0.495\textwidth]{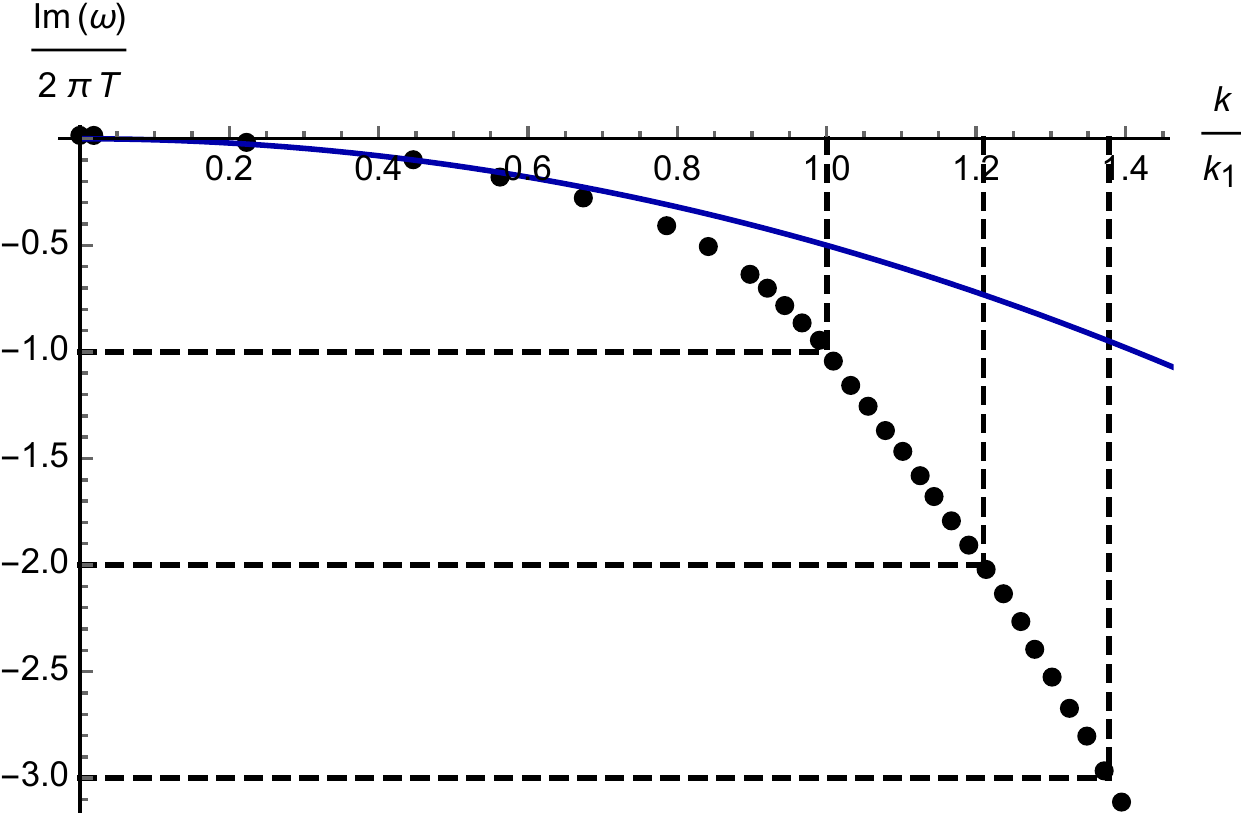}
\includegraphics[width=0.495\textwidth]{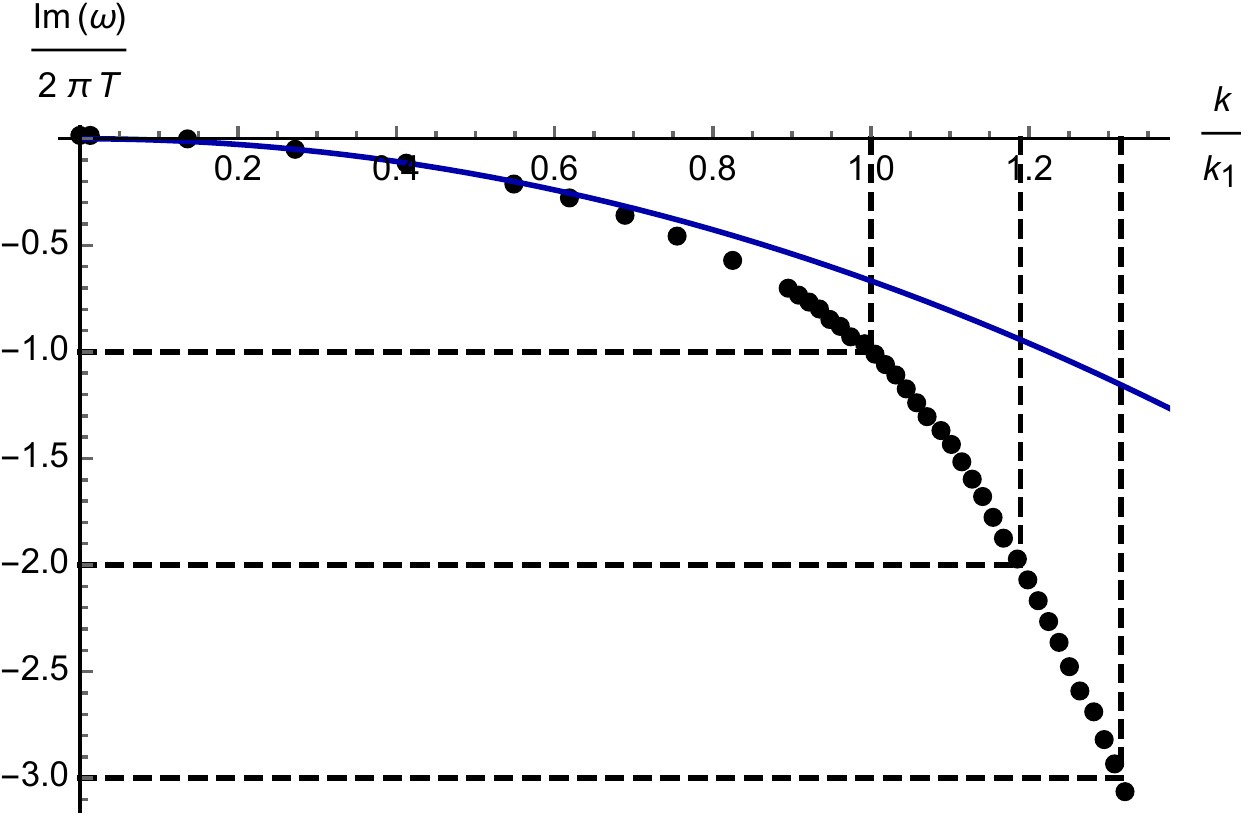}
\end{tabular}
\caption{Dispersion relation of the charge diffusion mode from AdS$_5$-Schwarzschild (left) and of the momentum diffusion mode from AdS$_4$-Schwarzschild (right). Black dots show the exact dispersion relation determined by numerical integration of the appropriate perturbation equations in $(t,r)$ coordinates, solid blue lines show the diffusive hydrodynamic dispersion relations (\eqref{chargehydro} and \eqref{viscdisp} respectively), and the intersections of the black dashed lines correspond to the (real $k$) pole-skipping points (\eqref{eq:ads5gaugepoints} and \eqref{eq:poleskippinglocstransversemom} respectively). The short-distance corrections to diffusive hydrodynamics are such that the pole passes through a succession of pole-skipping points.}
\label{fig:diffusion}
\end{figure}

\paragraph{} In the left hand panel of Figure~\ref{fig:diffusion} we plot (for real $k$) the exact dispersion relation $\omega_h(k)$ of the pole of $G^R_{J^xJ^x}(\omega,k)$ that is hydrodynamic at small $k$, with the pole-skipping points \eqref{eq:ads5gaugepoints} overlaid. This shows that the dispersion relation $\omega_h(k)$ passes through a succession of pole-skipping points as real $k$ is increased, and we expect the same to be true in higher dimensional AdS-Schwarzschild. The special case of $d=2$ is discussed in Appendix \ref{sec:gaugeappendix}. We have not checked whether the poles passing through the pole-skipping points with non-real $k_n$ are also related to the hydrodynamic dispersion relation $\omega_h(k)$, but it would be interesting to do so.

\paragraph{} Figure~\ref{fig:diffusion} is quite remarkable from the point of view of hydrodynamics, in which the dispersion relation \eqref{chargehydro} is normally calculated order-by-order in a small $k$ expansion. The requirement that the pole passes through the locations \eqref{eq:ads5gaugepoints}  provides exact (non-perturbative in $k$) information about this dispersion relation at $\omega\sim T$. In this way, the corrections to diffusive hydrodynamics \eqref{chargehydro} can potentially be constrained by a very simple analysis of near-horizon dynamics, and and we discuss this further in Section \ref{sec:discussion}.

\subsection{Energy-momentum tensor Green's functions}
\label{sec:hydrogravitycorrelators}
\paragraph{} We now turn to the case of the retarded Green's functions of boundary energy-momentum tensor operators $T^{\mu\nu}$. There are again multiple independent Green's functions depending on the relative orientation of the component of $T^{\mu\nu}$ and the wavevector $k$ \cite{Kovtun:2005ev}. We will focus on the two independent Green's functions which contain hydrodynamic poles at small $k$: those of transverse momentum density (i.e. the components of the momentum density perpendicular to $k$) and of longitudinal momentum density (i.e. the components of the momentum density parallel to $k$).\footnote{The components of the $T^{\mu\nu}$ Green's functions that do not support any hydrodynamic modes are controlled by bulk fields obeying equations of the same form as the scalar equation \eqref{scalareqEF}. It is therefore straightforward to show (using the techniques of the preceding sections) that these Green's functions exhibit pole-skipping at $\omega_n$, although we will not present the results here.} The latter example is related by a simple Ward identity to the retarded Green's function of energy density, one of whose pole-skipping properties was explored in \cite{Grozdanov:2017ajz, Blake:2017ris, Blake:2018leo,Grozdanov:2018kkt}.

\subsection*{Transverse momentum density}
\paragraph{} We firstly study the Green's function of transverse momentum density. In the main text, we will focus on the action
\begin{equation}
S = \int d^{d+2} x \sqrt{-g}\bigg(R - 2 \Lambda - \frac{1}{2}g^{\mu \nu} \partial_{\mu} \Phi \partial_{\nu} \Phi + V(\Phi)\bigg),
\label{viscaction}
\end{equation}
where we have allowed for a scalar field $\Phi(r)$ supporting a background metric of the form \eqref{backgroundmetric}. In Appendix \ref{sec:chargedtransverseapp}, we discuss the generalisation to charged black branes.

\paragraph{} We choose the wavenumber $k$ to point in the $x$-direction, and thus the transverse momentum density is dual to the perturbation $\delta g_{vy}$ of the metric where $y$ is a field theory spatial direction perpendicular to $x$. $\delta g_{vy}$ couples to the perturbations $\delta g_{xy}$ and $\delta g_{ry}$ of the metric. After Fourier transforming and solving algebraically for $\delta g_{ry}$, we are left with a single equation
\begin{equation}
\label{eq:transversemetriceqn}
\frac{d}{dr}\left[\frac{h^{d/2+1}}{\omega^2h-k^2r^2f}\left(r^2f\psi_2'-i\omega\psi_2\right)\right]+\frac{h^{d/2}}{\omega^2h-k^2r^2f}\left(-i\omega h\psi_2'-k^2\psi_2\right)=0,
\end{equation}
for the field
\begin{equation}
\label{eq:psi2defn}
\psi_2\equiv\frac{1}{h(r)}\left(\omega \delta g_{xy}+k\delta g_{vy}\right).
\end{equation}
This equation is very similar to the equation \eqref{eq:GIlongeq} for Maxwell field perturbations. Note that the scalar profile $\Phi(r)$ does not enter explicitly in this equation, and thus the pole-skipping points can be expressed in terms of the metric functions only.

\paragraph{} Performing a very similar analysis to that for the Maxwell field (see Appendix \ref{sec:transversemetricappendix}), we again find that there is pole-skipping at the frequencies $\omega_n=-i2\pi Tn$ for appropriate values of the wavenumber $k_n$. The first instance of pole-skipping occurs when
\begin{equation}
\label{locationvisc}
k_1^2 = d \pi T h'(r_0).
\end{equation}

\paragraph{}As in the previous subsection, we will again demonstrate that the dispersion relation $\omega_h(k)$ of a hydrodynamic mode passes through pole-skipping points. The transverse momentum correlator $G^R_{T^{ty}T^{ty}}(\omega,k)$ has a hydrodynamic pole corresponding to the diffusion of momentum with the small-$k$ dispersion relation
\begin{equation}
\label{viscdisp}
\omega_h(k) = - i D_{p} k^2 + \dots,
\end{equation}
where the shear viscosity $\eta$ sets the momentum diffusion constant such that $D_p=\eta/(sT)=1/(4\pi T)$ (see e.g.~\cite{Iqbal:2008by}).

\paragraph{} For the simplest case of the AdS$_{d+2}$-Schwarzschild metric \eqref{eq:schwarzsec2} with $\Phi=0$, one of the $n$ values of $k_n^2$ is positive for each frequency $\omega_n$ (see Appendix \ref{sec:transversemetricappendix}). For the particular case of AdS$_4$-Schwarzschild, the first few pole-skipping locations with real $k_n$ are
\begin{equation}
\begin{aligned}
\label{eq:poleskippinglocstransversemom}
k_1^2=3r_0^2,\quad\quad\quad k_2^2=3\sqrt{2}r_0^2,\quad\quad\quad k_3^2=3\sqrt{3}r_0^2.
\end{aligned}
\end{equation}
In the right hand panel of Figure~\ref{fig:diffusion} we show the exact dispersion relation $\omega_h(k)$ for real $k$, overlaid with the diffusive approximation  \eqref{viscdisp} and the pole-skipping locations \eqref{eq:poleskippinglocstransversemom}. This again shows that the $\omega_h(k)$ passes through a succession of pole-skipping locations as $k$ is increased, and thus that a simple analysis of near-horizon boundary conditions provides a series of non-perturbative constraints on how the hydrodynamic mode behaves at energy scales $\omega\sim T$.

\paragraph{}While we have only presented numerical results for the AdS$_4$-Schwarzschild black brane, we expect these results are representative of those for higher-dimensional generalisations of this solution.\footnote{Note added: Numerical results analogous to ours were found in \cite{Grozdanov:2019uhi} for the AdS$_5$-Schwarzschild black brane.} It would again be worthwhile to investigate whether the pole-skipping points with non-real values of $k_n$ are also related to the dispersion relation $\omega_h(k)$ of the hydrodynamic mode.

\subsection*{Energy density and longitudinal momentum density}
\label{sec:poleskippingchaos}

\paragraph{} Finally, we turn to the Green's function of the longitudinal momentum density $G^R_{T^{tx}T^{tx}}(\omega,k)$, which is related to the retarded Green's function of energy density by the Ward identity
\begin{equation}
G^R_{T^{tx}T^{tx}}=\frac{\omega^2}{k^2}G^R_{T^{tt}T^{tt}}(\omega,k).
\end{equation}
The pole-skipping properties of these correlators were studied in \cite{Grozdanov:2017ajz, Blake:2017ris, Blake:2018leo,Grozdanov:2018kkt}, motivated by their close relation to the many-body quantum chaotic properties of the system. In particular, it was shown in \cite{Blake:2018leo} that for gravity coupled to very general matter fields they exhibit pole skipping in the upper half of the complex $\omega$ plane at the location
\begin{equation}
\label{skippinggrav}
\omega_* = +i2\pi T,\quad\quad\quad k_*^2 = -\left(\frac{2\pi T}{v_B}\right)^2=-d\pi T h'(r_0),
\end{equation}
where $v_B$ is the butterfly velocity associated to many-body chaos. This pole-skipping arises due to the non-uniqueness of ingoing solutions to the relevant equations of motion at this special point in Fourier space, as in the other examples we have discussed in this paper.

\paragraph{} We will not repeat the arguments of \cite{Blake:2018leo} here, but instead we will show that $G^R_{T^{tt}T^{tt}}(\omega,k)$ and $G^R_{T^{tx}T^{tx}}(\omega,k)$ also exhibit pole-skipping in the lower half of the complex plane at frequencies $\omega_n$ and appropriate wavenumbers $k_n$, as for all of the other examples described in this paper.

\paragraph{}For simplicity, we will consider the AdS$_{d+2}$-Schwarzschild solutions \eqref{eq:schwarzsec2} to Einstein-Hilbert gravity with a negative cosmological constant. The relevant metric perturbations are $\delta g_{vv},\delta g_{vx}$ and those that they couple to. After Fourier transforming and solving algebraically for $\delta g_{rr}, \delta g_{vr}, \delta g_{xr}$, the dynamics of these fields reduce to the single second-order differential equation
\begin{equation}
\begin{aligned}
\label{eq:longitudinalscalareq}
&\frac{d}{dr}\left[\frac{r^d\left(r^2f\psi_3'-i\omega\psi_3\right)}{\left(\omega^2-k^2f-\frac{k^2}{2d}rf'(r)\right)^2}\right]+\frac{r^{d-2}}{\left(\omega^2-k^2f-\frac{k^2}{2d}rf'(r)\right)^2}\left(-i\omega r^2\psi_3'-k^2\psi_3\right)\\
&-\frac{\left(d-1\right)k^2r^{d+2}{f'(r)}^2}{2d\left(\omega^2-k^2f-\frac{k^2}{2d}rf'(r)\right)^3}\psi_3=0,
\end{aligned}
\end{equation}
for the field
\begin{equation}
\psi_3\equiv \frac{1}{r^2}\left(2\omega k\delta g_{vx}+\omega^2 \delta g_{xx}+k^2\delta g_{vv}-\frac{\left(\omega^2-k^2f-\frac{1}{2}k^2rf'(r)\right)}{d-1}\delta g_{x^ix^i}\right),
\end{equation}
where $i=1,\ldots,d$.
Having written the relevant equation \eqref{eq:longitudinalscalareq} in a form similar to that of the scalar equation \eqref{scalareqEF}, we can perform similar analyses to that of Sections \ref{sec:scalarfield} and \ref{sec:higherpoleskipping} to uncover the pole-skipping locations. This is described in Appendix \ref{sec:longitudinalmetricappendix}.

\paragraph{}The results are that, in addition to the pole-skipping point \eqref{skippinggrav} in the upper half plane, there is also pole-skipping in the lower half plane at $\omega_n$ and $k=k_n$, where the first few values of $k_n$ obey the polynomial equations
\begin{equation}
\begin{aligned}
\label{eq:longitudinalmetrickn}
0=&\;k_1^4-\left(d-2\right)\left(d+1\right)k_1^2r_0^2+\frac{d^2\left(d+1\right)^2}{4}r_0^4,\\
0=&\;k_2^4-2\left(d-2\right)\left(d+1\right)k_2^2r_0^2+\left(d+1\right)^2d\left(d-1\right)r_0^4,\\
0=&\;k_3^6-\frac{1}{2}\left(d-12\right)\left(d+1\right)k_3^4r_0^2-\frac{1}{4}\left(d+1\right)^2\left(21d^2-56d+16\right)k_3^2r_0^4\\
&+\frac{3}{8}d\left(d+1\right)^3\left(15d^2-28d+16\right)r_0^6.
\end{aligned}
\end{equation}
The order of the polynomial equation for $k_1$ is different to the previous cases we have discussed due to the more complicated equation of motion for $\psi_3$. \RD{I have not really thought exactly about where this comes from in terms of the near-horizon expansion}

\paragraph{}We emphasise that while the pole-skipping in the upper half of the complex $\omega$ plane at \eqref{skippinggrav} is obscured by formulating the dynamics in terms of the field $\psi_3$, it also arises due to the non-uniqueness of ingoing solutions and can be seen transparently in the fundamental form of the Einstein equations themselves (as described in \cite{Blake:2018leo}). See Appendix \ref{sec:longitudinalmetricappendix} for how the upper half-plane pole-skipping point \eqref{skippinggrav} can be derived from a careful near-horizon analysis of equation \eqref{eq:longitudinalscalareq}.

\paragraph{}In \cite{Grozdanov:2017ajz} it was shown numerically for the case of AdS$_5$-Schwarzschild that the dispersion relation of hydrodynamic sound passes through the upper half-plane pole skipping point \eqref{skippinggrav}. It would be interesting to determine whether the poles passing through the pole-skipping points \eqref{eq:longitudinalmetrickn} in the lower half-plane are related to the hydrodynamic poles, and also how the locations \eqref{eq:longitudinalmetrickn} change upon the inclusion of bulk matter fields.

\section{Discussion}
\label{sec:discussion}

\paragraph{} In this paper we have shown that a simple analysis of the near-horizon properties of classical perturbations leads to a series of non-trivial constraints on the properties of holographic Green's functions at frequencies $\omega\sim T$. In particular, we have demonstrated that at the negative Matsubara imaginary frequencies $\omega_n=-i2\pi Tn$ $(n=1,2,3,\ldots)$ and appropriate complex wavenumbers $k_n$, the retarded Green's functions of generic bosonic operators typically have the `pole-skipping' form \eqref{greensfunctionnear}. As a consequence, the dispersion relations $\omega(k)$ of poles and zeroes of the retarded Green's functions are constrained such that one of each must pass through every pole-skipping point $(\omega_n,k_n)$. In a number of simple examples, we illustrated that short-distance properties of the dispersion relations of hydrodynamic modes (at real values of $k$) are captured by our pole-skipping analysis.

\paragraph{}To close our paper we will now place our results in the context of the previous work \cite{Grozdanov:2017ajz, Blake:2017ris, Blake:2018leo,Grozdanov:2018kkt} that studied instances of pole-skipping in the context of many-body quantum chaos, and also outline a number of interesting open questions that deserve further study.

\subsection*{Field theory interpretation}

\paragraph{} As we mentioned in the introduction, the instances of pole-skipping described in this paper are qualitatively different to those discovered in the retarded Green's function of the energy density in \cite{Grozdanov:2017ajz, Blake:2017ris, Blake:2018leo,Grozdanov:2018kkt}. Unlike the cases discussed in this paper, the pole-skipping point identified in \cite{Grozdanov:2017ajz, Blake:2017ris, Blake:2018leo,Grozdanov:2018kkt} is universally related to the exponential growth observed in out-of-time-ordered correlators of the theory, a feature that is also predicted by the hydrodynamic effective theory of chaos proposed in \cite{Blake:2017ris}. Specifically the pole-skipping frequency $\omega$ identified in \cite{Grozdanov:2017ajz, Blake:2017ris, Blake:2018leo,Grozdanov:2018kkt} lies at a location in the upper half of the complex plane related to the Lyapunov exponent, while the pole-skipping wavenumber $k$ is universally related to the butterfly velocity.\footnote{In holographic theories, this happens because the Einstein equation responsible for determining the gravitational shock wave profile that controls the out-of-time-ordered correlators is the same equation that controls the location of the pole-skipping point \cite{Blake:2018leo,Grozdanov:2018kkt}.} The pole-skipping points described in this paper are in general unrelated to the exponentially growing mode and the butterfly velocity present in out-of-time-ordered correlators\footnote{While in some cases $k_n$ is related to $v_B$ (e.g.~\eqref{locationvisc}), this is only true for sufficiently simple bulk theories. As we demonstrate in Appendix~\ref{sec:chargedtransverseapp}, the pole-skipping wavenumber \eqref{locationvisc} is not robust to the generalisation to charged black holes.} and so we do not expect these cases are directly linked to chaos. Nevertheless, our general analysis of pole-skipping here provides context for appreciating the remarkable robustness of the results in \cite{Blake:2018leo}.

\paragraph{}It is clearly important to work to place our pole-skipping results in the context of quantum field theories more generally. With a better understanding of pole-skipping in quantum field theories, our conclusion that pole-skipping at $\omega_n$ is generic in thermal states with classical black hole descriptions could be used to help deduce when and why gravitational descriptions of quantum field theories exist. In this direction, further study of thermal states of CFTs would be very useful. In (1+1)d CFTs, pole-skipping occurs even in non-gravitational theories: for integer $\Delta$ it was shown that the thermal retarded Green's functions of scalar operators of a (1+1)d CFT in general are equivalent to those computed from the BTZ black hole \cite{Son:2002sd}, and so the pole-skipping properties are present even if there is not a gravitational description of the CFT.\footnote{Furthermore, in \cite{Haehl:2018izb} it was shown that the upper half-plane pole-skipping predicted in \cite{Blake:2017ris} is also present in all (1+1)d CFTs, although a large $c$ limit is required to identify the butterfly velocity $v_B$.} In order to more directly understand the field theory origin of these pole-skipping properties, it would also be very interesting to determine what pole-skipping properties are exhibited by higher-dimensional CFTs (see e.g.~\cite{Iliesiu:2018fao}) and the SYK chain model of \cite{Gu:2016oyy}.

\subsection*{Implications for hydrodynamics and transport}

\paragraph{} In Section \ref{sec:hydrocorrelators} we showed that the dispersion relations of hydrodynamic modes pass through pole-skipping points in simple holographic examples. It would be very advantageous to understand in general when it is the dispersion relation of the hydrodynamic modes that are constrained in this way as this would open a number of paths for a greater understanding of hydrodynamics and transport in holographic systems.

\paragraph{} First, it would allow us to determine whether the pole-skipping can be interpreted as arising due to underlying symmetries in a quantum effective action for hydrodynamic degrees of freedom. This was the case for the pole-skipping of the hydrodynamic mode in the energy density correlator studied in \cite{Blake:2017ris}, which is produced by the imposition of a non-perturbative shift symmetry in a quantum theory of hydrodynamics \cite{Crossley:2015evo,Glorioso:2017fpd}. The further study of higher-dimensional CFTs advocated previously would also be helpful in this regard.

\paragraph{} Second, the constraints imposed on the dispersion relations $\omega_h(k)$ of the hydrodynamic modes by the pole-skipping analysis could potentially be used to obtain constraints on the thermodynamic and transport coefficients of holographic systems. Within the realm of validity of the hydrodynamic gradient expansion, it is these coefficients that control the dispersion relations $\omega_h(k)$ and thus this may be possible if the pole-skipping points lie within this realm of validity (see \cite{Grozdanov:2019kge,Grozdanov:2019uhi} for work in this direction).

\paragraph{} We can already use the results of this paper to better understand the observations in \cite{Blake:2016wvh,Blake:2017qgd,Blake:2016jnn,Blake:2016sud,Davison:2018ofp,Davison:2018nxm} relating the diffusivities, $D$, of certain strongly interacting quantum field theories to horizon data. The pole-skipping arguments developed in \cite{Blake:2018leo} and Section~\ref{sec:hydrocorrelators} provide a more precise and general relationship between the dispersion relations of hydrodynamic poles in boundary Green's function and properties of the near-horizon geometry, that provides a new perspective on these previous results.  Assuming the dispersion relation of the hydrodynamic mode is relatively smooth up until $\omega \sim T$ then we can use the first pole-skipping location $(\omega_H,k_H)$ of this mode to obtain a natural speed $v = |\omega_H|/|k_H|$ and timescale $\tau = |\omega_H|^{-1}$ to characterise the diffusivity (i.e.~ $D\sim v^2\tau$ \cite{Hartnoll:2014lpa}). This reasoning (see also \cite{Blake:2017ris,Blake:2018leo}), combined with the result \eqref{skippinggrav} for the pole-skipping in the energy density retarded Green's function, therefore explains the form of the thermal diffusivity $D_T\sim v_B^2/T$ near a large variety of holographic quantum critical points \cite{Blake:2017qgd}. Furthermore, it was shown in \cite{Blake:2016wvh} that the diffusivity of transverse momentum also takes the form $D_p\sim v_B^2/T$ near quantum critical points of neutral holographic theories. This can now be similarly understood from the more precise pole-skipping condition \eqref{locationvisc} of the retarded transverse momentum correlation function. The fact that it is only the energy density pole-skipping point studied in \cite{Blake:2018leo} that is robustly related to $v_B$ is therefore consistent with the observations that the only diffusivity that is robustly related to $v_B$ is the thermal diffusivity \cite{Lucas:2016yfl,Baggioli:2016pia,Patel:2016wdy,Davison:2016ngz,Blake:2017qgd,Werman:2017abn,Guo:2019csw}. Further study of the regime of applicability of diffusive hydrodynamics and of pole-skipping in charged black holes (where a single Green's function has multiple hydrodynamic poles) would be helpful to sharpen these arguments.

\subsection*{Further constraints from near-horizon perturbations}

\paragraph{} Whilst we have given a thorough overview of the constraints on retarded Green's functions resulting from the properties of perturbations near the horizon, it has certainly not been exhaustive and there a number of related phenomena that we sketch below which are worthy of fuller investigation.

\paragraph{}The first concerns the properties of perturbations at frequencies $\omega_n$ but away from the pole-skipping momenta $k_n$. As we discussed in Section~\ref{sec:scalarmultisols}, at these points the solution proportional to $\phi_0$ contains logarithmic terms near the horizon and so the general ingoing solution depends on the single coefficient $\phi_n$ (see Appendix~\ref{app:logarithms} and specifically equation \eqref{inoutmode}). Furthermore, this solution is also the only regular solution in outgoing coordinates. That is there is one solution \eqref{inoutmode} which is regular in both ingoing and outgoing coordinates, and a second solution that (because of the logarithms) is not regular in either coordinate system. Analogous statements also hold at Matsubara frequencies in the upper half plane $\omega = i 2 \pi T n$ and general $k$. This implies that in general there is a non-trivial relationship between the retarded and advanced correlation functions $G^R_{\cal O \cal O}(\omega,k)$ and $G^A_{\cal O \cal O}(\omega,k)$ of holographic theories
\begin{equation}
\begin{aligned}
\label{constraint}
&\;G^R_{\cal O \cal O}( \omega, k) = G^A_{\cal O \cal O}(\omega, k) + \dots,   \;\;\;\;\;\;\;\;\;\;\;\;\;\; (\omega= \pm i 2 \pi T n, \;\;\;\;\; k^2 \neq k^2_n),\\
\end{aligned}
\end{equation}
where $\dots$ denote potential contact terms that may differ between the retarded and advanced functions.\footnote{It is simple to check that the analytic expressions for the boundary Green's function of scalar fields in BTZ studied in Section~\ref{sec:btz} exactly satisfy the identity \eqref{constraint} (without any extra contact terms).} For the exceptional case $k=k_n$ there is still a solution that is regular in both coordinate systems, but it is not the only regular solution and thus the Green's functions are not both uniquely defined there. From this we can conclude that the only poles of $G^R_{\cal O \cal O}(\omega,k)$ that pass through $\omega_n = - i 2 \pi T n$ at real $k$ are those found at pole-skipping points $k=k_n$, as $G^A_{\cal O \cal O}(\omega,k)$ has no poles in the lower half plane for real $k$. It would be interesting to investigate further consequences of the relation \eqref{constraint}, and also to determine the appropriate generalisations for $U(1)$ current and energy-momentum tensor Green's functions.\footnote{The relation \eqref{constraint} is not true for the small $\omega$ limit of the retarded Green's function of energy density in the SYK chain \cite{Gu:2016oyy}.}

\paragraph{} The second are the meaning of the `anomalous points' described in Sections~\ref{sec:matching} and~\ref{sec:anomaloussection2}. These are points $(\omega,k)$ at which the ingoing solution to the perturbation equations (with appropriate asymptotic boundary conditions) is not uniquely defined, but where nevertheless the retarded Green's function does not take the `pole-skipping' form \eqref{greensfunctionnear}. The pole-skipping form is not realised because whilst there is a unique solution slightly away from the anomalous point, this solution does not depend continuously on the direction $\delta\omega/\delta k$. We have encountered examples of anomalous points in both the BTZ and AdS-Schwarzschild spacetimes, and saw empirically that these points coincided with unusual analytic structures in the corresponding boundary retarded Green's functions. In the BTZ case discussed we found that two distinct poles of the Green's function intersected at the anomalous points (Appendix~\ref{app:integerdeltaBTZ}) while for a conserved U(1) current Green's function in Schwarzschild-AdS$_4$ (Appendix~\ref{app:anomalousgaugefield}) we found that a pole and a zero intersected. It would be interesting to calculate the generic form of Green's functions near anomalous points and to determine what implications this has for their analytic properties. As an immediate application, such analysis could tell us what is happening at the anomalous point identified for the retarded Green's function of energy density in Schwarzschild-AdS spacetimes (see Appendix~\ref{sec:longitudinalmetricappendix}).

\paragraph{}Thirdly, while in this paper we have exploited the one-sided prescription of \cite{Son:2002sd} for calculating retarded Green's functions in holographic theories, it would be illuminating to rephrase our discussion in terms of the more general real-time holography prescriptions of \cite{Herzog:2002pc,Skenderis:2008dh,Skenderis:2008dg,Son:2009vu,Glorioso:2018mmw,deBoer:2018qqm}. In addition to potentially giving us a clearer perspective on the origin of pole-skipping, this formulation would also be the starting point for a generalisation to higher-order correlation functions.

\paragraph{}Fourth, while we have shown that pole-skipping occurs for a variety of different operators there remain further interesting examples that we did not address. One natural extension would be to the case of fermionic operators. For example, the boundary retarded Green's function $G^R_{\psi \psi}(\om, k)$ dual to a bulk Dirac fermion of (non-half-integer) mass $m$ propagating in the BTZ spacetime \eqref{backgroundmetricbtz} is \cite{Iqbal:2009fd} \be
\label{fermion}
G^R_{\psi \psi}(\om, k) \propto {
 \Gamma\le({m\ov 2}+{1\ov 4}+{i(k-\om)\ov 4\pi T}\ri)
 \Gamma\le({m\ov 2}+{3\ov 4}-{i(k+\om)\ov 4\pi T}\ri) \ov
 \Gamma\le(-{m\ov 2}+{3\ov 4}+{i(k-\om)\ov 4\pi T}\ri)
 \Gamma\le(-{m\ov 2}+{1\ov 4}-{i(k+\om)\ov 4\pi T}\ri) }.
\ee
In a similar manner to our discussion in Section~\ref{sec:btz} then for non-half-integer $m$ the various Gamma functions in \eqref{fermion} give rise to lines of poles and zeroes in $G^R_{\psi \psi}(\omega,k)$ that intersect at locations
\begin{eqnarray}
\label{locationsferm}
  \om_n = -i\pi T (2n+1), \qquad k_{n,q_1} &=&  2\pi i T(n-2q_1+m), \nonumber \\  k_{n,q_2} &=&  2\pi i T(n + 1 -2q_2 - m),
\end{eqnarray}
for any $n \in \{0,1,\ldots\}$ and with $q_1 \in \{ 0, \ldots, n \}$, $q_2 \in \{ 1, \ldots, n \}$.\footnote{For $n=0$ there are no solutions in the $k_{n,q_2}$ branch of \eqref{locationsferm}.} We therefore find that this Green's function again exhibits pole-skipping, this time at fermionic Matsubara frequencies, and so we expect that the locations~\eqref{locationsferm} can similarly be derived from a near-horizon expansion of the fermionic bulk wave-equation. Another extension is to study correlation functions of higher spin operators: for instance it was observed in \cite{Haehl:2018izb} that the Green's function of a spin-3 current operator in (1+1)d CFTs exhibits pole-skipping at frequencies $\omega = \pm i 2 \pi T, \pm i 4 \pi T$.

\paragraph{}Finally, it would be interesting to uncover the implications of our reasoning when generalised to other types of spacetimes. One interesting generalisation would be to the spinning BTZ solution: its out-of-time-ordered correlators depend on both horizon radii \cite{Reynolds:2016pmi,Stikonas:2018ane,Poojary:2018esz,Jahnke:2019gxr} and so could be used to further clarify the relation between pole-skipping and chaos. A second area worthy of exploration would be spacetimes that are not asymptotically AdS. As it is the horizon of the spacetime (rather than the asymptotics) that is key in our analysis, our approach may prove useful for constraining the quasinormal mode spectra of more general spacetimes and of understanding general features of possible holographic field theory duals. However we note that the analogue of our continuous parameter $k$ is in many cases a discrete angular momentum number, and obtaining constraints may require us to treat this as a complex number.\footnote{See \cite{MaassenvandenBrink:2000iwh} for a related discussion of the subtleties of imposing ingoing boundary conditions on metric perturbations of the Schwarzschild black hole in (3+1)-dimensions.}

\acknowledgments{We are grateful to Nejc Ceplak, Saso Grozdanov, Hong Liu, and Andrei Starinets for helpful discussions. M.~B.~ received support from the Office of High Energy Physics of U.S. Department of Energy under grant Contract Number  DE-SC0012567. The work of R.~D.~is supported by the STFC Ernest Rutherford Grant ST/R004455/1 and by the STFC Consolidated Grant ST/P000681/1. The work of D.~V.~is supported by the STFC Ernest Rutherford Grant ST/P004334/1.

\appendix

\section{Ingoing and outgoing solutions at $\omega=\pm\omega_n$}
\label{app:logarithms}
\paragraph{} As we discussed in Section~\ref{sec:scalarfield}, the phenomenon of pole-skipping for a minimally coupled scalar field is intimately connected to the fact that at frequencies $\omega_n = - i 2 \pi T n$ the two naive power-law exponents in the near-horizon solution \eqref{powerlaws} both appear to give regular solutions. However, as is well known, this does not necessarily mean that both solutions to the wave-equation \eqref{scalareqEF} are regular at these frequencies. Since at $\omega_n = - i 2 \pi T n$ the two asymptotic power laws in \eqref{powerlaws} differ by an integer, one generically expects that there will be additional subleading logarithms in one of these solutions. Such logarithms result in derivatives of $\phi(r)$ diverging at the horizon and hence only one of the solutions to \eqref{powerlaws} (the one without logarithms) is really a regular solution of the form \eqref{regularexp} near the horizon.

\paragraph{} These logarithmic terms can be seen explicitly by constructing series solutions to \eqref{scalareqEF} at $\omega _n$ in an expansion around the horizon without directly imposing an ansatz of the form \eqref{regularexp}. The general solutions are of the form
\begin{equation}
\begin{aligned}
\label{solutionslogs}
\phi &=\phi_0\left[1+c_1(r-r_0)+\ldots+(r-r_0)^n\log(r-r_0)\det\mathcal{M}^{(n)}(\omega_n,k^2) \left(\tilde{c}_0+\tilde{c}_1(r-r_0)+\ldots\right)\right]\\
&+\phi_n(r-r_0)^n\left[1+d_1(r-r_0)+d_2(r-r_0)^2+\ldots\right],
\end{aligned}
\end{equation}
where $\phi_0$ and $\phi_n$ are free parameters and $\det\mathcal{M}^{(n)}(\omega_n,k^2)$ is the determinant of the matrix introduced in Section~\ref{sec:higherpoleskipping}. The coefficients $c_i,d_i,\tilde{c}_i$ have a fixed dependence on $n$, $k$, the background metric and the scalar mass.

\paragraph{}  For a generic choice of $k$ there is only one solution in \eqref{solutionslogs} that is regular (the one proportional to $\phi_n$), and a second solution which is not regular to due the logarithms (the one proportional to $\phi_0$). The solution that is regular in ingoing coordinates therefore generically takes the form\footnote{The fact that at $\omega_n = - i 2 \pi T n$ series solutions to the minimally coupled scalar wave-equation can have the leading near horizon behaviour $\phi = \phi_n (r- r_0)^n$ of an `outgoing' wave was previously observed in \cite{Horowitz:1999jd}. It was observed for metric perturbations of the Schwarzschild black hole in \cite{MaassenvandenBrink:2000iwh}, where the potential absence of logarithmic corrections was also discussed.}
\begin{equation}
\label{inoutmode}
\phi = \phi_n (r-r_0)^{n}\left[1+d_1 (r-r_0)+d_2 (r-r_0)^2+\ldots\right], \;\;\;\;\;\; (\omega= - i 2 \pi T n \;\;\;\;\; k^2 \neq k^2_n),
\end{equation}
which agrees precisely with our discussion below \eqref{horizonequation2}.

\paragraph{} However, for the purposes of pole-skipping, the key point is that at special values of $k^2=k_n^2$ then there are no logarithmic terms at all in the near-horizon expansion. This can explicitly be seen from the form of the general near horizon solutions in \eqref{solutionslogs}. Precisely at the wavenumbers $k^2 = k_n^2$ in \eqref{multiplelocations}, both solutions in \eqref{powerlaws} really do give rise to regular solutions of the form \eqref{regularexp} and there is therefore a two-parameter family of ingoing solutions of the form
\begin{equation}
\begin{aligned}
\label{solutionslogs2}
\phi &=\phi_0\left[1+c_1(r-r_0)+\ldots \right]+\phi_n(r-r_0)^n\left[1+d_1(r-r_0)+d_2(r-r_0)^2+\ldots\right],
\end{aligned}
\end{equation}
near the horizon. This is the origin of pole-skipping, as explained in Sections~\ref{sec:scalarfield} and \ref{sec:higherpoleskipping}.

\paragraph{} Note that for the case of $n=0$, there are always logarithmic terms irrespective of the value of $k$ and so for a scalar field there is no pole-skipping at this frequency.

\paragraph{} Whilst we focus mostly on the retarded Green's function in this paper, it is also of interest to consider the advanced Green's function $G^{A}_{\cal O O}(\omega,k)$ at frequencies $\omega_n = - i 2 \pi T n$. This can be extracted by constructing the outgoing solution to \eqref{scalareom}. In this case one finds that there is aways a unique outgoing solution to \eqref{scalareom} at $\omega_n$, which is simply the solution in \eqref{solutionslogs} proportional to $\phi_n$. Away from pole-skipping wavenumbers $k_n$ this solution coincides with the ingoing solution which results in the interesting identity between retarded and advanced Green's functions discussed in Section~\ref{sec:discussion}.

\paragraph{} Even at the special pole-skipping wavevectors $k_n^2$ there is still only a single outgoing solution, since the solution proportional to $\phi_0$ in \eqref{solutionslogs} is never regular in outgoing coordinates for any choice of $k$. The advanced Green's function $G^A_{\cal O \cal O}(\omega,k)$ therefore does not show pole-skipping in the lower half-plane. However by studying \eqref{scalareom} in outgoing coordinates it is simple to see that the entire pole-skipping analysis will be mirrored in outgoing coordinates if we swap $\omega \to - \omega$. $G^A_{\cal O \cal O}(\omega,k)$ will therefore exhibit pole-skipping in the upper half plane at the positive imaginary Matsubara frequencies $\omega= i 2 \pi T n$ and at the same wavevectors $k_n^2 $ in \eqref{multiplelocations}.

\section{Pole-skipping form of Green's functions}
\label{app:greensfunction}

\paragraph{} In this Appendix we wish to show explicitly how the matching argument in Section~\ref{sec:matching} leads to the pole-skipping form of the Green's function $G^R_{\cal O \cal O}(\omega,k)$ presented in \eqref{greensfunctionnear}. As we have argued in Section \ref{sec:scalarfield}, at the special locations in \eqref{kscalar} both linearly independent solutions to \eqref{scalareqEF} are consistent with ingoing boundary conditions. In particular there are ingoing solutions that are normalisable in the UV (i.e. have $\phi_A(\omega_1, k_1) = 0$) and also ingoing solutions with no normalisable component (i.e. with $\phi_B(\omega_1,k_1) = 0$).

\paragraph{} To be precise we define a normalisable solution $\phi^{(n)}$ as the solution to \eqref{scalareqEF} at \eqref{kscalar} such that we have $\phi_A(\omega_1, k_1) = 0$ and $\phi_B(\omega_1, k_1) = 1$. Similarly we define a solution with no normalisable component  $\phi^{(nn)}$ as the solution to \eqref{scalareqEF} at \eqref{kscalar} such that $\phi_A(\omega_1, k_1) = 1$ and $\phi_B(\omega_1, k_1) = 0$. Precisely at \eqref{kscalar} both of these solutions are consistent with ingoing boundary conditions and hence can be expanded near the horizon $r=r_0$ as series solutions of the form \eqref{regularexp}
\begin{eqnarray}\label{irexpappendix}
\phi^{(n)} &=& \phi_0^{(n)} + \phi_1^{(n)} (r - r_0) + \dots \nonumber, \\
\phi^{(nn)} &=& \phi_0^{(nn)} + \phi_1^{(nn)} (r - r_0) + \dots.
\end{eqnarray}

\paragraph{} Since all solutions to \eqref{scalareqEF} are consistent with ingoing boundary conditions the retarded Green's function $G^R_{\cal O \cal O}(\omega,k)$ is not well-defined at \eqref{kscalar}. To get a well-defined Green's function it is necessary to move infinitesimally away from \eqref{kscalar} to $\omega = \omega_1 + \epsilon \delta \omega$ and $k = k_1 + \epsilon \delta k$. After doing so there is a unique ingoing solution $\phi_{ig}(r)$ from which we can extract the Green's function $G^R_{\cal O \cal O}(\omega,k)$ near \eqref{kscalar}. In order to compute this Green's function we note that to leading order in $\epsilon$ we can express $\phi_{ig}(r)$ as a linear combination of $\phi^{(nn)}$ and $\phi^{(n)}$. Hence after choosing a convenient normalisation for $\phi_{ig}(r)$ we can write
\begin{equation}
\label{ingoingappendix}
\phi_{ig}(r) = \phi^{(nn)}(r) + {\cal B}(\delta \omega/\delta k) \phi^{(n)}(r),
\end{equation}
from which the Green's function can be extracted as
\begin{equation}
\label{greensfunctionnearapp}
G^R_{\cal O \cal O}(\omega_1 + \epsilon \delta \omega, k_1 + \epsilon \delta k) = (2 \Delta - d - 1 ) {\cal B}(\delta \omega/\delta k).
\end{equation}

All that remains is to determine the coefficient ${\cal B}(\delta \omega/\delta k)$. This can be achieved by inserting the expansions in \eqref{irexpappendix} into the equation \eqref{nearlocation}. This yields an explicit expression for ${\cal B}(\delta \omega/\delta k)$ in terms of the expansion parameters \eqref{irexpappendix} of the solutions $\phi^{(n)}$ and $\phi^{(nn)}$ to \eqref{scalareqEF} at \eqref{kscalar}
\begin{equation}
\label{bomegak}
{\cal B}(\delta \omega/\delta k) = - \frac{(i \delta \omega d h'(r_0) + 4 k_1 \delta k ) \phi_0^{(nn)} + 4 i h(r_0) \delta \omega \phi_1^{(nn)}}{(i \delta \omega d h'(r_0) + 4 k_1 \delta k) \phi_0^{(n)} + 4 i h(r_0) \delta \omega \phi_1^{(n)}},
\end{equation}
from which one can see that $G^R_{\cal O \cal O}(\omega,k)$ has both a line of poles and a line of zeroes passing through \eqref{kscalar}. Through simple algebra then \eqref{bomegak} or equivalently \eqref{greensfunctionnearapp} can be written in pole-skipping form \eqref{greensfunctionnear} with the slope $(\delta \omega/\delta k)_p$ of the line of poles given by \eqref{slope} and the slope $(\delta \omega/\delta k)_z$ of the line of zeroes given by an expression involving $\phi^{(nn)}_0, \phi^{(nn)}_1$. To explicitly determine the coefficients in \eqref{irexpappendix} which control the slopes $(\delta \omega/\delta k)_p$ and $(\delta \omega/\delta k)_z$, one must know the radial evolution of the normalisable and non-normalisable solutions. Therefore these slopes cannot be determined from just our near-horizon analysis.

\paragraph{} Whilst for the sake of clarity we have presented this explicit argument only for the case of $n=1$ pole-skipping points, an entirely analogous discussion can be applied to the higher order pole-skipping examples discussed in Section~\ref{sec:higherpoleskipping} so long as $\det\mathcal{{M}}^{(n)}(\omega_n, k_n^2) \neq 0$. In this case \eqref{horizonmultipleexp} can now be used to determine ${\cal B}(\delta \omega/\delta k)$ in terms of the expansion parameters $\phi^{(nn)}_0, \phi^{(nn)}_n, \phi^{(n)}_0, \phi^{(n)}_n$ of solutions to \eqref{kscalar} at \eqref{multiplelocations}. As such the Green's function near \eqref{multiplelocations}  again generically takes the pole-skipping form \eqref{greensfunctionnear}, albeit with more complicated expressions for the slopes $(\delta \omega/\delta k)_{p}, (\delta \omega/\delta k)_{z}$.

\section{Pole-skipping in BTZ with integer $\Delta$}
 \label{app:integerdeltaBTZ}

\paragraph{} Whilst the discussion in Section~\ref{sec:btz} holds for non-integer $\Delta$ it is necessary to perform a more careful analysis when $\Delta$ is an integer. In this case not all solutions $k_n^2$ to $\det\mathcal{M}^{(n)}(\omega_n, k^2) = 0$ necessarily correspond to conventional pole-skipping locations: for sufficiently large $n$ there are now also examples of `anomalous points' for which $\partial_{k} \det\mathcal{M}^{(n)}(\omega_n, k_n^2) = 0$.

\paragraph{} Whether or not there will be such anomalous points depends on the relative size of $n$ and $\Delta$. We will consider cases with $\Delta>0$. For $n < \Delta$ we find that there are no anomalous points, and that the solutions to $\det\mathcal{M}^{(n)}(\omega_n, k^2) = 0 $  give rise to conventional pole skipping at the $2 n$ wavevectors in \eqref{locations} exactly in the same manner as in the non-integer case. In contrast for $n \geq \Delta$ we find that the form of \eqref{detbtz} implies that there is only conventional pole-skipping at wavevectors corresponding to the largest $\Delta - 1$ values of $k^2_{n,q}$ in \eqref{detbtz}. For integer $\Delta$ we therefore expect conventional pole-skipping at the locations
\be
\label{locations2}
\om_n = -i 2\pi  T n, \qquad k_{n,q} = \pm 2\pi i T(n-2q+\Delta),
\ee
for $n \in \{1,2,\ldots\}$ and $q \in \{ 1, \ldots, \mathrm{min}( n, \Delta -1) \}$.

\paragraph{} For the case of $n \geq \Delta$ the anomalous points arise due to solutions to $\det\mathcal{M}^{(n)}(\omega_n, k^2) = 0 $ which correspond either to repeated roots for this equation or to $k_{n}^2 = 0$. Such solutions satisfy $\partial_{k} \det\mathcal{M}^{(n)}(\omega_n, k_n^2) = 0$ and therefore the matching procedure of Sections~\ref{sec:matching} and~\ref{sec:anomaloussection2} breaks down. In general for $n \geq \Delta$ we find there will be $(n - \Delta + 1)$ such anomalous points  $(\omega_n,  k_n)$ and that the locations of these points depends on whether $n - \Delta$ is zero, an odd integer or an even integer. For odd $n - \Delta $ then our near horizon analysis predicts there should be anomalous points at
\be
\label{locations3}
\om_n = - i 2\pi  T n, \qquad k_{n,q} = \pm 2 \pi i T (n-2q+\Delta),
\ee
with $n \in \{1,2,\ldots\}$ and $q \in \{ \Delta, \ldots, (n + \Delta - 1)/2  \}$. In contrast if $n - \Delta $ is an even positive integer then there should be anomalous points at
\be
\label{locations4}
\om_n = -i 2\pi  T n, \qquad k_{n,q} = 0, \pm 2\pi i T(n-2q+\Delta),
\ee
for  $n \in \{1,2,\ldots\}$ and $q \in \{ \Delta, \ldots, (n + \Delta - 2)/2   \}$. Finally if $n - \Delta = 0$ there will be a single anomalous point at the location
\be
\label{locations5}
\om_n = -i 2\pi  T n, \qquad k_{n} = 0.
\ee
\paragraph{} For integer $\Delta$ we can again compare the predictions of our near-horizon analysis to an exact analytic expression for the Green's function. In this case there are additional logarithmic terms in the bulk scalar wavefunction and the expression for the boundary Green's function $G^R_{\cal O \cal O}(\omega,k)$ in \eqref{btzgreensfunction} is modified to
{\footnotesize
\be
\label{integerdelta}
  G^R_{\cal O \cal O}(\om, k) \propto {
 \Gamma\le({\Delta \ov 2}+{i(k-\om)\ov 4\pi T}\ri)
 \Gamma\le({\Delta \ov 2}-{i(k+\om)\ov 4\pi T}\ri) \ov
 \Gamma\le(1 - { \Delta \ov 2}+{i(k-\om)\ov 4\pi T}\ri)
 \Gamma\le(1 - { \Delta  \ov 2}-{i(k+\om)\ov 4\pi T}\ri) }
   \le[  \psi\le({\Delta \ov 2}+{i(k-\om)\ov 4\pi T}\ri) + \psi\le({\Delta \ov 2}-{i(k+\om)\ov 4\pi T}\ri) \ri],
\ee
}
where $\psi(z)$ is the digamma function. One needs to be careful in analysing the lines of zeroes and poles in \eqref{integerdelta} because the arguments of the Gamma functions in the prefactor of \eqref{integerdelta} now differ by an integer and hence there can be cancellations between poles in the numerator and denominator. Specifically for integer $\Delta$  the ratio of Gamma functions in \eqref{integerdelta} can be simplified using the identity $\Gamma(z +1 ) =z \Gamma(z) $ to write
\begin{equation}
\label{eq:gammaprefactors}
\hskip -0.3cm
 \frac{\Gamma\le({\Delta \ov 2}+{i(k-\om)\ov 4\pi T}\ri)}{
 \Gamma\le(1 - { \Delta \ov 2}+{i(k-\om)\ov 4\pi T}\ri)}
 = \underbrace{\bigg(1 - { \Delta \ov 2}+{i(k-\om)\ov 4\pi T} \bigg)\bigg(2 - { \Delta \ov 2}+{i(k-\om)\ov 4\pi T} \bigg) \dots \bigg({ \Delta \ov 2} - 1+{i(k-\om)\ov 4\pi T} \bigg)}_{\Delta -1 \;\; \mathrm{factors}},
\end{equation}
and similarly one can obtain an analogous expression for the ratio of the other two Gamma functions in \eqref{integerdelta}.

\paragraph{}As such for integer $\Delta$ the ratios of Gamma functions in \eqref{integerdelta} does not contribute any poles, but just gives rise to $2 \Delta - 2$ lines of zeroes with dispersion relations
\be
\label{integerzeroes}
  \om^z_{L, m} = k-2\pi i T (2 - \Delta + 2 m),   \qquad  \om^z_{R, m} = -k-2\pi i T (2 - \Delta + 2 m), \ee
for $m \in \{0,1,\ldots\ \Delta - 2\}$.

\paragraph{} In addition to these lines of zeroes there are also lines of poles in \eqref{integerdelta} which now come from the digamma functions in \eqref{integerdelta}. These give rise to infinitely many lines of poles along
\be
\label{integerpoles}
  \om^{p}_{L, m} = k-2\pi i T (\Delta + 2 m),   \qquad  \om^{p}_{R, m} = -k-2\pi i T (\Delta  + 2 m), \ee
for $m \in \{0,1, 2 \ldots \}$. These lines of poles intersect with the lines of zeroes in \eqref{integerzeroes} at
\be
\label{locationsint}
\om_n = -i 2\pi  T n, \qquad k_{n,q} = \pm 2\pi i T(n-2q+\Delta),
\ee
for $n \in \{1,2,\ldots\}$ and $q \in \{ 1, \ldots, \mathrm{min}(n, \Delta -1) \}$ and hence we have pole-skipping exactly at the locations predicted by our near horizon analysis \eqref{locations2}. This pattern of pole-skipping is demonstrated in Figure~\ref{fig:intersectionsdelta3} in which we have plotted the lines of poles and zeroes in \eqref{integerzeroes} and \eqref{integerpoles} for the special case of $\Delta = 3$. Note that when $\Delta=1$ there are no zeroes coming from the prefactors \eqref{eq:gammaprefactors} and thus no pole-skipping points, which is consistent with our near-horizon analysis.

 \begin{figure}
 \begin{tabular}{cc}
\includegraphics[width=.45\textwidth]{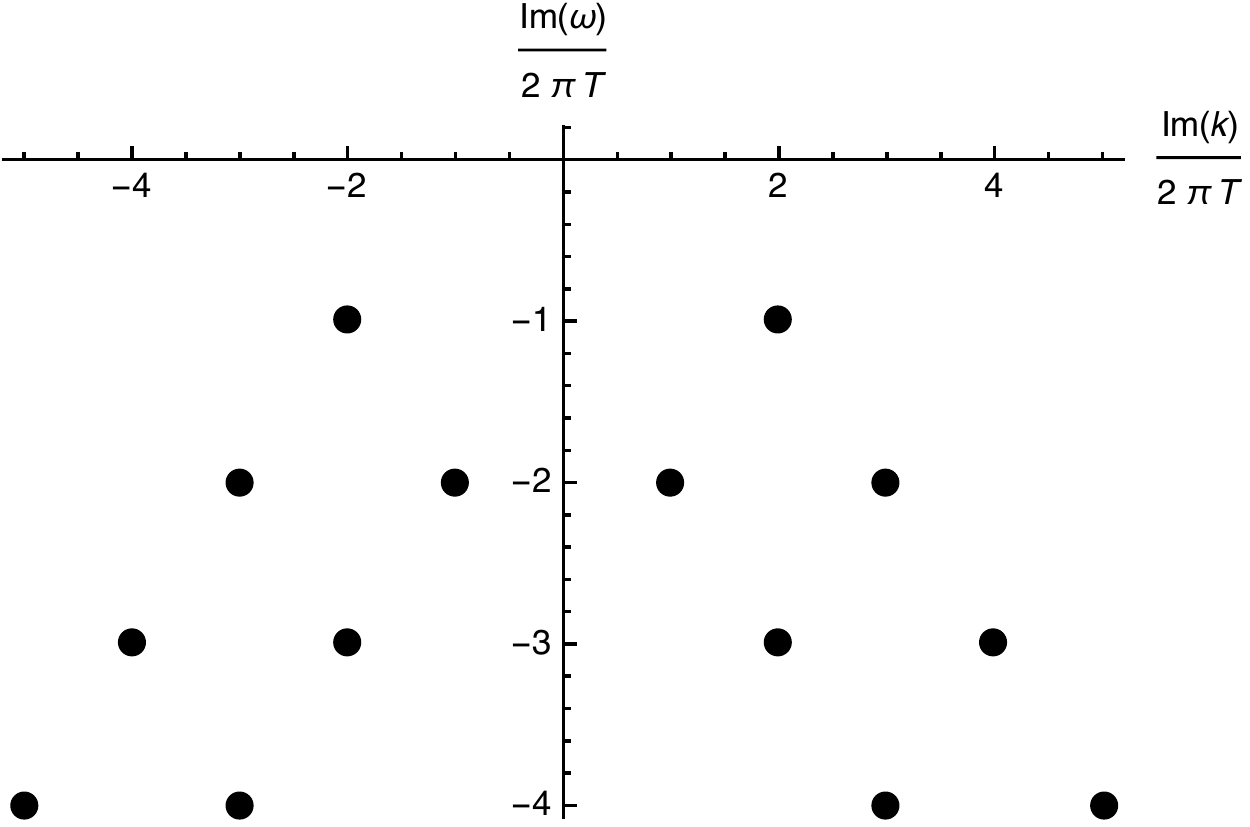}&
\includegraphics[width=.45\textwidth]{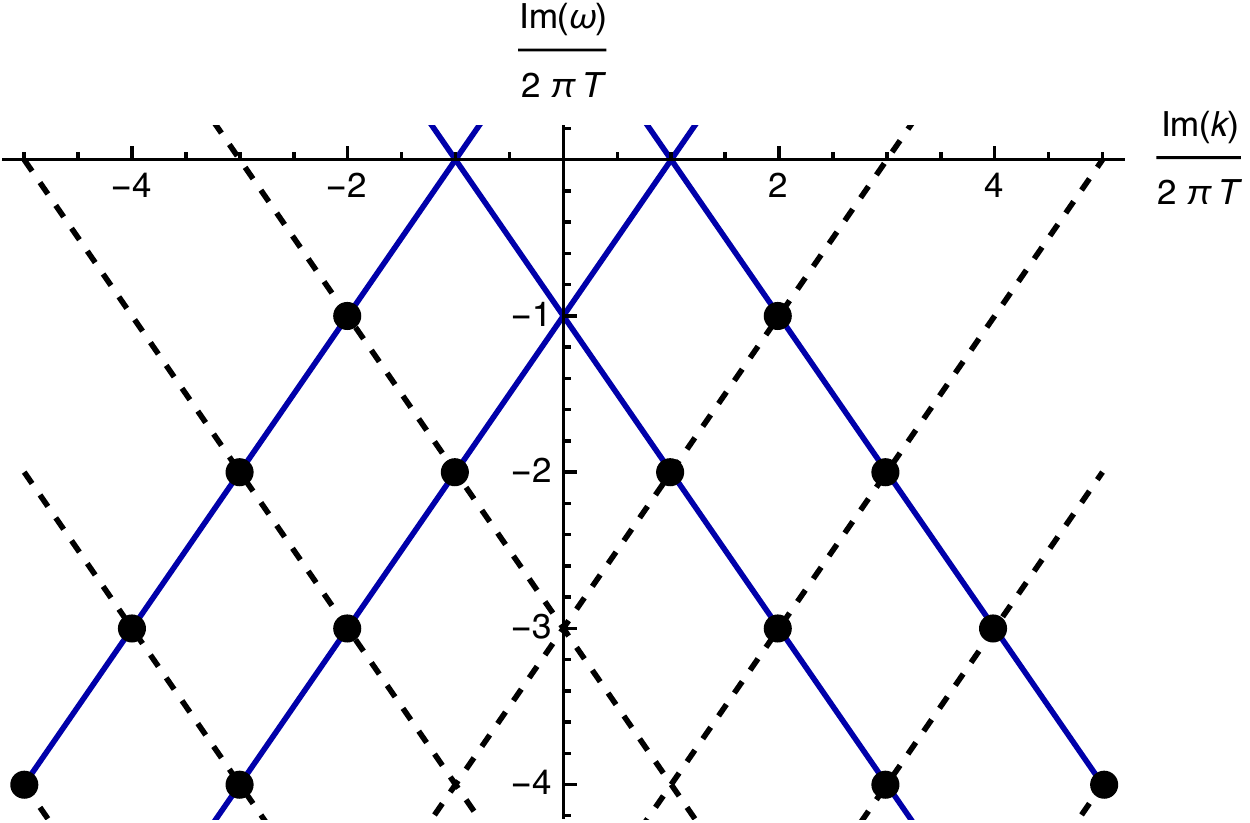}
\end{tabular}
\caption{The left hand plot shows the pole-skipping locations \eqref{locations2} predicted from our near horizon analysis for $\Delta = 3$ and $n=1,2,3,4$. The right hand plots shows the lines of zeroes (blue) and poles (dashed) in the $\Delta = 3$ Green's function \eqref{integerpoles}. The intersections of these lines give rise to $\mathrm{min}(2n, 2 \Delta - 2 )$ instances of pole skipping at frequencies $\omega_n  = - i 2 \pi T n$.}
\label{fig:intersectionsdelta3}
\end{figure}

Finally we will examine what happens in the expression \eqref{integerdelta} near the locations \eqref{locations3} \eqref{locations4} and \eqref{locations5} at which our near horizon analysis found `anomalous points'. Intriguingly we find that at each of these locations there is an intersection of one of the left-moving poles and one of the right-moving poles of \eqref{integerpoles}. This can be seen in the plot of the lines poles and zeroes in the $\Delta = 3$ Green's function in Figure~\ref{fig:intersectionsdelta3}, for which there are anomalous points at $(\omega, k) = (-  i 6 \pi T, 0)$ and $(\omega, k) = (-  i 8 \pi T,  \pm i 2 \pi  T)$. As we discuss in Section~\ref{sec:discussion}, it would be interesting to further study these anomalous points in future work.

\section{Exact scalar Green's functions in BTZ}

\label{app:btz}

For completeness, in this Appendix, we rederive the real-time Green's function of a scalar field in the three-dimensional BTZ black hole background. These calculations were originally done in \cite{Son:2002sd} (see also \cite{Birmingham:2001hc, Cardoso:2001hn, Birmingham:2001pj, Herzog:2002pc, Skenderis:2008dg, vanRees:2009rw}).

The metric of the non-extremal BTZ black hole \cite{Banados:1992wn, Banados:1992gq} is given by
\be
  \nonumber
  ds^2 = - {(r^2-r_+^2) (r^2-r_-^2) \ov r^2}  dt^2 +
  {r^2 dr^2 \ov (r^2-r_+^2) (r^2-r_-^2) } + r^2 \le(d \theta - {r_+ r_- \ov r^2} dt \ri)^2 ,
\ee
where $r=r_\pm$ are the locations of the inner and outer horizons. The geometry is locally AdS$_3$. The mass and angular momentum of the black hole are related to the horizon radii via
\be
  \nonumber
M = {r_+^2 + r_-^2 \ov 8 G_N}, \qquad J = {r_+ r_- \ov 4 G_N} ,
\ee
where $G_N$ is Newton's constant.
The dual 2d CFT has non-vanishing left and right temperatures
\be
  \nonumber
 T_L = \frac{r_+ - r_-}{2\pi}, \qquad T_R = \frac{r_+ + r_-}{2\pi} .
\ee
It is convenient to switch to another coordinate system $(t,\theta, r) \to (T, X, \rho)$ defined by
\bea
  \nonumber
  r^2 &=& r_+^2 \cosh^2 {\rho} - r_-^2 \sinh^2 {\rho} , \\
  \nonumber
  T + X &=& (r_+ + r_-)(t + \theta) , \\
  \nonumber
  T - X &=& (r_+ - r_-)(t-\theta) .
\eea
The metric in terms of these coordinates simplifies considerably
\be
  \nonumber
   ds^2 = - \sinh^2 \rho \,  dT^2 + \cosh^2 \rho \, dX^2  + d \rho^2 .
\ee
We will now consider a massive scalar field $\varphi$ on this rigid background and take a plane wave ansatz on constant $\rho$ slices. The plane wave can be written in either the new $(T,x)$ or the old $(t,\theta)$ coordinates
\be
  \nonumber
  \varphi(T, X, \rho)  = e^{-ik_T T +i  k_X X} \varphi (\rho)  = e^{-i\omega t + i k\theta}\varphi(\rho) ,
\ee
where the momenta $(\om,k)$ are related to $(k_T, k_X)$ by
\be
  \nonumber
k_T + k_X = \frac{\omega + k}{2\pi T_R}, \qquad k_T - k_X = \frac{\omega - k}{2\pi T_L} .
\ee
Even though $\theta$ is an angular  variable, in the following we will view the conjugate momentum $k$ as a continuous parameter.
The wave equation for $\varphi (\rho)$ turns out to be
\be
  \nonumber
  \varphi''(\rho)+ 2 \coth 2\rho \, \varphi'(\rho) + \le( {k_T^2 \ov \sinh^2 \rho}-{k_X^2 \ov \cosh^2 \rho} - m^2 \ri) \varphi(\rho)= 0 .
\ee
After changing to a new radial coordinate given by $z = \tanh^2 \rho$, we get
\be
  \nonumber
  \varphi''(z) + {\varphi'(z) \ov z} + \le[{k_T^2 \ov 4z^2(1-z)} -{k_X^2 \ov 4z(1-z)} -{m^2 \ov 4z(1-z)^2}  \ri] \varphi(z) = 0 .
\ee
In this coordinate system, the event horizon is located at $z=0$ while the boundary of spacetime is at $z=1$.

\subsection*{The general case}

For generic values of $k_T$ and $k_X$ the ingoing solution is given by
\be
  \label{eqn:ing}
  \varphi_\textrm{in}(z) = z^{-{i k_T \ov 2}} (1-z)^{\Delta_- \ov 2} {}_2F_1\le({k_T-k_X\ov 2i}+{\Delta_- \ov 2},{k_T+k_X\ov 2i}+{\Delta_- \ov 2}; \, 1 - i k_T; \, z \ri) ,
\ee
while the outgoing solution is
\be
  \label{eqn:outg}
  \varphi_\textrm{out}(z) = z^{+{i k_T \ov 2}} (1-z)^{\Delta_- \ov 2} {}_2F_1\le(-{k_T-k_X\ov 2i}+{\Delta_- \ov 2},-{k_T+k_X\ov 2i}+{\Delta_- \ov 2}; \, 1 + i k_T; \, z \ri) .
\ee
Near the event horizon these solutions are ingoing or outgoing waves
\be
  \nonumber
  \varphi_\textrm{in}(z) \propto z^{-{i k_T \ov 2}}, \qquad
  \varphi_\textrm{out}(z) \propto z^{+{i k_T \ov 2}} .
\ee
Near the boundary they generically behave as
\be
  \label{eqn:behave}
  \varphi(z) \approx (1-z)^{\Delta_+ \ov 2}\le[c_+^{(0)} + c_+^{(1)}(1-z) + \ldots\ri] +  (1-z)^{\Delta_- \ov 2} \le[c_-^{(0)} +  c_-^{(1)}(1-z) + \ldots\ri] ,
\ee
where $\Delta_\pm = 1\pm \sqrt{1+m^2}$. In normal (alternative) quantization, $\Delta_+$ ($\Delta_-$) is the dimension of the bosonic operator dual to the bulk scalar field. In the following, we will consider $\Delta>0$.

Up to an unimportant constant factor, the retarded (advanced) Green's function is computed by taking the ratio $c_+^{(0)} /c_-^{(0)}$ for the ingoing (outgoing) solution. The expansion of the hypergeometric functions near $z\approx 1$ gives
\be
  \label{eqn:genericret}
 G_R(k_T, k_X) \propto {
 \Gamma\le({\Delta_+ \ov 2}-i{k_T+k_X\ov 2}\ri)
 \Gamma\le({\Delta_+ \ov 2}-i{k_T-k_X\ov 2}\ri) \ov
 \Gamma\le({\Delta_- \ov 2}-i{k_T+k_X\ov 2}\ri)
 \Gamma\le({\Delta_- \ov 2}-i{k_T-k_X\ov 2}\ri) } ,
\ee
for the retarded Green's function, and
\be
  \nonumber
 G_A(k_T, k_X) \propto {
 \Gamma\le({\Delta_+ \ov 2}+i{k_T+k_X\ov 2}\ri)
 \Gamma\le({\Delta_+ \ov 2}+i{k_T-k_X\ov 2}\ri) \ov
 \Gamma\le({\Delta_- \ov 2}+i{k_T+k_X\ov 2}\ri)
 \Gamma\le({\Delta_- \ov 2}+i{k_T-k_X\ov 2}\ri) },
\ee
for the advanced Green's function.
In alternative quantization \cite{Klebanov:1999tb}, one obtains the reciprocal of these functions, which exchanges poles and zeroes.

\subsection*{Pole-skipping}

If we set $T = T_L = T_R$, then the black hole is static. In this case, one obtains the following retarded Green's function in terms of $\om$ and $k$.\footnote{The Green's function is proportional to (4.16) in \cite{Son:2002sd} if $\om$ and $k$ are both real.}
\be
  \nonumber
 G_R(\om, k) \propto {
 \Gamma\le({\Delta_+ \ov 2}+{i(k-\om)\ov 4\pi T}\ri)
 \Gamma\le({\Delta_+ \ov 2}-{i(k+\om)\ov 4\pi T}\ri) \ov
 \Gamma\le({\Delta_- \ov 2}+{i(k-\om)\ov 4\pi T}\ri)
 \Gamma\le({\Delta_- \ov 2}-{i(k+\om)\ov 4\pi T}\ri) } .
\ee
As discussed in the main text, pole-skipping occurs at special values of the frequency and wavenumber where poles of the Gamma functions in the numerator and the denominator coincide. This gives a series of pole-skipping points
\be
  \label{eqn:pskip}
  \om_n = -i 2\pi T n, \qquad k_{n,q} = \pm i 2\pi T(n-2q+\Delta_+) ,
\ee
for any $n \in \{1,2,\ldots\}$ and $q \in \{ 1, \ldots, n \}$. In alternative quantization one simply needs to exchange $\Delta_+ \leftrightarrow \Delta_-$.


\subsection*{At Matsubara frequencies}

At generic $k_X$, the hypergeometric function in (\ref{eqn:ing}) is well-defined unless its third argument $ 1 - i k_T$ is a non-positive integer. Let us now investigate what happens at such points by taking the limit $i k_T \to n$ where $n$ is a positive integer. For the non-spinning black hole, these values correspond precisely to the Matsubara frequencies $\om_n= -i 2\pi T n$.
As we take $i k_T \to n$, the ingoing solution blows up. A finite limit can be defined by dividing by another infinite factor (which gives a regularized hypergeometric function)
\be
 \nonumber
   \tilde \varphi_\textrm{in}(z) \equiv  \lim_{i k_T \to n} {\varphi_\textrm{in}(z) \ov\Gamma(1 - i k_T)} .
\ee
Although this is well-defined, the two solutions are now degenerate
\be
  \nonumber
  \tilde \varphi_\textrm{in}(z) =
  {
  \Gamma\le({\Delta_+ + n - i k_X \ov 2}\ri)
  \Gamma\le({\Delta_+ + n + i k_X \ov 2}\ri)
  \ov
   \Gamma(1+n) \
  \Gamma\le({\Delta_+ - n - i k_X \ov 2}\ri)
  \Gamma\le({\Delta_+ - n + i k_X \ov 2}\ri)
  }
  {\varphi_\textrm{out}(z) } .
\ee
Consequently, the retarded and advanced Green's functions are equal at these frequencies.
Another, independent, solution is provided by the Meijer G-function
\be
 \nonumber
   \tilde\varphi_\textrm{out}(z) =  z^{-{n \ov 2}} (1-z)^{\Delta_- \ov 2} G_{2,2}^{2,0}
   \left( \le. \begin{array}{cc}
{\Delta_+ + n - i k_X \ov 2} & {\Delta_+ + n + i k_X \ov 2} \\
0 & n
\end{array} \right|
z
\ri) .
\ee
The near-horizon expansion of the G-function contains a logarithm at the expected order (see Appendix \ref{app:logarithms}). Its coefficient vanishes at pole-skipping $k_X$ values and the function becomes regular.
In fact, the solutions drastically simplify at such points. For instance, if we pick $n=k=1$ then (\ref{eqn:pskip}) gives $k_T = -i$ and $k_X = \pm i (\Delta_+ -1)$. The two independent solutions can be chosen to be
\be
  \varphi_\pm(z) = { (1-z)^{\Delta_\pm / 2} \ov \sqrt{z}} ,
\ee
which is related to our earlier basis in (\ref{eqn:ing}), (\ref{eqn:outg}) via
\be
\nonumber
  \varphi_\textrm{in}(z)|_\textrm{pole-skipping} = \varphi_+(z) , \qquad
  \varphi_\textrm{out}(z)|_\textrm{pole-skipping} = {\varphi_-(z) - \varphi_+(z)\ov  \Delta_+-1 } .
\ee

\newpage

\subsection*{At integer $\Delta$}

At integer $\Delta_+$ values, the calculation of the Green's function is slightly more involved.
In this case, the exponents in (\ref{eqn:behave}) differ by an integer and logarithms appear in the near-boundary expansion. This is related to matter conformal anomalies.

The following expansion of the hypergeometric function is relevant in this case\footnote{The expansion above is valid for $\Delta_+ \ge 2$. At $\Delta_+ = 1$ one can instead use
\be
  \nonumber
  {}_2F_1(a, b; \, a+b; \, z) =  {  \Gamma(a+b) \ov \Gamma(a)\Gamma(b)}
   \sum_{j=0}^\infty
  {(a)_j (b)_j  \ov (j!)^2}\le[-\log(1-z) + 2\psi(j+1) -\psi(a+j)-\psi(b+j) \ri] (1-z)^j .
\ee
},
{\footnotesize
\bea
  \label{eqn:expansion}
  && {}_2F_1(a, b; \, a+b+n; \, z) = {(n-1)! \Gamma(a+b+n) \ov \Gamma(a+n)\Gamma(b+n)}
  \sum_{j=0}^{n-1} {(a)_j (b)_j (1-z)^j \ov j! (1-n)_j} + {  \Gamma(a+b+n) \ov \Gamma(a)\Gamma(b)}(z-1)^n  \times \\
  \nonumber
  && \times  \sum_{j=0}^\infty
  {(a+n)_j (b+n)_j  \ov j! (j+n)!}\le[-\log(1-z) +\psi(j+1) + \psi(j+n+1)-\psi(a+j+n)-\psi(b+j+n) \ri] (1-z)^j ,
\eea
}
\noindent
where $n$ is an integer, $(x)_j \equiv {\Gamma(x+j)\ov \Gamma(x)}$ is the Pochhammer symbol, and $\psi(x)$ is the digamma function.
In the case of the ingoing solution the constants are
\be
  \nonumber
  a={k_T-k_X\ov 2i}+{\Delta_- \ov 2}, \qquad b={k_T+k_X\ov 2i}+{\Delta_- \ov 2}, \qquad n=\Delta_+ -1 .
\ee
In order to compute the retarded Green's function, we will need terms up to order $(1-z)^n$ in the expansion. 
The source is the leading term, while the expectation value is the sum of non-logarithmic terms multiplying $(1-z)^n$ \cite{deHaro:2000vlm}. The Green's function can be computed by taking their ratio (up to a normalization factor). There are no integer values of $\Delta_+$ for which alternative quantization is possible with positive conformal dimension.
Note that the prefactor $z^{\pm{i k_T \ov 2}}$ in (\ref{eqn:ing}) also contributes and thus it has to be expanded near $z\approx 1$,
\be
  z^{-{i k_T \ov 2}} = \sum_{j=0}^\infty {1\ov j!} {\Gamma(-{i k_T \ov 2}+1) \ov \Gamma(-{i k_T \ov 2}-j+1)} (z-1)^j .
\ee
Terms in this expansion multiply terms in the first sum in (\ref{eqn:expansion}) and contribute to the finite piece at order $(1-z)^n$. However, these contributions turn out to be contact terms.\footnote{Note that including such contact terms is important if one wants to check the identity $G_R = G_A$ at Matsubara frequencies discussed in Section~\ref{sec:discussion}.}
The final result for the Green's function at integer $\Delta$ is (up to contact terms)
{\footnotesize
\be
  \nonumber
 G_R(k_T, k_X) \propto {
 \Gamma\le({\Delta_+ \ov 2}-i{k_T+k_X\ov 2}\ri)
 \Gamma\le({\Delta_+ \ov 2}-i{k_T-k_X\ov 2}\ri) \ov
 \Gamma\le({\Delta_- \ov 2}-i{k_T+k_X\ov 2}\ri)
 \Gamma\le({\Delta_- \ov 2}-i{k_T-k_X\ov 2}\ri) }
 \le[
  \psi\le({\Delta_+ \ov 2}-i{k_T+k_X\ov 2}\ri) +
  \psi\le({\Delta_+ \ov 2}-i{k_T-k_X\ov 2}\ri)
 \ri],
\ee
}
which differs from the generic case (\ref{eqn:genericret}) by the extra factor in the square brackets.

\section{Details of near-horizon expansions}
\label{app:higherorderdetails}

In this Appendix, we present the details of the near-horizon expansions of the equations of motion discussed in Sections \ref{sec:higherpoleskipping} and \ref{sec:hydrocorrelators}.

\subsection{Minimally massless scalar field}
\label{sec:scalarappendix}

As explained in the main text, a Taylor series solution to the minimally coupled scalar equation of motion \eqref{scalareqEF} exists when the matrix equation \eqref{eq:matrixequation} is satisfied. The first few elements of this matrix are
\begin{equation}
\begin{aligned}
\label{eq:scalarmatrixelements}
M_{11}&=-\frac{1}{4h(r_0)}\left(2k^2+2m^2h(r_0)+i\omega dh'(r_0)\right),\\
M_{21}&=\frac{1}{16h(r_0)^2}\left[-2h'(r_0)\left\{\left(d-2\right)k^2+dm^2h(r_0)\right\}-i\omega d\left\{\left(d-2\right)h'(r_0)^2+2h(r_0)h''(r_0)\right\}\right],\\
M_{22}&=\frac{1}{8h(r_0)}\left[-2k^2+dh'(r_0)\left\{2r_0^2f'(r_0)-3i\omega\right\}+h(r_0)\left\{2r_0^2f''(r_0)+8r_0f'(r_0)-2m^2\right\}\right],\\
M_{31}&=\frac{1}{96h(r_0)^3}\bigl[-2\left(d-2\right)h'(r_0)^2\left\{\left(d-4\right)k^2+dm^2h(r_0)\right\}-i\omega d\left(d^2-6d+8\right)h'(r_0)^3\\
&-6i\omega d\left(d-2\right)h(r_0)h'(r_0)h''(r_0)-4h(r_0)h''(r_0)\left\{\left(d-2\right)k^2+dm^2h(r_0)\right\}\\
&-4i\omega dh(r_0)^2h'''(r_0)\bigr],\\
M_{32}=&\frac{1}{48h(r_0)^2}\bigl[\left(d-2\right)h'(r_0)\left\{-4k^2+dh'(r_0)\left(3r_0^2f'(r_0)-4i\omega\right)\right\}\\
&+2dh(r_0)h'(r_0)\left\{3r_0^2f''(r_0)+12r_0f'(r_0)-2m^2\right\}+2dh(r_0)h''(r_0)\left\{3r_0^2f'(r_0)-4i\omega\right\}\\
&+24h(r_0)^2f'(r_0)+4r_0h(r_0)^2\left\{6f''(r_0)+r_0f'''(r_0)\right\}\bigr],\\
M_{33}=&\frac{1}{12h(r_0)}\left[-2k^2+dh'(r_0)\left\{6r_0^2f'(r_0)-5i\omega\right\}+h(r_0)\left\{6r_0^2f''(r_0)+24r_0f'(r_0)-2m^2\right\}\right].
\end{aligned}
\end{equation}
It is straightforward to calculate further elements, but the expressions are lengthy and so we will not write them explicitly. The explicit results for pole-skipping locations in BTZ and AdS$_{d+2}$-Schwarzschild spacetimes presented in Section \ref{sec:scalarexamples} can be calculated from \eqref{eq:scalarmatrixelements} as described in the main text.

\subsection{Gauge field perturbations}
\label{sec:gaugeappendix}

\paragraph{} Perturbations of the gauge field parallel to the wavenumber $k$ are described by the equation \eqref{eq:GIlongeq}. Assuming that $Z(\Phi)$ is normalised such that $Z\rightarrow1$ near the $r\rightarrow\infty$ boundary of the spacetime, the retarded Green's functions of the dual field theory are related to the ingoing solutions $\psi_1$ of \eqref{eq:GIlongeq} by
\begin{equation}
\label{eq:currentgreensdefns}
G^R_{J^tJ^t}(\omega,k)=\frac{k^2}{\omega^2-k^2}\frac{\psi_1^{(d-1)}}{\psi_1^{(0)}},\quad\quad\quad\quad G^R_{J^xJ^x}(\omega,k)=\frac{\omega^2}{k^2}G^R_{J^tJ^t}(\omega,k),
\end{equation}
up to an overall prefactor and contact terms. This can be shown by an analysis analogous to that in \cite{Kovtun:2005ev}. $\psi_1^{(m)}$ here denotes the coefficient of the $r^{-m}$ term in the near-boundary expansion of the solution $\psi_1(r)$.

\paragraph{} The equation \eqref{eq:GIlongeq} obeyed by $\psi_1$ is structurally similar to the scalar equation \eqref{scalareqEF} we studied previously. The main difference is the $(\omega^2h-k^2r^2f)$ terms appearing in denominators, but provided $\omega\ne0$ these denominators are non-zero at the horizon and thus the near-horizon expansion of \eqref{eq:GIlongeq} has a similar form to that of the scalar equation. In particular, making a Taylor series ansatz for $\psi_1$ near the horizon, the near-horizon equations of motion can again be written in the matrix form \eqref{eq:matrixequation} where the first few non-trivial elements are
\begin{equation}
\begin{aligned}
\label{eq:gaugematrixelements}
M_{11}=&-\frac{i}{4\omega h(r_0)}\left[2k^2r_0^2f'(r_0)+\omega\left\{-2ik^2+\left(d-2\right)\omega h'(r_0)\right\}+2\omega^2 h(r_0)\frac{Z'(r_0)}{Z(r_0)}\right],\\
M_{21}=&-\frac{i}{16\omega^3h(r_0)^2}\Biggl[8k^4r_0^4f'(r_0)^2+4k^2\omega r_0f'(r_0)\left\{-ik^2r_0+4\omega h(r_0)+\left(d-4\right)\omega r_0h'(r_0)\right\}\\
&+\omega^2\bigl\{-2i\omega k^2\left(d-4\right)h'(r_0)+\omega^2\left(d^2-6d+8\right)h'(r_0)^2+4k^2r_0^2f''(r_0)h(r_0)\\
&+2\left(d-2\right)\omega^2 h(r_0)h''(r_0)\bigr\}+4\omega^4h(r_0)^2\frac{Z''(r_0)}{Z(r_0)}\\
&+4\omega^2h(r_0)\frac{Z'(r_0)}{Z(r_0)}\left\{2k^2r_0^2f'(r_0)-i\omega k^2+\left(d-2\right)\omega^2 h'(r_0)\right\}\Biggr],\\
M_{22}=&\frac{1}{8\omega^2 h(r_0)}\Biggl[4k^2r_0^4f'(r_0)^2+2r_0\omega f'(r_0)\left\{-3ik^2r_0+4\omega h(r_0)+\left(d-2\right)\omega r_0h'(r_0)\right\}\\
&+\omega^2\left\{-2k^2-3i\omega\left(d-2\right)h'(r_0)+2r_0^2h(r_0)f''(r_0)\right\}\\
&+2\omega^2h(r_0)\left(2r_0^2f'(r_0)-3i\omega\right)\frac{Z'(r_0)}{Z(r_0)}\Biggr].
\end{aligned}
\end{equation}
The presence of $\omega$ in the denominators is because the near-horizon expansion is different when $\omega=0$ as mentioned above. We will address the $\omega=0$ case at the end of the subsection.

\paragraph{} With the equation in this form, we can repeat the arguments of Sections \ref{sec:scalarfield} and \ref{sec:higherpoleskipping} and conclude that for frequencies $\omega_n$ and appropriate choices of $k=k_n$ (satisfying $\det\mathcal{M}^{(n)}(\omega_n,k_n^2)=0$), the retarded Green's functions of the conserved charge and current \eqref{eq:currentgreensdefns} exhibit pole skipping at the special points $(\omega_n,k_n)$. The location of the first pole skipping point $k_1$ is given in equation \eqref{eq:longitudinalskipping}. It is straightforward to calculate $k_n$ for higher values of $n$ but for conciseness we will not present them here.

\paragraph{} Instead we will focus on the results for the simplest non-trivial spacetimes: the AdS$_{d+2}$-Schwarzschild metric \eqref{eq:schwarzsec2} with $Z(\Phi)=1$, holographically dual to non-zero temperature conformal field theories in $d$ spatial dimensions. For these cases, the first few pole-skipping wavenumbers $k_n$ are given by the solutions to the equations
\begin{equation}
\begin{aligned}
\label{eq:gaugefieldspecialKs}
0=&\;k_1^2-\frac{\left(d-2\right)\left(d+1\right)}{2}r_0^2,\\
0=&\;k_2^4+2\left(d+1\right)k_2^2r_0^2-\left(d+1\right)^2\left(d-2\right)\left(d-1\right)r_0^4,\\
0=&\;k_3^6+\frac{1}{2}\left(5d^2+11d+6\right)k_3^4r_0^2-\frac{1}{4}\left(d+1\right)^2\left(9d^2-64d+36\right)k_3^2r_0^4\\
&-\frac{9}{8}\left(d+1\right)^3\left(5d^3-18d^2+20d-8\right)r_0^6.
\end{aligned}
\end{equation}

\paragraph{} As in the scalar field examples, for each $\omega_n$ there are generically $n$ values of the wavenumber $k_n^2$ at which pole-skipping occurs. One notable difference from the examples of scalar fields in BTZ and AdS$_{d+2}$-Schwarzschild spacetimes (Section~\ref{sec:scalarexamples}) is that for each $n$ there is now one $k_n^2$ corresponding to real $k_n$. In the main text (Section~\ref{sec:gaugefieldpoleskip}) we show the connection between hydrodynamics and pole skipping at real $k_n$.

\paragraph{}As mentioned above, the $\omega=0$ case is special and must be treated separately. An explicit calculation shows that the general Taylor series solution for $\psi_1$ near the horizon is unique (up to an overall prefactor) provided that $k\ne0$ and hence the only potential pole-skipping point is at $\omega=k=0$.  The retarded Green's function at the origin of Fourier space is subtle in our formulation due to the $\omega$ and $k$ dependence in equation \eqref{eq:currentgreensdefns}. As hydrodynamic arguments already tell us the precise form of the retarded Green's function near the origin (see e.g.~\cite{Kovtun:2012rj}), we will not attempt to re-derive this form using pole-skipping arguments here.

\subsection*{The AdS$_4$-Schwarzschild spacetime}
\label{app:anomalousgaugefield}

\paragraph{} Electric-magnetic duality of a gauge field in (3+1)-dimensions implies that the charge current retarded Green's function in AdS$_4$-Schwarzschild is exactly $G^R_{J^xJ^x}(\omega,k=0)=i\omega$ \cite{Herzog:2007ij}. This particular case also has special pole-skipping properties: the results in \eqref{eq:gaugefieldspecialKs} (for $d=2$) imply that there are potential pole-skipping points at $k_n=0$ for every $\omega_n$. However these points are anomalous (in the sense discussed at the end of Section~\ref{sec:scalarfield}) because if we solve the equation of motion for $\psi_1$ at the location $\omega=-i2\pi Tn+\epsilon\delta\omega$ and $k=\epsilon\delta k$, then the solution at leading order in $\epsilon$ is unique. It does not depend on the ratio $\delta\omega/\delta k$ and thus the Green's function does not take the pole-skipping form \eqref{greensfunctionnear}.

\paragraph{} While in Appendix~\ref{app:integerdeltaBTZ}, anomalous points for a scalar field in the BTZ background were shown to correspond to the intersection of two lines of poles, in this case we can show that they do in fact correspond to an intersection of lines of poles and zeroes but in such a way that the Green's function takes a more complicated form than \eqref{greensfunctionnear}. Specifically, by performing a procedure similar to that of Appendix \ref{app:greensfunction} but scaling the deviations from the special location as $\omega=-i2\pi Tn+\epsilon^2\delta\omega$, $k=\epsilon\delta k$ (i.e.~such that $\delta\omega/\delta k^2\sim\epsilon^0$), one finds
\begin{equation}
G^R_{J^xJ^x}(-i2\pi Tn+\epsilon^2\delta\omega,\epsilon\delta k)=\frac{A_n\delta k^2}{-i\delta\omega+B_n\delta k^2}+O(\epsilon),
\end{equation}
at leading order in $\epsilon$, where $A_n$ and $B_n$ are $n$-dependent constants that can be computed explicitly but for conciseness we omit. The fact that poles and zeroes pass through these points in Fourier space was observed numerically in \cite{WitczakKrempa:2013ht}. Similarly, in \cite{Andrade:2015hpa} it was observed that there are normalisable, ingoing solutions for perturbations of `axion' black branes at $k=0$ and $\omega_n$ for certain $n$ and we think it is likely this property can be more directly seen by the type of near-horizon analysis presented here.

\subsection{Transverse metric perturbations}
\label{sec:transversemetricappendix}

\paragraph{} The retarded Green's functions of the transverse momentum operator $T^{ty}$ are captured by the bulk field $\psi_2$ which obeys the equation of motion \eqref{eq:transversemetriceqn}. Specifically, up to an overall prefactor and ignoring contact terms,
\begin{equation}
G^R_{T^{ty}T^{ty}}(\omega,k)=\frac{k^2}{\omega^2-k^2}\frac{\psi_2^{(d+1)}}{\psi_2^{(0)}},\quad\quad\quad G^R_{T^{xy}T^{xy}}(\omega,k)=\frac{\omega^2}{k^2}G^R_{T^{ty}T^{ty}}(\omega,k),
\end{equation}
where $\psi_2^{(m)}$ denotes the coefficients of the $r^{-m}$ term in the near-boundary expansion of $\psi_2$. This can be shown by an analysis analogous to that in \cite{Kovtun:2005ev}. As in the previous subsection, provided that $\omega\ne0$ a Taylor series ansatz for $\psi_2$ near the horizon yields near-horizon equations of motion of the matrix form \eqref{eq:matrixequation}, where the first few non-trivial elements are
\begin{equation}
\begin{aligned}
\label{eq:transversematrixelements}
M_{11}=&-\frac{i}{4\omega h(r_0)}\left[2k^2r_0^2f'(r_0)+\omega\left\{-2ik^2+d\omega h'(r_0)\right\}\right],\\
M_{21}=&-\frac{i}{16\omega^3h(r_0)^2}\bigl[8k^4r_0^4f'(r_0)^2+4k^2\omega r_0f'(r_0)\left\{-ik^2r_0+4\omega h(r_0)+\left(d-2\right)\omega r_0h'(r_0)\right\}\\
&+\omega^2\bigl\{-2\left(d-2\right)i\omega k^2h'(r_0)+d\left(d-2\right)\omega^2h'(r_0)^2+4k^2h(r_0)r_0^2f''(r_0)\\
&+2d\omega^2 h(r_0)h''(r_0)\bigr\}\bigr],\\
M_{22}=&\frac{1}{8\omega^2 h(r_0)}\bigl[4k^2r_0^4f'(r_0)^2+2\omega r_0f'(r_0)\left\{-3ik^2r_0+4\omega h(r_0)+d\omega r_0h'(r_0)\right\}\\
&+\omega^2\left\{-2k^2-3di\omega h'(r_0)+2r_0^2h(r_0)f''(r_0)\right\}\bigr].\\
\end{aligned}
\end{equation}
Repeating again the arguments of Sections \ref{sec:scalarfield} and \ref{sec:higherpoleskipping}, we find that there is generically pole skipping in $G^R_{T^{ty}T^{ty}}(\omega,k)$ at frequencies $\omega_n$ and wavenumbers $k=k_n$ satisfying $\det\mathcal{M}^{(n)}(\omega_n,k_n^2)=0$. The first pole-skipping point is located at \eqref{locationvisc}, and it is straightforward to compute the appropriate expressions for higher $n$.

\paragraph{} For the explicit case of the AdS$_{d+2}$-Schwarzschild metric \eqref{eq:schwarzsec2} (i.e.~$\Phi=0$) dual to a non-zero temperature conformal field theory, the first few $k_n$ obey
\begin{equation}
\begin{aligned}
0=&\;k_1^2-\frac{d\left(d+1\right)}{2}r_0^2,\\
0=&\;k_2^4-\left(d+1\right)^2d\left(d-1\right)r_0^4,\\
0=&\;k_3^6+\frac{5}{2}d\left(d+1\right)k_3^4r_0^2-\frac{3}{4}d\left(d+1\right)^2\left(3d-4\right)k_3^2r_0^4-\frac{3}{8}d\left(d+1\right)^3\left(15d^2-28d+16\right)r_0^6.
\end{aligned}
\end{equation}
For each $n$, there is one value of $k_n^2$ for which $k_n$ is real. The relation between these pole-skipping points and the hydrodynamic poles is shown in Section~\ref{sec:hydrogravitycorrelators}.

\paragraph{} As in the previous subsection, the $\omega=0$ point is special and requires a more careful analysis. Due to the very similar form of the equations of motion for $\psi_1$ and $\psi_2$, we again find that the only potential pole-skipping point of $G^R_{T^{ty}T^{ty}}(\omega,k)$ at $\omega=0$ is when $k=0$. As hydrodynamics fixes the form of $G^R_{T^{ty}T^{ty}}(\omega,k)$ near this point (see e.g.~\cite{Kovtun:2012rj}), we will not pursue this special case further.

\subsection{Longitudinal metric perturbations}
\label{sec:longitudinalmetricappendix}

\paragraph{}The retarded Green's functions of the longitudinal momentum density $T^{tx}$ and energy density $T^{tt}$ are related to the solutions of the equation \eqref{eq:longitudinalscalareq} for $\psi_3$ by (up to an overall prefactor, and neglecting contact terms)
\begin{equation}
G^R_{T^{tt}T^{tt}}(\omega,k)=\frac{k^4}{\left(\omega^2-k^2\right)^2}\frac{\psi_3^{(d+1)}}{\psi_3^{(0)}},\quad\quad\quad G^R_{T^{tx}T^{tx}}(\omega,k)=\frac{\omega^2}{k^2}G^R_{T^{tt}T^{tt}}(\omega,k),
\end{equation}
where $\psi_3^{(m)}$ denotes the coefficient of the $r^{-m}$ term in the near-boundary expansion of $\psi_3$. This can be shown by an analysis analogous to that in \cite{Kovtun:2005ev}. To identify pole-skipping locations we examine the properties of \eqref{eq:longitudinalscalareq} near the horizon, where there are two distinct possibilities. In the generic case, where $\omega^2\ne k^2(d+1)/2d$, the denominators in \eqref{eq:longitudinalscalareq} are non-zero at the horizon and the near-horizon equations of motion have a similar structure to those of the minimally coupled scalar field. Specifically, by making a Taylor series ansatz for $\psi_3$ near the horizon, one finds that the near-horizon equations can be written in the matrix form \eqref{eq:matrixequation} where the elements of the matrix are
\begin{equation}
\begin{aligned}
M_{11}=&\frac{1}{2r_0^2\left[\left(d+1\right)k^2-2d\omega^2\right]}\bigl[-\left(d+1\right)k^4+2i\omega^3d^2r_0+k^2\bigl\{\left(d-1\right)\left(d+1\right)^2r_0^2\\
&+i\omega r_0\left(d-2\right)\left(d+1\right)+2d\omega^2\bigr\}\bigr],\\
M_{21}=&\frac{1}{4r_0^3\left[\left(d+1\right)k^2-2d\omega^2\right]^2}\bigl[\left(d+1\right)^2k^4\left\{dk^2-r_0^2\left(d+1\right)\left(d-1\right)\left(4d-1\right)\right\}\\
&-i\omega k^4r_0\left(d+1\right)^2\left(d-1\right)\left(5d-2\right)+8i\omega^3k^2r_0d\left(d^2-1\right)-4i\omega^5r_0d^3\left(d-1\right)\\
&+2d\left(d+1\right)k^2\omega^2\bigl\{-2k^2+r_0^2\left(d-1\right)\left(d+1\right)\left(d+2\right)\bigr\}-4d^2\left(d-2\right)k^2\omega^4\bigr],\\
M_{22}=&\frac{1}{4r_0^2\left[\left(d+1\right)k^2-2d\omega^2\right]}\bigl[-\left(d+1\right)k^4-2dr_0\omega^2\bigl\{\left(d+1\right)\left(d+2\right)r_0-3di\omega\bigr\}\\
&+k^2\bigl\{-\left(d+1\right)^2\left(2d-5\right)r_0^2+3ir_0\omega\left(d+1\right)\left(d-2\right)+2d\omega^2\bigr\}\bigr].
\end{aligned}
\end{equation}
As a consequence, we can apply the arguments of Sections \ref{sec:scalarfield} and \ref{sec:higherpoleskipping} and conclude that there is pole-skipping at frequencies $\omega_n$ and wavenumbers $k_n$ obeying $\det\mathcal{M}^{(n)}(\omega_n,k_n^2)=0$. The explicit equations determining the first few values of $k_n$ are given in the main text in equation \eqref{eq:longitudinalmetrickn} (with the implicit assumption that $k_n^2\ne2d\omega_n^2/(d+1)$).

\paragraph{} From \cite{Blake:2018leo} we know that there must also be pole-skipping in the upper half of the complex $\omega$ plane. While this upper half plane pole-skipping can easily be seen by a direct analysis of the Einstein equations \cite{Blake:2018leo}, this feature is obscured by formulating the dynamics in terms of the scalar degree of freedom $\psi_3$. To observe it, we must consider the special case
\begin{equation}
\label{eq:longspecialcase}
k^2=\frac{2d}{d+1}\omega^2,
\end{equation}
of the equation of motion \eqref{eq:longitudinalscalareq}, where the vanishing of the denominators at the horizon changes the near-horizon structure of the equation of motion. Specifically, after imposing \eqref{eq:longspecialcase} on the equations of motion and looking for power law solutions  $\psi_3(r)=(r-r_0)^\alpha$
near the horizon, we find that the allowed powers are
\begin{equation}
\label{eq:longspecialpowers}
\alpha = 1,\;1+\frac{i\omega}{2\pi T}.
\end{equation}
\eqref{eq:longspecialpowers} suggests that there are three cases in which it is possible that both independent solutions for $\psi_3(r)$ are regular at the horizon: $\omega=\pm i2\pi T, \omega=0$. For the cases $\omega=\pm i2\pi T$, an analysis of the near-horizon equations of motion confirms that at the general Taylor series solution for $\psi_3$ around the horizon has two free parameters. As in the previous two subsections, we will not explore the case of $\omega=0,k=0$ because hydrodynamics already dictates the exact form of $G^R_{T^{tx}T^{tx}}(\omega\rightarrow0,k\rightarrow0)$ \cite{Kovtun:2012rj}.

\paragraph{}We will therefore now focus on the potential pole-skipping points at $\omega=\pm i2\pi T$ and $k^2=-2d(2\pi T)^2/(d+1)$. Recall from the discussion in Section \ref{sec:anomaloussection2} that for pole-skipping to occur, it is not sufficient for there to be two independent solutions for $\psi_3$ that are regular at the horizon. Additionally, we require that moving slightly away from the potential pole skipping point picks out a unique ingoing solution (up to overall normalisation) that depends on the slope $\delta\omega/\delta k$. To check this condition, we take
\begin{equation}
\omega=\pm i2\pi T+\epsilon\delta\omega,\quad\quad\quad k^2=\frac{2d}{d+1}\omega^2+r_0\epsilon\delta k,
\end{equation}
make a near-horizon Taylor series ansatz for the field $\psi_3$ and then solve the equation of motion \eqref{eq:longitudinalscalareq} in an expansion near the horizon. At lowest order in $\epsilon$, the result is that for the case $\omega=+i2\pi T$,
\begin{equation}
\psi_3(r)\propto\left[1-\frac{d\left(i\delta k+(3d-1)\delta\omega\right)}{2\left(i\delta k+2d\delta\omega\right)}\frac{(r-r_0)}{r_0}+\frac{d\left(i\delta k+(d^2+2d-1)\delta\omega\right)}{2\left(i\delta k+2d\delta\omega\right)}\frac{(r-r_0)^2}{r_0^2}+\ldots\right]
\end{equation}
while for the case $\omega=-i2\pi T$
\begin{equation}
\begin{aligned}
\psi_3(r)\propto(r-r_0)\Bigl[1-\frac{(r-r_0)}{r_0}&-\frac{\left(d^2-d-12\right)}{12r_0^2}(r-r_0)^2+\ldots\Bigr].
\end{aligned}
\end{equation}
Thus there is pole skipping at the point \eqref{eq:longspecialcase} with $\omega=+i2\pi T$ but not with $\omega=-i2\pi T$. This latter case is in fact an example of an anomalous point as described in Section \ref{sec:anomaloussection2}. For a scalar field in the BTZ spacetime, we showed in Appendix \ref{app:integerdeltaBTZ} that anomalous points correspond to intersections of multiple poles, and it would be very interesting to examine whether that is also the case for the example presented here.

\subsection{Transverse metric perturbations in a charged black hole}
\label{sec:chargedtransverseapp}

\paragraph{}In this subsection, we will briefly describe how to identify the existence of pole-skipping at $\omega=-i2\pi T$ in $G^R_{T^{ty}T^{ty}}(\omega,k)$ of the charged state dual to the AdS$_4$-Reissner-Nordstrom black brane. In addition to further exemplifying the generic nature of pole-skipping in holographic theories, this also illustrates that unlike in the case of energy density correlators \cite{Blake:2018leo}, the pole skipping location $k_n$ for generic hydrodynamic correlators is in general \textit{not} related in a simple way to the butterfly velocity $v_B$.

\paragraph{}The AdS$_4$ Reissner-Nordstrom solution
\begin{equation}
f(r)=1-\left(1+\frac{\mu^2}{4r_0^2}\right)\frac{r_0^3}{r^3}+\frac{\mu^2r_0^2}{4r^4},\quad\quad\quad h(r)=r^2,\quad\quad\quad A_v(r)=\mu\left(1-\frac{r_0}{r}\right),
\end{equation}
is a solution to the classical equations of the action
\begin{equation}
S=\int d^4x\sqrt{-g}\left(R+6-\frac{1}{4}F^2\right).
\end{equation}

\paragraph{} $G^R_{T^{ty}T^{ty}}(\omega,k)$ is controlled by the coupled perturbations of the metric $\delta g_{vy}, \delta  g_{xy}, \delta g_{ry}$ and $\delta A_y$. After Fourier transforming and solving algebraically for $\delta g_{ry}$, we are left with the following two coupled equations for the variables $\delta A_y(r)$ and $\psi_2(r)$ (defined in equation \eqref{eq:psi2defn})
\begin{equation}
\begin{aligned}
\label{eq:chargedmomeqs}
&\frac{d}{dr}\left[\frac{r^2}{\omega^2-k^2f}\left(r^2f\psi_2'-i\omega\psi_2\right)\right]+\frac{1}{\omega^2-k^2f}\left(-i\omega r^2\psi_2'-k^2\psi_2\right)\\
&+\frac{kr^2fA_v'}{\omega^2-k^2f}\delta A_y'+\frac{1}{\omega^2-k^2f}\left(-i\omega kA_v'+\frac{k\omega^2r^2f'A_v'}{\omega^2-k^2f}\right)\delta A_y=0,\\
&\frac{d}{dr}\left[r^2f\delta A_y'-i\omega\delta A_y\right]-i\omega\delta A_y'-\left(\frac{k^2}{r^2}+\frac{\omega^2{A_v'}^2}{\omega^2-k^2f}\right)\delta A_y-\frac{kr^2fA_v'}{\omega^2-k^2f}\psi_2'\\
&+\frac{ik\omega A_v'}{\omega^2-k^2f}\psi_2=0.
\end{aligned}
\end{equation}

\paragraph{} Making an ansatz of Taylor series solutions at $\omega=-i2\pi T$
\begin{equation}
\psi_2=\sum_{n=0}\psi_2^{(n)}(r-r_0)^n,\quad\quad\quad \delta A_y=\sum_{n=0}\delta A_y^{(n)}(r-r_0)^n,
\end{equation}
and solving the equations \eqref{eq:chargedmomeqs} order-by-order around the horizon, one finds that generically the solution is characterised by two free parameters $(\psi_2^{(1)},\delta A_y^{(1)})$. This results in a uniquely defined retarded Green's function $G^R_{T^{ty}T^{ty}}(\omega,k)$. However, when $\omega=-i2\pi T$ and $k=k_1$ with
\begin{equation}
\label{eq:chargedk1sol}
k_1^4-r_04\pi T\left(k_1^2+\mu^2\right)=0,
\end{equation}
the Taylor series solutions near the horizon are characterised by three free parameters ($\psi_2^{(0)}$ in addition to the two above). As a consequence, there is pole skipping in $G^R_{T^{ty}T^{ty}}(\omega,k)$ at $\omega=-i2\pi T, k=k_1$. The dispersion relation of the hydrodynamic pole of this Green's function was computed  numerically in \cite{Brattan:2010pq} and the value of $k$ at which it passes through $\omega=-i2\pi T$ is consistent with our equation for $k_1$ (after accounting for the different conventions for $\mu$ used in \cite{Brattan:2010pq}).

\paragraph{} We will now use this result to comment on the relation between pole-skipping and chaos in general. There is an upper half-plane pole-skipping point \eqref{skippinggrav} in the retarded Green's function of energy density that is fixed simply by the butterfly velocity $v_B$. This is also true for a wide variety of matter content of the gravitational theory and is evidence for an effective hydrodynamic description of chaos \cite{Blake:2018leo}. In our analysis of the pole-skipping points of $G^R_{T^{ty}T^{ty}}(\omega,k)$ for solutions to Einstein-scalar gravity in Section \ref{sec:hydrogravitycorrelators}, we found an instance of pole-skipping in the lower half-plane whose location \eqref{locationvisc} can be written as
\begin{equation}
\label{eq:transversepoleskippingans}
\omega=-i2\pi T,\quad\quad\quad k^2=(2\pi T/v_B)^2,
\end{equation}
using $v_B^2=4\pi T/dh'(r_0)$ \cite{Blake:2016wvh,Roberts:2016wdl}. This raises the question of whether the pole-skipping in $G^R_{T^{ty}T^{ty}}(\omega,k)$ is also intimately related to chaos. It is simple to check using \eqref{eq:chargedk1sol} that \eqref{eq:transversepoleskippingans} is only a pole-skipping point when $\mu=0$ i.e.~when the solution is uncharged. In other words, unlike for the the energy density correlator, the close relation between the pole-skipping location of $G^R_{T^{ty}T^{ty}}(\omega,k)$ and the butterfly velocity $v_B$ is not robust to the generalisation to charged black holes and thus we view it as unlikely that the pole-skipping in $G^R_{T^{ty}T^{ty}}(\omega,k)$ is in general related in a fundamental way to quantum chaos.

\bibliographystyle{JHEP}
\bibliography{GeneralPoleSkippingBib}

\providecommand{\href}[2]{#2}\begingroup\raggedright\begin{thebibliography}{10}

\bibitem{Son:2002sd}
D.~T. Son and A.~O. Starinets, {\it {Minkowski space correlators in AdS / CFT
  correspondence: Recipe and applications}},  {\em JHEP} {\bf 09} (2002) 042,
  [\href{http://arxiv.org/abs/hep-th/0205051}{{\tt hep-th/0205051}}].

\bibitem{Maldacena:1997re}
J.~M. Maldacena, {\it {The Large N limit of superconformal field theories and
  supergravity}},  {\em Int. J. Theor. Phys.} {\bf 38} (1999) 1113--1133,
  [\href{http://arxiv.org/abs/hep-th/9711200}{{\tt hep-th/9711200}}]. [Adv.
  Theor. Math. Phys.2,231(1998)].

\bibitem{Gubser:1998bc}
S.~S. Gubser, I.~R. Klebanov, and A.~M. Polyakov, {\it {Gauge theory
  correlators from noncritical string theory}},  {\em Phys. Lett.} {\bf B428}
  (1998) 105--114, [\href{http://arxiv.org/abs/hep-th/9802109}{{\tt
  hep-th/9802109}}].

\bibitem{Witten:1998qj}
E.~Witten, {\it {Anti-de Sitter space and holography}},  {\em Adv. Theor. Math.
  Phys.} {\bf 2} (1998) 253--291,
  [\href{http://arxiv.org/abs/hep-th/9802150}{{\tt hep-th/9802150}}].

\bibitem{Horowitz:1999jd}
G.~T. Horowitz and V.~E. Hubeny, {\it {Quasinormal modes of AdS black holes and
  the approach to thermal equilibrium}},  {\em Phys. Rev.} {\bf D62} (2000)
  024027, [\href{http://arxiv.org/abs/hep-th/9909056}{{\tt hep-th/9909056}}].

\bibitem{Herzog:2002pc}
C.~P. Herzog and D.~T. Son, {\it {Schwinger-Keldysh propagators from AdS/CFT
  correspondence}},  {\em JHEP} {\bf 03} (2003) 046,
  [\href{http://arxiv.org/abs/hep-th/0212072}{{\tt hep-th/0212072}}].

\bibitem{Skenderis:2008dh}
K.~Skenderis and B.~C. van Rees, {\it {Real-time gauge/gravity duality}},  {\em
  Phys. Rev. Lett.} {\bf 101} (2008) 081601,
  [\href{http://arxiv.org/abs/0805.0150}{{\tt arXiv:0805.0150}}].

\bibitem{Skenderis:2008dg}
K.~Skenderis and B.~C. van Rees, {\it {Real-time gauge/gravity duality:
  Prescription, Renormalization and Examples}},  {\em JHEP} {\bf 05} (2009)
  085, [\href{http://arxiv.org/abs/0812.2909}{{\tt arXiv:0812.2909}}].

\bibitem{Son:2009vu}
D.~T. Son and D.~Teaney, {\it {Thermal Noise and Stochastic Strings in
  AdS/CFT}},  {\em JHEP} {\bf 07} (2009) 021,
  [\href{http://arxiv.org/abs/0901.2338}{{\tt arXiv:0901.2338}}].

\bibitem{Glorioso:2018mmw}
P.~Glorioso, M.~Crossley, and H.~Liu, {\it {A prescription for holographic
  Schwinger-Keldysh contour in non-equilibrium systems}},
  \href{http://arxiv.org/abs/1812.08785}{{\tt arXiv:1812.08785}}.

\bibitem{Liu:2018crr}
H.~Liu and J.~Sonner, {\it {Holographic systems far from equilibrium: a
  review}},  \href{http://arxiv.org/abs/1810.02367}{{\tt arXiv:1810.02367}}.

\bibitem{Kovtun:2004de}
P.~Kovtun, D.~T. Son, and A.~O. Starinets, {\it {Viscosity in strongly
  interacting quantum field theories from black hole physics}},  {\em Phys.
  Rev. Lett.} {\bf 94} (2005) 111601,
  [\href{http://arxiv.org/abs/hep-th/0405231}{{\tt hep-th/0405231}}].

\bibitem{Iqbal:2008by}
N.~Iqbal and H.~Liu, {\it {Universality of the hydrodynamic limit in AdS/CFT
  and the membrane paradigm}},  {\em Phys. Rev.} {\bf D79} (2009) 025023,
  [\href{http://arxiv.org/abs/0809.3808}{{\tt arXiv:0809.3808}}].

\bibitem{Blake:2018leo}
M.~Blake, R.~A. Davison, S.~Grozdanov, and H.~Liu, {\it {Many-body chaos and
  energy dynamics in holography}},  {\em JHEP} {\bf 10} (2018) 035,
  [\href{http://arxiv.org/abs/1809.01169}{{\tt arXiv:1809.01169}}].

\bibitem{Roberts:2014isa}
D.~A. Roberts, D.~Stanford, and L.~Susskind, {\it {Localized shocks}},  {\em
  JHEP} {\bf 03} (2015) 051, [\href{http://arxiv.org/abs/1409.8180}{{\tt
  arXiv:1409.8180}}].

\bibitem{Roberts:2016wdl}
D.~A. Roberts and B.~Swingle, {\it {Lieb-Robinson Bound and the Butterfly
  Effect in Quantum Field Theories}},  {\em Phys. Rev. Lett.} {\bf 117} (2016),
  no.~9 091602, [\href{http://arxiv.org/abs/1603.09298}{{\tt
  arXiv:1603.09298}}].

\bibitem{Blake:2016wvh}
M.~Blake, {\it {Universal Charge Diffusion and the Butterfly Effect in
  Holographic Theories}},  {\em Phys. Rev. Lett.} {\bf 117} (2016), no.~9
  091601, [\href{http://arxiv.org/abs/1603.08510}{{\tt arXiv:1603.08510}}].

\bibitem{Grozdanov:2017ajz}
S.~Grozdanov, K.~Schalm, and V.~Scopelliti, {\it {Black hole scrambling from
  hydrodynamics}},  {\em Phys. Rev. Lett.} {\bf 120} (2018), no.~23 231601,
  [\href{http://arxiv.org/abs/1710.00921}{{\tt arXiv:1710.00921}}].

\bibitem{Blake:2017ris}
M.~Blake, H.~Lee, and H.~Liu, {\it {A quantum hydrodynamical description for
  scrambling and many-body chaos}},  {\em JHEP} {\bf 10} (2018) 127,
  [\href{http://arxiv.org/abs/1801.00010}{{\tt arXiv:1801.00010}}].

\bibitem{Grozdanov:2018kkt}
S.~Grozdanov, {\it {On the connection between hydrodynamics and quantum chaos
  in holographic theories with stringy corrections}},  {\em JHEP} {\bf 01}
  (2019) 048, [\href{http://arxiv.org/abs/1811.09641}{{\tt arXiv:1811.09641}}].

\bibitem{Grozdanov:2019uhi}
S.~Grozdanov, P.~K. Kovtun, A.~O. Starinets, and P.~Tadić, {\it {The complex
  life of hydrodynamic modes}},  \href{http://arxiv.org/abs/1904.12862}{{\tt
  arXiv:1904.12862}}.

\bibitem{InceBook}
E.~L. Ince, {\em Ordinary Differential Equations (Section 16.4)}.
\newblock Dover Publications, New York, 1~ed., 1956.

\bibitem{Morgan:2009pn}
J.~Morgan, V.~Cardoso, A.~S. Miranda, C.~Molina, and V.~T. Zanchin, {\it
  {Gravitational quasinormal modes of AdS black branes in d spacetime
  dimensions}},  {\em JHEP} {\bf 09} (2009) 117,
  [\href{http://arxiv.org/abs/0907.5011}{{\tt arXiv:0907.5011}}].

\bibitem{Denef:2009yy}
F.~Denef, S.~A. Hartnoll, and S.~Sachdev, {\it {Quantum oscillations and black
  hole ringing}},  {\em Phys. Rev.} {\bf D80} (2009) 126016,
  [\href{http://arxiv.org/abs/0908.1788}{{\tt arXiv:0908.1788}}].

\bibitem{Kovtun:2012rj}
P.~Kovtun, {\it {Lectures on hydrodynamic fluctuations in relativistic
  theories}},  {\em J. Phys.} {\bf A45} (2012) 473001,
  [\href{http://arxiv.org/abs/1205.5040}{{\tt arXiv:1205.5040}}].

\bibitem{Kovtun:2005ev}
P.~K. Kovtun and A.~O. Starinets, {\it {Quasinormal modes and holography}},
  {\em Phys. Rev.} {\bf D72} (2005) 086009,
  [\href{http://arxiv.org/abs/hep-th/0506184}{{\tt hep-th/0506184}}].

\bibitem{Haehl:2018izb}
F.~M. Haehl and M.~Rozali, {\it {Effective Field Theory for Chaotic CFTs}},
  {\em JHEP} {\bf 10} (2018) 118, [\href{http://arxiv.org/abs/1808.02898}{{\tt
  arXiv:1808.02898}}].

\bibitem{Iliesiu:2018fao}
L.~Iliesiu, M.~Kologlu, R.~Mahajan, E.~Perlmutter, and D.~Simmons-Duffin, {\it
  {The Conformal Bootstrap at Finite Temperature}},  {\em JHEP} {\bf 10} (2018)
  070, [\href{http://arxiv.org/abs/1802.10266}{{\tt arXiv:1802.10266}}].

\bibitem{Gu:2016oyy}
Y.~Gu, X.-L. Qi, and D.~Stanford, {\it {Local criticality, diffusion and chaos
  in generalized Sachdev-Ye-Kitaev models}},  {\em JHEP} {\bf 05} (2017) 125,
  [\href{http://arxiv.org/abs/1609.07832}{{\tt arXiv:1609.07832}}].

\bibitem{Crossley:2015evo}
M.~Crossley, P.~Glorioso, and H.~Liu, {\it {Effective field theory of
  dissipative fluids}},  {\em JHEP} {\bf 09} (2017) 095,
  [\href{http://arxiv.org/abs/1511.03646}{{\tt arXiv:1511.03646}}].

\bibitem{Glorioso:2017fpd}
P.~Glorioso, M.~Crossley, and H.~Liu, {\it {Effective field theory of
  dissipative fluids (II): classical limit, dynamical KMS symmetry and entropy
  current}},  {\em JHEP} {\bf 09} (2017) 096,
  [\href{http://arxiv.org/abs/1701.07817}{{\tt arXiv:1701.07817}}].

\bibitem{Grozdanov:2019kge}
S.~Grozdanov, P.~K. Kovtun, A.~O. Starinets, and P.~Tadic, {\it {On the
  convergence of the gradient expansion in hydrodynamics}},
  \href{http://arxiv.org/abs/1904.01018}{{\tt arXiv:1904.01018}}.

\bibitem{Blake:2017qgd}
M.~Blake, R.~A. Davison, and S.~Sachdev, {\it {Thermal diffusivity and chaos in
  metals without quasiparticles}},  {\em Phys. Rev.} {\bf D96} (2017), no.~10
  106008, [\href{http://arxiv.org/abs/1705.07896}{{\tt arXiv:1705.07896}}].

\bibitem{Blake:2016jnn}
M.~Blake and A.~Donos, {\it {Diffusion and Chaos from near AdS$_2$ horizons}},
  {\em JHEP} {\bf 02} (2017) 013, [\href{http://arxiv.org/abs/1611.09380}{{\tt
  arXiv:1611.09380}}].

\bibitem{Blake:2016sud}
M.~Blake, {\it {Universal Diffusion in Incoherent Black Holes}},  {\em Phys.
  Rev.} {\bf D94} (2016), no.~8 086014,
  [\href{http://arxiv.org/abs/1604.01754}{{\tt arXiv:1604.01754}}].

\bibitem{Davison:2018ofp}
R.~A. Davison, S.~A. Gentle, and B.~Goutéraux, {\it {Slow relaxation and
  diffusion in holographic quantum critical phases}},
  \href{http://arxiv.org/abs/1808.05659}{{\tt arXiv:1808.05659}}.

\bibitem{Davison:2018nxm}
R.~A. Davison, S.~A. Gentle, and B.~Goutéraux, {\it {Impact of irrelevant
  deformations on thermodynamics and transport in holographic quantum critical
  states}},  \href{http://arxiv.org/abs/1812.11060}{{\tt arXiv:1812.11060}}.

\bibitem{Hartnoll:2014lpa}
S.~A. Hartnoll, {\it {Theory of universal incoherent metallic transport}},
  {\em Nature Phys.} {\bf 11} (2015) 54,
  [\href{http://arxiv.org/abs/1405.3651}{{\tt arXiv:1405.3651}}].

\bibitem{Lucas:2016yfl}
A.~Lucas and J.~Steinberg, {\it {Charge diffusion and the butterfly effect in
  striped holographic matter}},  {\em JHEP} {\bf 10} (2016) 143,
  [\href{http://arxiv.org/abs/1608.03286}{{\tt arXiv:1608.03286}}].

\bibitem{Baggioli:2016pia}
M.~Baggioli, B.~Gouteraux, E.~Kiritsis, and W.-J. Li, {\it {Higher derivative
  corrections to incoherent metallic transport in holography}},  {\em JHEP}
  {\bf 03} (2017) 170, [\href{http://arxiv.org/abs/1612.05500}{{\tt
  arXiv:1612.05500}}].

\bibitem{Patel:2016wdy}
A.~A. Patel and S.~Sachdev, {\it {Quantum chaos on a critical Fermi surface}},
  {\em Proc. Nat. Acad. Sci.} {\bf 114} (2017) 1844--1849,
  [\href{http://arxiv.org/abs/1611.00003}{{\tt arXiv:1611.00003}}].

\bibitem{Davison:2016ngz}
R.~A. Davison, W.~Fu, A.~Georges, Y.~Gu, K.~Jensen, and S.~Sachdev, {\it
  {Thermoelectric transport in disordered metals without quasiparticles: The
  Sachdev-Ye-Kitaev models and holography}},  {\em Phys. Rev.} {\bf B95}
  (2017), no.~15 155131, [\href{http://arxiv.org/abs/1612.00849}{{\tt
  arXiv:1612.00849}}].

\bibitem{Werman:2017abn}
Y.~Werman, S.~A. Kivelson, and E.~Berg, {\it {Quantum chaos in an
  electron-phonon bad metal}},  \href{http://arxiv.org/abs/1705.07895}{{\tt
  arXiv:1705.07895}}.

\bibitem{Guo:2019csw}
H.~Guo, Y.~Gu, and S.~Sachdev, {\it {Transport and chaos in lattice
  Sachdev-Ye-Kitaev models}},  \href{http://arxiv.org/abs/1904.02174}{{\tt
  arXiv:1904.02174}}.

\bibitem{deBoer:2018qqm}
J.~de~Boer, M.~P. Heller, and N.~Pinzani-Fokeeva, {\it {Holographic
  Schwinger-Keldysh effective field theories}},  {\em JHEP} {\bf 05} (2019)
  188, [\href{http://arxiv.org/abs/1812.06093}{{\tt arXiv:1812.06093}}].

\bibitem{Iqbal:2009fd}
N.~Iqbal and H.~Liu, {\it {Real-time response in AdS/CFT with application to
  spinors}},  {\em Fortsch. Phys.} {\bf 57} (2009) 367--384,
  [\href{http://arxiv.org/abs/0903.2596}{{\tt arXiv:0903.2596}}].

\bibitem{Reynolds:2016pmi}
A.~P. Reynolds and S.~F. Ross, {\it {Butterflies with rotation and charge}},
  {\em Class. Quant. Grav.} {\bf 33} (2016), no.~21 215008,
  [\href{http://arxiv.org/abs/1604.04099}{{\tt arXiv:1604.04099}}].

\bibitem{Stikonas:2018ane}
A.~Stikonas, {\it {Scrambling time from local perturbations of the rotating BTZ
  black hole}},  {\em JHEP} {\bf 02} (2019) 054,
  [\href{http://arxiv.org/abs/1810.06110}{{\tt arXiv:1810.06110}}].

\bibitem{Poojary:2018esz}
R.~R. Poojary, {\it {BTZ dynamics and chaos}},
  \href{http://arxiv.org/abs/1812.10073}{{\tt arXiv:1812.10073}}.

\bibitem{Jahnke:2019gxr}
V.~Jahnke, K.-Y. Kim, and J.~Yoon, {\it {On the Chaos Bound in Rotating Black
  Holes}},  \href{http://arxiv.org/abs/1903.09086}{{\tt arXiv:1903.09086}}.

\bibitem{MaassenvandenBrink:2000iwh}
A.~Maassen van~den Brink, {\it {Analytic treatment of black hole gravitational
  waves at the algebraically special frequency}},  {\em Phys. Rev.} {\bf D62}
  (2000) 064009, [\href{http://arxiv.org/abs/gr-qc/0001032}{{\tt
  gr-qc/0001032}}].

\bibitem{Birmingham:2001hc}
D.~Birmingham, {\it {Choptuik scaling and quasinormal modes in the AdS / CFT
  correspondence}},  {\em Phys. Rev.} {\bf D64} (2001) 064024,
  [\href{http://arxiv.org/abs/hep-th/0101194}{{\tt hep-th/0101194}}].

\bibitem{Cardoso:2001hn}
V.~Cardoso and J.~P.~S. Lemos, {\it {Scalar, electromagnetic and Weyl
  perturbations of BTZ black holes: Quasinormal modes}},  {\em Phys. Rev.} {\bf
  D63} (2001) 124015, [\href{http://arxiv.org/abs/gr-qc/0101052}{{\tt
  gr-qc/0101052}}].

\bibitem{Birmingham:2001pj}
D.~Birmingham, I.~Sachs, and S.~N. Solodukhin, {\it {Conformal field theory
  interpretation of black hole quasinormal modes}},  {\em Phys. Rev. Lett.}
  {\bf 88} (2002) 151301, [\href{http://arxiv.org/abs/hep-th/0112055}{{\tt
  hep-th/0112055}}].

\bibitem{vanRees:2009rw}
B.~C. van Rees, {\it {Real-time gauge/gravity duality and ingoing boundary
  conditions}},  {\em Nucl. Phys. Proc. Suppl.} {\bf 192-193} (2009) 193--196,
  [\href{http://arxiv.org/abs/0902.4010}{{\tt arXiv:0902.4010}}].

\bibitem{Banados:1992wn}
M.~Banados, C.~Teitelboim, and J.~Zanelli, {\it {The Black hole in
  three-dimensional space-time}},  {\em Phys. Rev. Lett.} {\bf 69} (1992)
  1849--1851, [\href{http://arxiv.org/abs/hep-th/9204099}{{\tt
  hep-th/9204099}}].

\bibitem{Banados:1992gq}
M.~Banados, M.~Henneaux, C.~Teitelboim, and J.~Zanelli, {\it {Geometry of the
  (2+1) black hole}},  {\em Phys. Rev.} {\bf D48} (1993) 1506--1525,
  [\href{http://arxiv.org/abs/gr-qc/9302012}{{\tt gr-qc/9302012}}]. [Erratum:
  Phys. Rev.D88,069902(2013)].

\bibitem{Klebanov:1999tb}
I.~R. Klebanov and E.~Witten, {\it {AdS / CFT correspondence and symmetry
  breaking}},  {\em Nucl. Phys.} {\bf B556} (1999) 89--114,
  [\href{http://arxiv.org/abs/hep-th/9905104}{{\tt hep-th/9905104}}].

\bibitem{deHaro:2000vlm}
S.~de~Haro, S.~N. Solodukhin, and K.~Skenderis, {\it {Holographic
  reconstruction of space-time and renormalization in the AdS / CFT
  correspondence}},  {\em Commun. Math. Phys.} {\bf 217} (2001) 595--622,
  [\href{http://arxiv.org/abs/hep-th/0002230}{{\tt hep-th/0002230}}].

\bibitem{Herzog:2007ij}
C.~P. Herzog, P.~Kovtun, S.~Sachdev, and D.~T. Son, {\it {Quantum critical
  transport, duality, and M-theory}},  {\em Phys. Rev.} {\bf D75} (2007)
  085020, [\href{http://arxiv.org/abs/hep-th/0701036}{{\tt hep-th/0701036}}].

\bibitem{WitczakKrempa:2013ht}
W.~Witczak-Krempa and S.~Sachdev, {\it {Dispersing quasinormal modes in 2+1
  dimensional conformal field theories}},  {\em Phys. Rev.} {\bf B87} (2013)
  155149, [\href{http://arxiv.org/abs/1302.0847}{{\tt arXiv:1302.0847}}].

\bibitem{Andrade:2015hpa}
T.~Andrade, S.~A. Gentle, and B.~Withers, {\it {Drude in D major}},  {\em JHEP}
  {\bf 06} (2016) 134, [\href{http://arxiv.org/abs/1512.06263}{{\tt
  arXiv:1512.06263}}].

\bibitem{Brattan:2010pq}
D.~K. Brattan and S.~A. Gentle, {\it {Shear channel correlators from hot
  charged black holes}},  {\em JHEP} {\bf 04} (2011) 082,
  [\href{http://arxiv.org/abs/1012.1280}{{\tt arXiv:1012.1280}}].

\end{thebibliography}\endgroup

 \end{document}